\documentclass[twoside,12pt]{article}
\usepackage{epsfig}

\newcommand{\be}{\begin{equation}}
\newcommand{\ee}{\end{equation}}
\newcommand{\bea}{\begin{eqnarray}}
\newcommand{\eea}{\end{eqnarray}}

\begin{document}

\title{ \vspace{1cm} Geo-neutrinos}

\author{G.\ Bellini,$^1$ A.\ Ianni,$^2$ L.\ Ludhova,$^1$ F.\ Mantovani,$^3$ W.F.\ McDonough$^4$\\
\\
$^1$Dip. di Fis. dell'Universit\`a degli Studi and INFN, Milano, Italy\\
$^2$INFN, Laboratori Nazionali di Gran Sasso, Italy\\
$^3$Dip. di Fis. e Sc. della Terra dell'Universit\`a degli Studi and INFN, Ferrara, Italy\\
$^4$Department of Geology, University of Maryland, College Park, USA\\
}
\maketitle
\begin{abstract} 
We review a new interdisciplinary field between Geology and Physics: the study of the Earth's geo-neutrino flux.  We describe competing models for the composition of the Earth, present geological insights into the make up of the continental and oceanic crust, those parts of the Earth that concentrate Th and U, the heat producing elements, and provide details of the regional settings in the continents and oceans where operating and planned detectors are sited. Details are presented for the only two operating detectors that are capable of measuring the Earth's geo-neutrinos flux: Borexino and KamLAND; results achieved to date are presented, along with their impacts on geophysical and geochemical models of the Earth. Finally, future planned experiments are highlighted.  
\end{abstract}

\section{Introduction}
\label{Sec:Intro}

	Geo-neutrinos are electron antineutrinos that come from radioactive decays in the Earth's interior. Their sources are natural $\beta^{-}$-decays of nuclides including the three most heat producing elements, $^{238}$U and $^{232}$Th families, and $^{40}$K, following the scheme:
\begin{equation}
^{238}\mathrm{U} \rightarrow ^{206}\mathrm{Pb} + 8\alpha + 8 e^{-} + 6 \bar{\nu}_e + 51.7 ~~~\mathrm{MeV} 
\label{Eq:geo1}
\end{equation}
 \begin{equation}
^{232}\mathrm{Th} \rightarrow ^{208}\mathrm{Pb} + 6\alpha + 4 e^{-} + 4 \bar{\nu}_e + 42.7 ~~~\mathrm{MeV} 
\label{Eq:geo2}
\end{equation}
 \begin{equation}
^{40}\mathrm{K} \rightarrow ^{40}\mathrm{Ca}  +  e^{-} +  \bar{\nu}_e + 1.31 ~~~\mathrm{MeV}
\label{Eq:geo3}
\end{equation}

\begin{figure}[tb]
\begin{center}
\centering{\epsfig{file=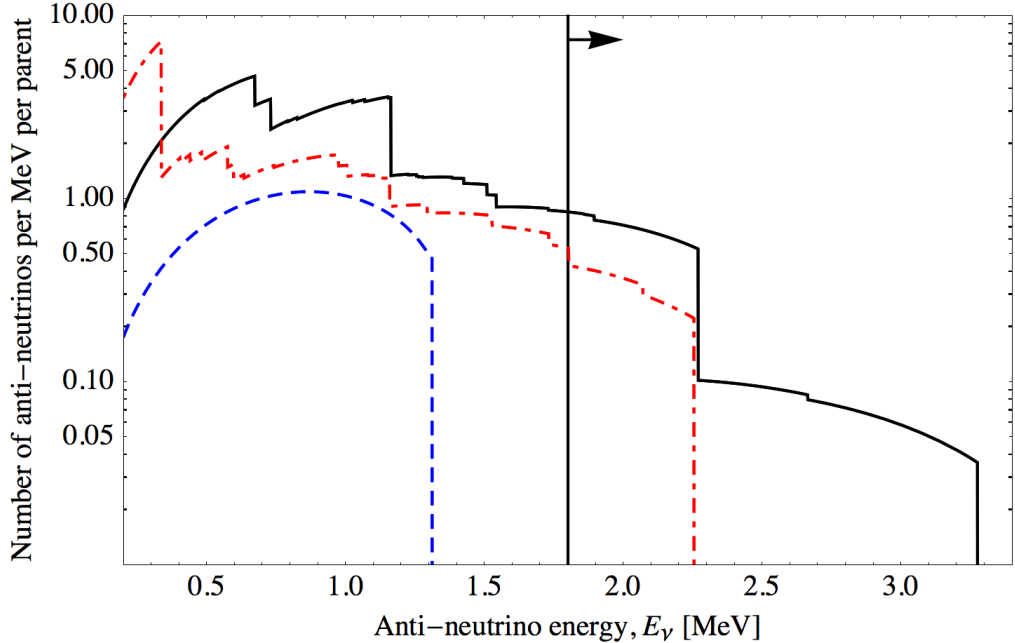, angle = -90,scale=0.5}}
\begin{minipage}[t]{16.5 cm}
\caption{Energy spectra of geo-neutrinos released in the reactions Eq.~\ref{Eq:geo1} ($^{238}$U chain, solid black line), Eq~\ref{Eq:geo2} ($^{232}$Th chain, dashed-dotted red line), and Eq.~\ref{Eq:geo3} ($^{40}$K, dashed blue line). The vertical dashed line shows the kinematic threshold (1.806 MeV) of the inverse beta decay interaction.
\label{Fig:GeonuSpectrum}}
\end{minipage}
\end{center}
\end{figure}

The geo-neutrino flux and the radiogenic heat, released during radioactive decays, are in a well-fixed ratio. Therefore, by measuring the total geo-neutrino flux, it is possible in principle, to determine the contribution of radiogenic heat released in radioactive decays, quoted in Eqs.~\ref{Eq:geo1}, \ref{Eq:geo2}, and \ref{Eq:geo3}, to the total terrestrial surface heat flux ($\sim$46\,TW).  The energy spectra of geo-neutrinos released in these reactions are shown in Fig.~\ref{Fig:GeonuSpectrum}. The U, Th, and K spectra are normalized to 6, 4, and 1 antineutrino, respectively, according to Eqs.~\ref{Eq:geo1}, \ref{Eq:geo2}, \ref{Eq:geo3}.

Even though the geo-neutrino flux at the Earth surface is some $10^6$\,cm$^{-2}$\,s$^{-1}$, their detection is challenging as antineutrinos interact with matter only through the weak interaction, thus the probability of such interactions, and possible detection, is very small. The cross section of the main detectable interaction of electron-flavor antineutrinos, the inverse beta decay interaction:
\begin{equation}
\bar{\nu}_e + p \rightarrow e^+ + n,
\label{Eq:InvBeta}
\end{equation}
is $3.3 \times 10^{-44}$\,cm$^2$ at 2\,MeV~\cite{strumia} and increases by about an order of magnitude at 3\,MeV.  The kinematic threshold of this interaction is 1.806\,MeV. Thus, the geo-neutrinos produced in the $^{40}$K decays cannot be detected, as their end-point energy spectrum of $\sim$1.31\,MeV (see Fig.~\ref{Fig:GeonuSpectrum}) is below this level.  A fraction of antineutrinos from $^{232}$Th decay chain with end-points energies of 2.1\,MeV ($^{228}$Ac) and 2.3\,MeV ($^{212}$Bi) and those from $^{238}$U with end-points 1.9, 2.7, and 3.3\,MeV ($^{214}$Bi) and 2.2 ($^{234}$Pa$^{m}$) can be detected via reaction in Eq.~\ref{Eq:InvBeta}.

Geo-neutrinos travel almost undisturbed through the Earth with a finite, albeit small, probability to interact in scintillation liquids contained in kiloton-scale detectors installed in underground laboratories. Therefore, these particles are unique direct probes, which bring information from the Earth's internal regions not accessible by any other means. Importantly, the measured flux at a detector is proportional to the abundance and distribution of U and Th in the Earth, critical inputs for many geological, geophysical, and geochemical models that describe the complex processes taking place inside the Earth. 

The geo-neutrino signal is especially useful for providing insights into the radiogenic power of the deep mantle, which is not directly obtainable from other methods. Some information on the chemical composition of the upper mantle can be obtained from samples brought to the surface through volcanic and tectonic processes. The chemical composition of such samples, however, can also be altered during their transport, particularly so for mobile elements like K, Th and U. The lower mantle is completely inaccessible by means of direct sampling. A systematic study of geo-neutrinos can provide constraints on a broad range of questions in Earth Sciences, including defining the energy available to drive plate tectonics, critically testing compositional models of the present-day mantle, determining the contribution of the radiogenic heat to the total terrestrial surface heat flux, providing insights into the power generating the geo-dynamo, the Earth's magnetic field, and establishing the relative abundance of Th and U in the silicate Earth and its Th/U ratio, as compared to that in some meteorites (useful contribution for the understanding of the Solar system and the Earth formation).  As it is understood, observations of the chemical behavior of Th and U over a wide range of conditions inside the Earth are consistent with radioactive elements being absent from the Earth's core. However, some authors~\cite{herndon} suggest the existence of a georeactor active in the Earth's central inner core and such theories can also be tested by means of detecting electron antineutrinos.

The very low neutrino cross section and the geo-neutrino relatively low energy require that the detectors have special properties, such as a large size and a very low radioactive background. These requirements necessitate advanced technologies and considerable efforts to achieve detection success. Therefore the experiments capable to study the Earth's geo-neutrino flux are few and it is not easy to set up other detectors in different regions of the world. Such detectors are placed in underground laboratories to shield the experimental setup from cosmic radiations, which can mimic antineutrino interactions

The results of geo-neutrino measurements can be expressed in several ways. One such expression is the normalized event rate. It can be expressed using the so called Terrestrial Neutrino Unit (TNU), which is the number of antineutrino events detected during one year on a target of  $10^{32}$ protons ($\sim$1\,kton of liquid scintillator) and 100\% detection efficiency. Conversion between the signal $S$ expressed in TNU and the oscillated, electron flavor flux $\phi$ is straightforward and requires a knowledge of the geo-neutrino energy spectrum and the interaction cross section, which scales with the energy of the electron antineutrino:
 \begin{equation}
S(^{232}\rm{Th}) [\rm{TNU}] = 4.07 \times \phi (^{232}\rm{Th})~~[10^6  \rm{cm}^{-2} \rm{s}^{-1}]
\label{Eq:TNUFluxTh}
\end{equation}
 \begin{equation}
S(^{238}\rm{U}) [\rm{TNU}] = 12.8 \times \phi (^{232}\rm{Th})~~ [10^6  \rm{cm}^{-2} \rm{s}^{-1}]
\label{Eq:TNUFluxU}
\end{equation}

This paper provides a state-of-the-art perspective on the new interdisciplinary field of Neutrino Geoscience, which involves, bringing together, communities of Earth scientists and particle physicists. In Sec.~\ref{Sec:GeoModels} of this paper geological and geophysical models of the Earth are reviewed; in Sec.~\ref{Sec:GeoSignal} we describe the models of the continental and oceanic crusts and the expected geo-neutrino signal at the sites where the experiments are placed; in Sec.~\ref{Sec:detectors} the presently running geo-neutrino detectors are presented; in Sec.~\ref{Sec:Results} the results already achieved and their impact on the Earth models are discussed; and finally in Sec.~\ref{Sec:Future} future experiments are highlighted.

\section{Models of the Earth}
\label{Sec:GeoModels}

This paper examines the nature of geo-neutrinos and how they relate to the Earth, its composition and energy budget, and how they define the power available to drive the Earth's engine.  Geo-neutrinos are electron antineutrinos that are naturally emitted during beta decays and their detection can in principle tell us about the amount of thorium and uranium inside the Earth.  In turn, measuring the Earth's flux of geo-neutrinos will constrain the nature of materials that were available to construct the planet some 4.5\,billion years ago at one astronomical unit (AU) out from the Sun.  

Collaboration between geologists and physicists has at least a 150\,year history that has yielded exciting new science discoveries.  When Lord Kelvin began to consider the age of the Earth and the rate of heat dissipation from the planet, a constrained but unsolved problem, he posed the problem as one of simple conductive dissipation of heat.  A central question in geology today is what proportion of the present day surface heat flux of the Earth is due to radiogenic heating and how much is due to the release of primordial heat left over from accretion and core differentiation?

The assembly of the Earth from the initial solar nebular involved the accretion of many cosmic gas-dust fragments that accreted into planetesimals and into an ever increasing hierarchical accumulation of mass.  There was a considerable amount of accretional energy that accompanied planet assembly. Likewise, the settling of metal into the center of the Earth during core formation involves the accumulation of gravitational energy that is later dissipated as thermal energy. Collectively, this formative period of the Earth produces a highly energetic thermal state whereby heat dissipation is regulated by the structure of the Earth involving a thermally conductive, metallic core surrounded by an insulating oxide shell.  

Understanding the age of the Earth from the perspective of a simple cooling solid is not an accurate reference frame, as was assumed by Lord Kelvin. The dissipation of heat from the Earth is in large part controlled by heat loss across a thermal boundary layer that surrounds at least two convective shells, the plastically deforming convecting mantle and the liquid outer core, shells with markedly different thermal properties.  As later recognized by Ernest Rutherford, radiogenic heating plays an additional, albeit minor role in this story. Thus, to understand the Earth's energy budget requires an assessment of the relative contributions of primordial and radiogenic heat and defining of the rate of heat dissipation.

Fortunately, geo-neutrino studies offer yet another opportunity for a fruitful collaboration between geology and physics to address the issue of how much radiogenic heat is contained in the Earth.  Early results from this field are overwhelmingly positive, in that they define the Earth as containing a complement of both primordial and radiogenic power, and are beginning to resolve the absolute amount of radiogenic power inside the Earth.  This latter information is important for critically evaluating models regarding the building blocks available to construct the Earth. The new, interdisciplinary field of Neutrino Geoscience is now providing critical insights into the bulk composition of the Earth, as well as the energy to power mantle convection, plate tectonic and geodynamo. Additional recent reviews of this field, particularly from the perspective of the geological inputs and how different models of the Earth influence the prediction of detected geo-neutrino signal, are also available~\cite{Sramek2013, Dye2012}.

\subsection{The origin of the Earth}
\label{SubSec:Origin}

There is considerable debate surrounding discussions on the age, origin and composition of the Earth. Increasingly, as we gain observational information from stellar nurseries, accretion disks, extra-solar planetary systems and meteorites, we are presented with a variety of potential processes and materials that can be envisaged for the building blocks and processes involved in planetary growth.  

\subsubsection{Chondrites: the building blocks of the planets}
\label{SubSubSec:Chondrites}

Today questions remain about whether or not the Earth has a chondritic composition, and if so, which of the chondrites were the essential building block of the Earth.  Chondrites are primitive, undifferentiated meteorites (i.e., a chaotic assemblage of rock and metal) that are a collection of the earliest formed material in the solar system.  Studies of meteorites add much to our understanding of the age of the solar system and the nature of the building blocks that makes up the planets. The earliest formed fragments of the solar system (i.e., calcium-aluminum inclusions), found in some of these chondrites, are high temperature ceramic grains up to a cm across and represent nebular condensates, which experienced a series of post-formational histories that include chemical exchange in the nebular between gas and condensate, shock heating and melting events and oftentimes secondary mineral formation.  From these fragments and other observations we know that the solar system formed 4.568 billion years ago~\cite{bouvier} due to the gravitational collapse of a portion of a large molecular cloud that formed a pre-solar nebula, a rotating and collapsing disk where much of the mass collects into the central portion that becomes increasingly hotter. Further out from the center of the disk the planets formed by collisional accretion and gravitational attraction, ultimately coalescing to form ever larger bodies.  

The how and when planets formed remains a subject of considerable debate.  Did Jupiter form first and other planets later?  Did the inner rocky and outer gas planets formed simultaneously or sequentially?  The Grand Tack model~\cite{walsh} envisages Jupiter and the outer gas giants as having accreted in the first few million years of the solar system: these bodies then moved inward towards the Sun, perturbed the region of the inner solar system, which lead to the formation of the inner rocky bodies, and later these gas giants returned to their more distant positions, after having gravitationally filtered this inner pool of planetesimals and initiating accretion of the rocky planets. Theoretical studies envisage the hierarchical growth of the inner rocky planets from the aggregation of planetesimals into planets with the exchange of different components (refractory versus volatile components, metals versus silicates, etc) across different heliocentric zones. These processes can be modeled and the results having attributes consistent with the compositional heterogeneities seen in the inner solar system and with acceptable solutions for the distribution of planetary masses~\cite{chambers}.  

\begin{figure}[tb]
\begin{center}
\centering{\epsfig{file=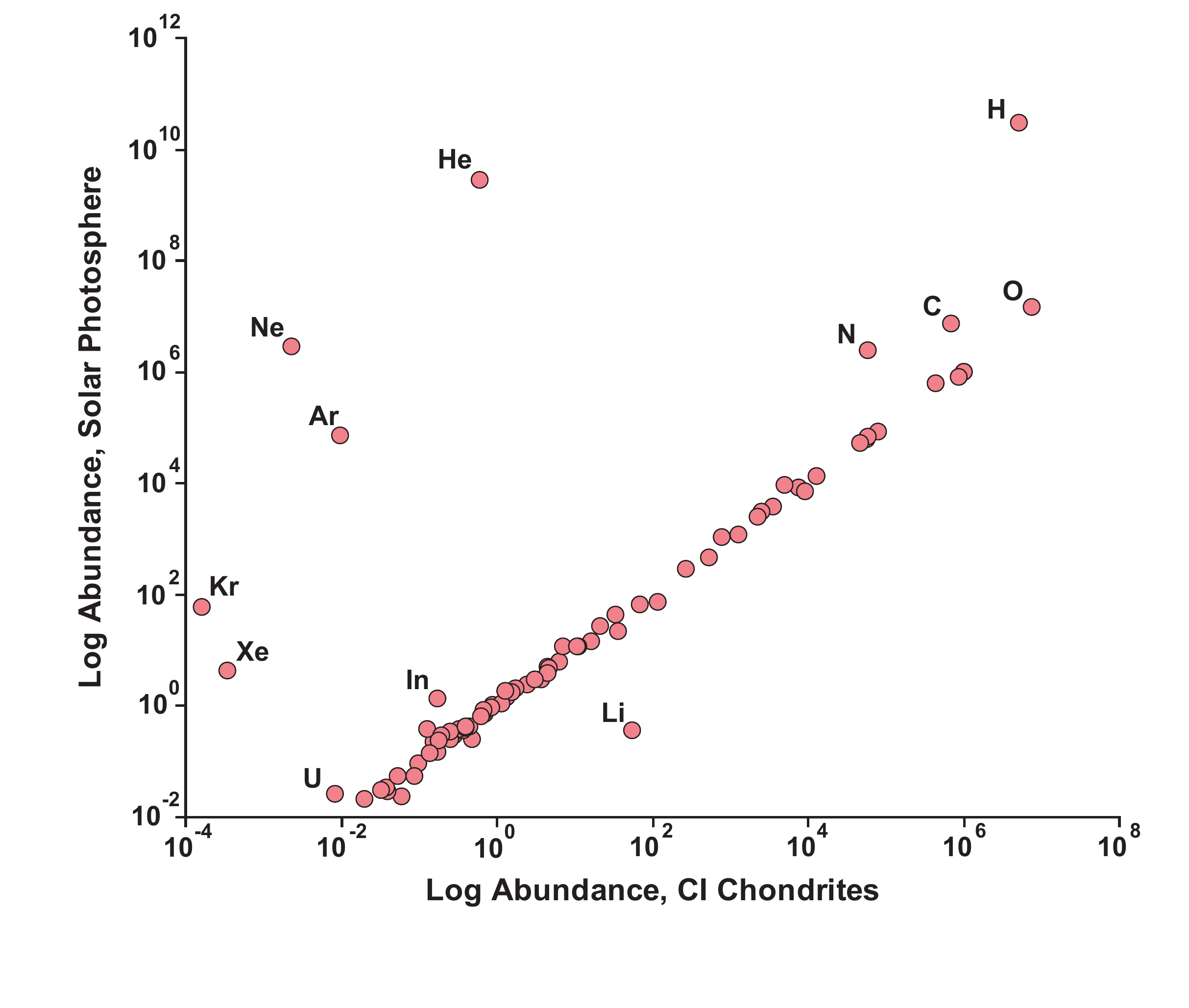,scale=0.5}}
\begin{minipage}[t]{16.5 cm}
\caption{Comparison of the composition of the solar photosphere and that of C1 carbonaceous chondrites. Mass abundance data, $A$(El) [kg/kg], from Palme and Jones~\cite{palme};  the number of atoms of an element, $N$(El) [kg$^{-1}$], was derived for chondrites to be $N$(El) = $10\exp{ (A(\rm{El}) - 1.55)}$ and for the solar photosphere to be $N$(El) = $10 \exp {(A(\rm{El}) - 1.54)}$. Both derived values are normalized, that is the number of silicon atoms is $N$(Si) = $10^6$, consistent with all other presentations of this data. Augmented data, when necessary, for the solar photosphere came from Lodders~\cite{lodders}, which was given in the form $N$(El) and did not have to be derived from $A$(El) data, while that for CI chondrites came from Asplund~\cite{asplund}, whose derived equation was $N$(El) = $10\exp{(A(\rm{El}) - 1.51)}$, which was normalized to $N$(Si) = $10^6$.
\label{Fig:SunCI}}
\end{minipage}
\end{center}
\end{figure}

Chondritic meteorites are a mixture of silicate and metal materials in proportions similar to that found in the terrestrial planet (i.e., the mass fraction of the metallic core and rocky shell of the mantle and crust).  It has also been shown that the compositions of the most primitive of the chondritic meteorites, the C1 carbonaceous chondrite, matches that of the solar photosphere~\cite{lodders}, the outer "surface" layer of the Sun, and that this match ranges over five orders of magnitude in element concentration (Fig.~\ref{Fig:SunCI}).  The comparison with the solar photosphere is significant, given that the mass of the solar system is the Sun, with Jupiter being a thousand times smaller than it and the Earth being another thousand times smaller still. Thus, chondrites represent a guide to the building blocks of the solar system.

\begin{figure}[tb]
\begin{center}
\centering{\epsfig{file=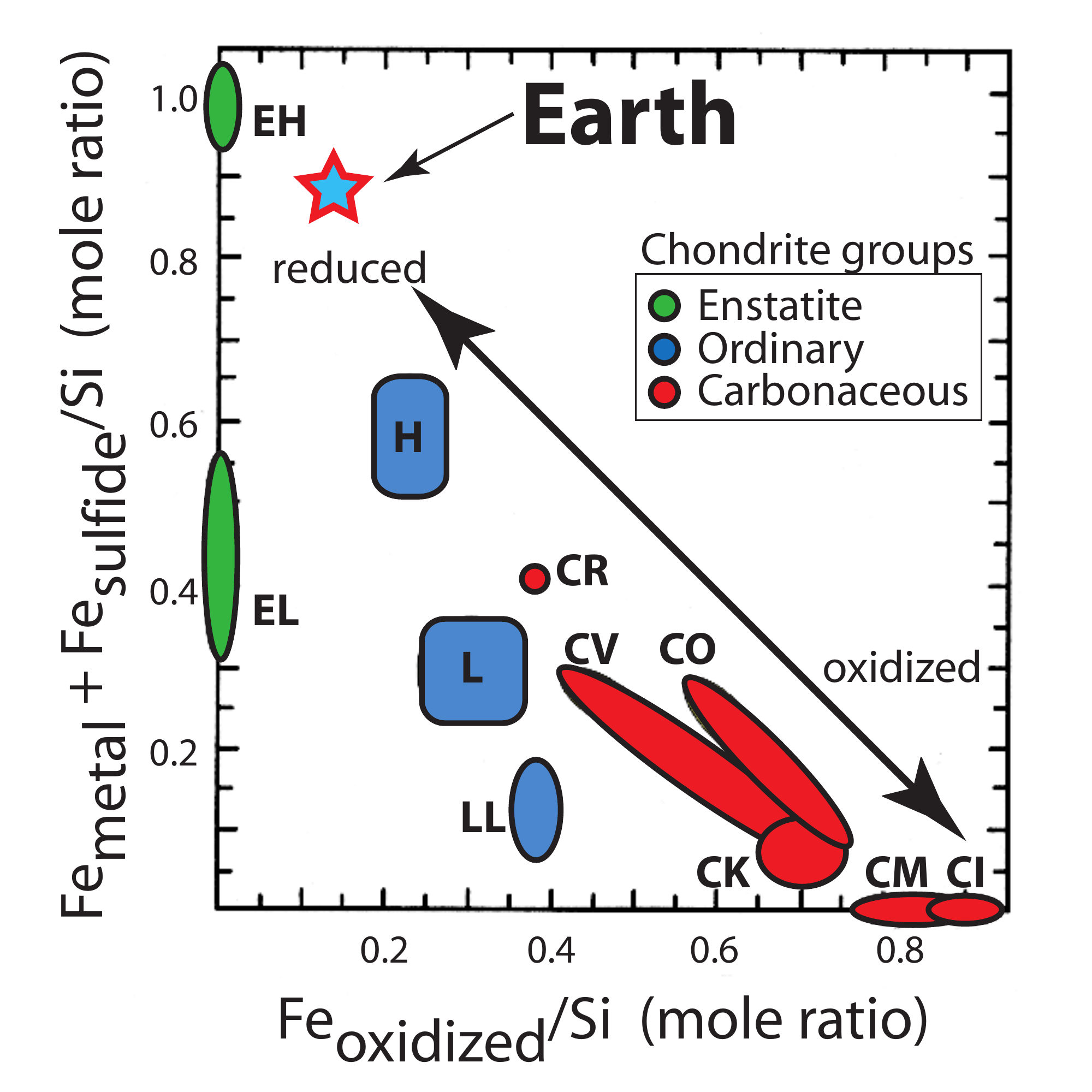,scale=0.5}}
\begin{minipage}[t]{16.5 cm}
\caption{A Urey-Craig diagram that separates chondritic meteorites according to the ratio of oxidized Fe to silicate ($x$-axis) relative to the ratio of reduced iron (and that in sulfide) to total silicon.  The Earth has a large metallic core and thus its bulk composition plots close to reduced end of the redox scale.
\label{Fig:UreyCraig}}
\end{minipage}
\end{center}
\end{figure}

There are a various types of chondritic meteorites, with their classification being established on petrographic grounds that relate to their oxidation state of iron, their amount of iron, and metamorphic grade (or degree of aqueous alteration) of their constituent minerals.  There are several varieties of chondritic meteorites, but the three dominant groups include the carbonaceous chondrites (e.g., with sub-groups labeled CI, CM, CV, CO, CR, CK, with distinctions being due to mineral attributes), the ordinary chondrites (i.e., H, L and LL, distinguished by their high, low and very low iron contents) and the enstatite chondrites (i.e., EH and EL, and again distinguished by their high and low iron contents) and these varieties are notable for their redox state of iron, being oxidized, intermediate and reduced, respectively.  This redox state of iron also couples to a number of chemical and isotopic attributes of these chondrites~\cite{warren}, where iron is in the oxide form and associated with silicates (the major mineral family) or iron is in the reduced or sulfide form, with enstatite chondrites having negligible iron in the silicates (Fig.~\ref{Fig:UreyCraig}).  A simple compositional model for the Earth (red star Fig.~\ref{Fig:UreyCraig}) compared to chondrites is plotted given the mass of the core and the amount of iron in the core and the Bulk Silicate Earth (i.e., the BSE is the crust plus the mantle, which is the primitive undifferentiated silicate fraction of the Earth, after core subtraction).

\subsubsection{Element behavior: differentiation and redox potentials in the disk and planets}
\label{SubSubSec:Elements}

Importantly for the Earth, the redox state of iron controls the size of the core, which is 1/3 the mass of the planet.  Moreover, the differentiation of the Earth into the core, silicate Earth (mantle plus crust) and hydrosphere/atmosphere, appears to have been a relative early Earth process (i.e., mostly completed in the first 50\,million years or thereabouts). Evidence for the timing of core formation comes from the short-lived $^{182}$Hf-$^{182}$W isotope system ($\beta^-$, with $t_{1/2}$ = 8.9\,million years), with both chondrites and iron meteorites having distinctly lower $^{182}$W/$^{184}$W isotopic compositions than the silicate Earth, implying a young formation age (order ten or a few tens of million of years after the formation of the solar system), which is based on the number and timing of separation steps of extracting W into the core and leaving Hf in the silicate Earth~\cite{kleine, yin}.  Although the models differ on the exact timing and the number of multi-stage steps involved in its evolution, there is increasing consensus that core formation effectively occurred as early as 11\,million years after solar system formation or as late as $\sim$4.50\,Ga, with the latter being constrained by the ages of the earlier minerals on Earth and the time of Moon formation, which created a Moon having an identical isotopic composition as that of the Earth. 

The segregation of iron (and Ni) into core is the single most important chemical differentiation event that occurred on Earth and it established to a first degree the distribution of elements in the planet.  Historically, Emil Wiechert, a German physicist and geophysicist, envisaged the first order structure of the Earth in 1897, with the Earth having a metallic core surrounded by a silicate shell. By 1913, his PhD student Beno Gutenberg defined the depth to the core mantle boundary at 2900 km, roughly the same depth to which it is known today (2893 $\pm$ 5\, km).  With this perspective, coupled with the planet's moment of inertia and an understanding of meteorites, scientists compared the Earth to the heavenly objects falling upon it, the chondrites.  However, these comparisons lead to much speculation about the appropriate analogs of planets.

At a simple level the bulk of the Earth can be described by four elements (i.e., O, Fe, Si and Mg), which make up about 93\% by mass of the planet. In combination with Al, Ca and Ni, these seven elements describe more than 98\% of the mass of the Earth and thus define the bulk of the system.  Of these seven elements, Ca and Al play a crucial role and their abundances can be directly linked to that of Th and U, as these four elements, along with some thirty other elements are considered refractory elements.  

The refractory elements are those elements that condense out of a nebular disk at high temperatures and empirically are observed in equal proportion in the chondrites.  Thus, chondritic ratios are conserved for the refractory elements, whereas the relative abundances of the other five abundant elements in the Earth (i.e., O, Fe, Si, Mg, Ni) and the remaining non-refractory elements vary markedly between different types of chondrites.  Consequently, if we can establish the absolute abundance of Th and U in the planet, we can use chondritic ratios of refractory elements to set their abundances and from that model the remaining abundances of the other elements.  Importantly, if we could determine the absolute abundances of potassium, a moderately volatile element, in the Earth, we could establish the volatility curve for the planet.  (Geo-neutrinos from $^{40}$K have energies below the kinematic threshold of the current detection interaction, the inverse beta decay.)

\begin{figure}[tb]
\begin{center}
\centering{\epsfig{file=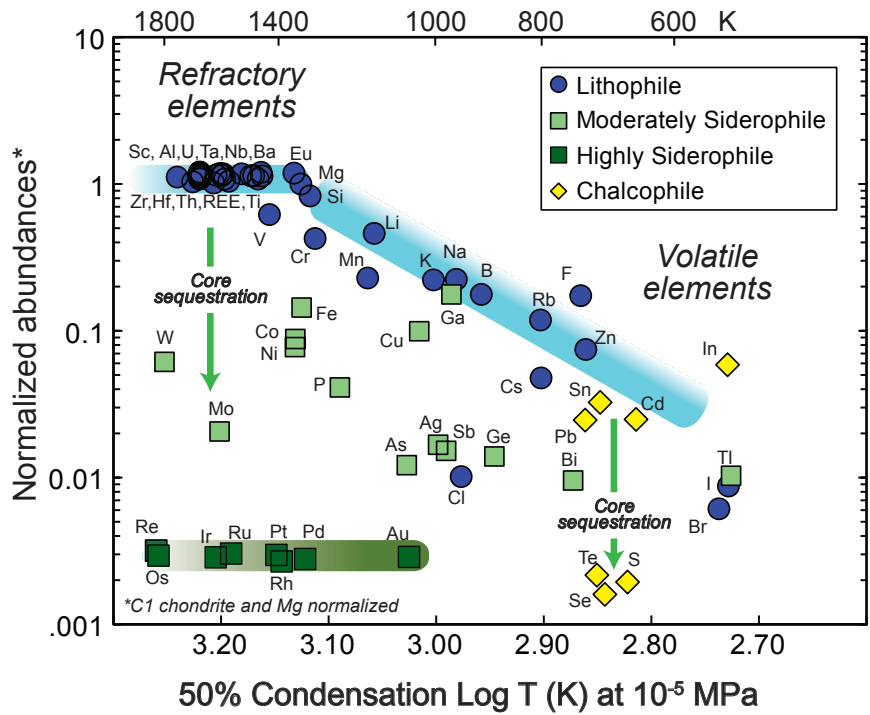,scale=0.5}}
\begin{minipage}[t]{16.5 cm}
\caption{The abundances of elements in the silicate Earth (i.e., the primitive mantle, which was then differentiated to the present-day crust and mantle) divide by their relative abundances in C1 carbonaceous chondrite and plotted against the half-mass condensation temperatures for a gas density in the solar system's nebular disk at approximately 1\,AU.  See~\cite{mcdonough} for details. 
\label{Fig:Volatility}}
\end{minipage}
\end{center}
\end{figure}

Collectively, the abundance of the volatile lithophile elements establishes the planet's volatility scale and provides a constraint on the time-integrated nature of material accreted at\,1 AU.  Relative to C1 carbonaceous chondrites the Earth is strongly depleted in volatile elements (Fig.~\ref{Fig:Volatility}).  Not only does this include the ices (i.e., compounds of H, C, N and O), but also included are the alkali metals, S, and other non-refractory elements.  The short-lived (3.7\,million years) radioactive system $^{53}$Mn, which decays to $^{53}$Cr, has been used to document that the Earth's volatile depletion signature, like that of various meteorites, was established within two million years of the start of the solar system~\cite{shukolyukov,moynier,trinquier}. Thus, the Earth's building blocks were likely volatile depleted and so much so that we do not have an analogous example among the chondritic meteorites.  

In addition to considering the behavior of elements (i.e., refractory versus volatile) in the nebular disk, geologists classify elements according to their geochemical affinities during geological processes, with elemental affinities cast according to partition functions with metal ({\it siderophile}), silicate ({\it lithophile}), sulfides ({\it chalcophile}), and water and gases ({\it atmophile}).  Hence, inventories of siderophile elements are stored in the Earth's core, with minor amounts in the mantle, while the chalcophile elements were divided between the core and mantle\cite{mcdonough}. Core formation was likely protracted over one to a few million year time scale and it occurred over a range of conditions, but, on average, appears to have been established at mid-mantle pressure and temperature and a dominantly reduced oxygen fugacity~\cite{oneil,rubie,wood}.  This scenario of core formation is derived from combining data on the absolute and relative abundances of elements in the silicate Earth and in chondrites, with experimental observations that establish the thermodynamic behavior of elements in analog material at controlled pressure, temperature and gas fugacity conditions in the laboratory. The depletion of siderophile and chalcophile elements in silicate Earth is accounted for by their sequestering into the core. To different degrees the siderophile and chalcophile elements have dissimilar geochemical affinities, as can be observed (Fig.~\ref{Fig:Volatility}) from the distinctive depletions of the two refractory elements Mo and W and the highly siderophile noble metals.

The lithophile elements, those that partition into silicates and other oxides, are excluded from the core forming metals due to their chemical affinities for oxygen. However, these potentials are established by the ambient oxygen fugacity at the time of metal silicate equilibrium and it is possible that during core formation some nominally lithophile elements may have been reduced to their metallic state. Therefore, a fraction of the inventory of some lithophile elements may be found in the core.  Experiments that mimic a range of core forming conditions even in the presence of sulfide bearing metal find that U has negligible affinities for a core forming metal~\cite{wheeler}; the conditions for forcing Th into a metal phase are more extreme than that for U~\cite{jones} and thus even less likely to be in the core. Although debate surrounds the potential for Th and/or U being partitioned into the core~\cite{malvergene}, three significant observations are inconsistent with such assertions, given the Earth has a chondritic Th/U value of 3.9 $\pm$ 0.3: (1) the average mantle Th/U ratio is $\sim$3 based on ocean island basalts~\cite{arevalo2013} and mid-ocean ridge basalts~\cite{arevalo2009,arevalo2010}, (2) the continents, the complementary reservoir to the mantle, has an average crustal Th/U ratio of between 4 and 5, based on studies of crustal rocks~\cite{rudnick, huang}, and (3) the time integrated $\kappa$ value (a measure of the $^{232}$Th/$^{238}$U from the slope of $^{208}$Pb/$^{206}$Pb) of mantle and crustal rocks is $\sim$4~\cite{galer,elliot,paul}.

\subsection{Structure of the Earth}
\label{SubSec:Structure}

The Earth is a differentiate planet made up of three shells:  a metallic core, overlain by a rocky layer of mantle and crust, which is in turn surrounded by an outermost fluid layer of hydrosphere and atmosphere (see Table~\ref{tab:Earth}) for further details).  This structure is a consequence of the physical and chemical processes that occurred early in Earth's history, particularly with an initial core formation. The fundamental result of planetary differentiation is that element distribution in the Earth is not random, but controlled by a combination of chemical and physical potentials.  Although dense iron is at the Earth's gravitational center, other heavy elements like uranium and thorium are concentrated upwards in the mantle and more so in the continental crust due to their chemical properties.

\begin{table}
\begin{center} 
\begin{minipage}[t]{16.5 cm}
\caption{Properties of the Earth. Data from~\cite{yoder,masters,mcdonough, huang}.}
\label{tab:Earth}
\end{minipage}
\begin{tabular}{ll}
\hline
{\bf Radii} [m] & \\
Mean radius of the Earth	 &  6,371,010 $\pm$ 20 \\
Equatorial radius	& 6,378,138 $\pm$ 2 \\
Polar axis & 	6,356,752 \\ 
Inner (solid) core radius	& $(1.220 \pm10) \times 10^6$\\
Outer (liquid) core radius	& $(3.483 \pm 5) \times 10^6$ \\
\hline
{\bf Thickness} [m] & \\
Continental crust &	(34 $\pm$ $4) \times 10^3$\\
Oceanic crust	& (8.0 $\pm$ $2.7) \times 10^3$\\
\hline
{\bf Mass}  [kg] & \\
Earth & 	$5.9736 \times 10^{24}$ \\ 
Inner (solid) core &	$9.675 \times 10^{22}$\\ 
Outer (liquid) core & $1.835 \times 10^{24}$ \\ 
Core	& $1.932 \times 10^{24}$\\ 
Mantle	& $4.043 \times 10^{24}$ \\ 
Oceanic crust	& $(0.67 \pm 0.23) \times 10^{22}$ \\
Continental crust &	$(2.06 \pm 0.25) \times 10^{22}$\\ 
Bulk crust	 & $(2.73 \pm 0.48) \times 10^{22}$ \\
Ocean &	$1.4 \times 10^{21}$ \\
Atmosphere &	$5.1 \times 10^{18}$ \\ 
\hline
{\bf Fractional mass contributions } & \\
{\it -Bulk silicate Earth} & \\
Oceanic crust &	0.17\% \\
Continental crust &	0.51\% \\
Mantle & 99.32\% \\
{\it - Earth} & \\
Silicate Earth	& 67.7\% \\
Core & 	32.3\% \\
Inner core to bulk core & 	5.0\% \\
\hline
{\bf Volume}  [m$^3$] & \\
Earth	& $1.083 \times 10^{21}$ \\ 
Inner (solid) core &	$7.606 \times 10^{18}$ \\ 
Outer (liquid) core	 & $1.694 \times 10^{20}$ \\ 
Bulk core	& $1.770 \times 10^{20}$ \\ 
Bulk silicate Earth &	$9.138 \times 10^{20}$ \\ 
\hline
\end{tabular}
\end{center}
\end{table}

\subsubsection{Seismic data and mineral constitution of the mantle}
\label{SubSubSec:Seismic}

\begin{figure}[tb]
\begin{center}
\centering{\epsfig{file=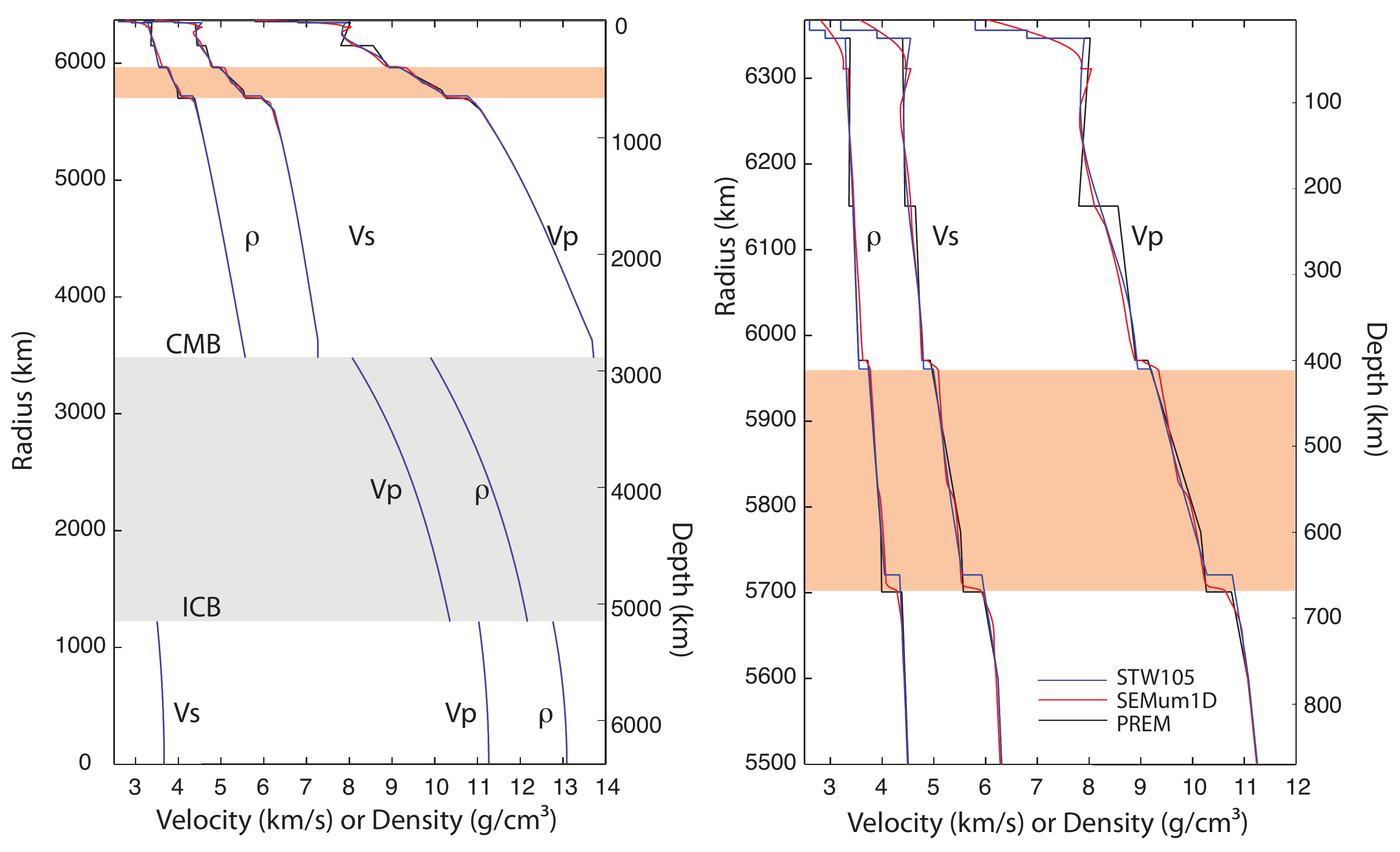,scale=0.5}}
\begin{minipage}[t]{16.5 cm}
\caption{The 1D seismological model of the Earth as reported in PREM: Preliminary Reference Earth Model~\cite{dziewonski}. $\rho$: density [g/cm$^3]$; $V_p$: velocity [km/s] of the primary, longitudinal waves; $V_s$: velocity [km/s] of the secondary, shear, transverse waves: note S-waves do not propagate through the liquid outer core; ICB: Inner Core Boundary, CMB: Core Mantle Boundary. The right panel, which focuses on the top 800\,km of the mantle, compares PREM data to other global models, STW105 from~\cite{kustowski}, shown in red, and SEMum 1D from~\cite{lekic}, shown in blue. The darker (orange) filled area, shown in both panels, delimited by the discontinuities at 410\,km and 660\,km depth, is the transition zone described in text.  
\label{Fig:PREM}}
\end{minipage}
\end{center}
\end{figure}

The first order structure of the Earth's interior is defined by the 1D seismological profile, called PREM: Preliminary Reference Earth Model~\cite{dziewonski}.  From the core outwards the Earth has a series of major, seismically defined features and discontinuities (Fig.~\ref{Fig:PREM}).  This seismic profile of the Earth describes its constitution and make up~\cite{birch}, given the equation of state of Earth materials at appropriate temperatures and conditions of the interior.  The seismic discontinuities or jumps are the result of mineralogical phase changes and/or crossing compositional boundaries.  

The most significant compositional boundary in the Earth is at the core-mantle boundary (Earth scientist's CMB) and defines the first discovered (circa 1907) and the most dramatic seismic discontinuity.  Here compressional wave velocities ($V_p$) drop substantially from the mantle values and then increase throughout the liquid outer core and solid inner core.  The absence of a shear wave in the outer core is consistent with it being liquid. By the 1920s seismologists~\cite{jeffreys} mapped out a series of discontinuities in the mantle and began discussing the nature of the seismic jumps as being either isochemical phase changes or compositional layering in the mantle.  

The major, seismically-defined, boundaries in the mantle are at 410\,km and 660 km\,depth and the D'' boundary near the CMB.  The D'' (pronounced D-double prime) layer is distinctively recognized as the region near the core mantle boundary (100-300\,km above the CMB) where there is a decrease in the vertical gradient of both compressional and shear wave velocities.  There is considerable community debate about the physical and chemical state of this irregularly shaped mantle domain, which is a thermal boundary layer between the hot core and cooler mantle.  Seismology provides an instantaneous picture of the Earth and so it is difficult to define the age of the D'' region.  There are several suggestions regarding the history of the D'' region, ranging from a long term (age of the Earth) or a shorter term ($10^8$ to $10^9$\,years) feature of the mantle that gets refreshed through time. The nature of the D'' region has been a source of great intellectual speculation, ranging from an early cumulate pile from an early Earth, global magma-ocean differentiation event~\cite{labrosse,lee}, to an early surface crust that was gravitationally sequestered to the base of the mantle and is representative of an Early Enriched Reservoir (EER from~\cite{boyet}), or is the final resting place for subducting slabs (surface tectonic plates) of recycled oceanic lithosphere (i.e., crust plus underlying mantle that is mechanically coupled to the plate).
	
\begin{figure}[tb]
\begin{center}
\centering{\epsfig{file=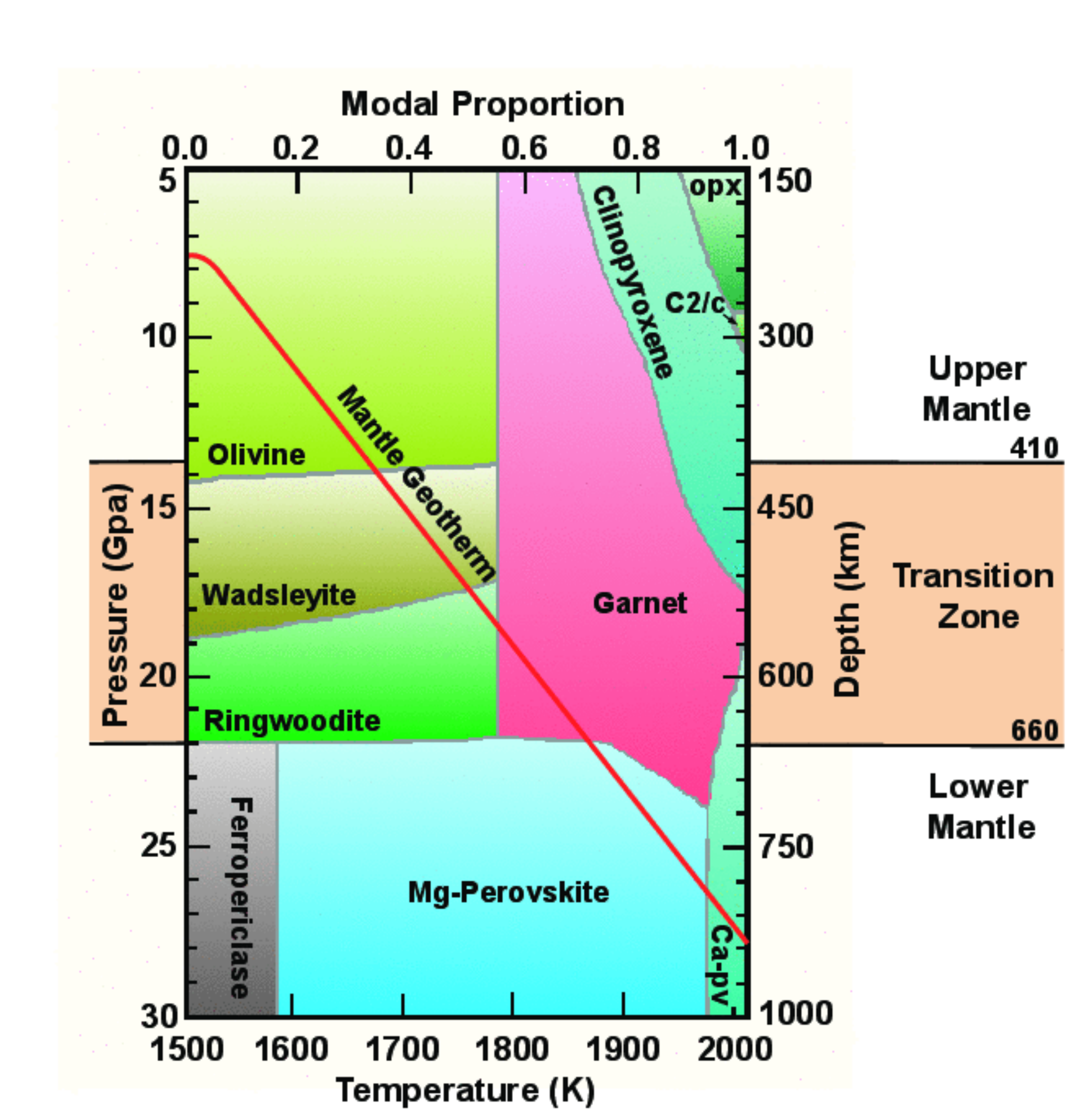,scale=0.5}}
\begin{minipage}[t]{16.5 cm}
\caption{A phase diagram of the mineralogy and thermal gradient of the top 1000\,km of the Earth's mantle.  The mode proportion of mantle minerals is presented on the top $x$-axis, whereas the temperature scale is shown on the bottom $x$-axis. The Transition Zone (see also Fig.~\ref{Fig:PREM}) is marked by the two major, seismic discontinuities at 410\,km and 660\,km, which are coincident with major phase changes.  Assuming a Mg to Fe mole proportion of 9 to 1 for the bulk composition, the olivine to wadsleyite transition occurs at $\sim$1670\,K and 410\,km and the disproportionation of ringwoodite to Mg-perovskite and ferropericlase occurs at $\sim$1870\,K and 660\,km~\cite{katsura, akaogi}.
\label{Fig:geotherm}}
\end{minipage}
\end{center}
\end{figure}

The 660\,km and 410\,km boundaries are well documented phases change boundaries (Fig.~\ref{Fig:geotherm}). It is uniformly agreed that the relatively sharp ($\leq$10\,km) 410\,km seismic boundary is due to the isochemical, positive Clapeyron slope ($dP/dT$), phase change of olivine to wadsleyite ($\alpha$-olivine to $\beta$-spinel structure, (Mg$_{0.9}$Fe$_{0.1}$)SiO$_4$), the dominant ($\geq$55 mole \%) mineral in the upper mantle. Consequently, the depth to the 410\,km seismic boundary appears to be thermally controlled with the phase transition occurring at shallower depth in cold regions (e.g., areas with subducting slabs) and deeper in higher temperature regions (e.g., upwelling hot mantle region, as in the Hawaiian plume).  

The 660\,km seismic discontinuity is a somewhat broader ($\leq$60\,km depth variation) transition that is also recognized as a phase change.  At these depth the bi-mineralic (majorite garnet and ringwoodite $\gamma$-spinel) assemblage breaks down into Mg-perovskite ((Mg$_{0.9}$Fe$_{0.1}$)SiO$_3$), ferropericlase ((Mg,Fe)O) and Ca-perovskite (CaSiO$_3$), with Al being distributed between the two perovskite phases and Fe$^{3+}$ distributed between all the phases.  At this depth the disproportionation of $\gamma$-spinel to Mg-perovskite and ferropericlase has a negative Clapeyron slope and so this transition should have an anti-correlation with that of the 410\,km discontinuity. Overall, however, there is little evidence for this anti-correlation~\cite{houser}.  There has been considerable debate in the community, however, regarding whether or not this phase change also defines a marked compositional change in the mantle. In large part models differ on the amount of ferropericlase in the lower mantle, or alternatively the difference in the amount of SiO$_2$ in the lower and upper mantle.  

The topography on the 410\,km and 660\,km discontinuities can be obtained from high resolution seismic images, which defines the mantle's Transition Zone thickness~\cite{lawrence}.  The average thickness of this zone is 242 $\pm$ 9\,km, with regions in the western Pacific being (also in the Red Sea to the Aegean region) as much as 35\,km thicker and up to 35\,km thinner in areas northwest of Hawaii and beneath Central Africa. Thinning and thickening of Transition Zone is mostly due to topography on the 660\,km discontinuity. There are regions with large amplitudes in boundary heights over narrow horizontal scales, which correlate with subducting slabs. Hot regions (e.g., upwelling plume, like Hawaii) of the mantle are correlated with anomalously thin transition zones and are also laterally narrow. Overall topography on the 410\,km and 660\,km discontinuities is generally correlated with temperature variations on small lateral scales (slabs and plumes).

There are two significant boundary layer structures in the Earth that are associated with the cooling of the Earth.  At the base of the mantle, the D'' layer is a structure that in part reflects the conductive thermal boundary between the hotter core and cooler mantle.  At the top of the mantle, the lithosphere is the mechanical plate that translates with mantle convection and its outermost conductive cooling layer.  The oceanic lithosphere, made of oceanic crust (8 $\pm$ 2\,km) and its subjacent lithospheric mantle (up to 80\,km thick), forms at mid-ocean spreading centers where adiabatic decompression leads to melting, crust production and basal accretion of residual mantle. Later and further afield from the spreading ridge, the base of this lithosphere continues to accrete ambient mantle due to conductive cooling processes.  The lithosphere beneath the continents is made up on average of 34\,km of crust underlain by lithospheric mantle that is estimated to reach down to 175 $\pm$ 75\,km~\cite{huang}.  Fragments of the deep lithosphere beneath are brought up in selected magma types and demonstrate that the roots of continents have comparable age distributions as those seen in their overlying crust. On average the oceanic lithosphere is about 50\,million years old (and everywhere $<$200\,million year old), whereas the continental lithosphere is on average about 2\,billion years old.

The oceanic crust is made up almost exclusively of basalts (SiO$_2$ $\sim$50\,wt\%) with limited compositional variation, whereas the continental crust is markedly different with a large diversity of rock types (igneous, metamorphic, and sedimentary) and a complete range of compositions (e.g., sandstones with $\sim$100\,wt\% SiO$_2$ and carbonates with $\sim$0 wt\% SiO$_2$).  On average the continents were made by extracting basaltic lavas from the mantle, followed by the burial, metamorphism, melting of these lavas and granite formation, which leads to the granites floating to the top of the crust and the loss of the mafic residue from continents.  This multi-stage, complex history produces this heterogeneous composition, which is on average andesitic in composition (SiO$_2$ $\sim$60\,wt\%)~\cite{rudnick}.  Consequently, these differentiation processes ultimately lead to a simplified distribution of U in the Earth with $\sim$1,000\,ng/g in the continental crust, $\sim$100 ng/g in the oceanic crust and $\sim$10 ng/g in the present-day mantle.

\subsubsection{Composition of the mantle and the existence of compositional layers}
\label{SubSubSec:mantle}

The observed variation in seismic velocity in the mantle and core can be used to interpret its composition based on the equation of state of materials at specific pressures and temperatures~\cite{birch}. In a pioneering study, using the bulk sound velocity information from seismic data and assuming a mantle geotherm, Birch~\cite{birch} concluded that the Transition Zone of the mantle was either a region of phase changes (Fig.~\ref{Fig:geotherm}), a compositional gradient or both and emphasized that the lower mantle would have a high-pressure modifications of the ferro-magnesian silicates, which are characteristics of upper mantle minerals.  He also speculated, rightly, that the Transition Zone would be "key to a number of major geophysical problems".

We can define a series of mineralogical and compositional models for the Earth, the core, and the mantle that are consistent with range of compositions seen in chondritic meteorites.  However, given that the results are non-unique, debate on what is the constitution of the deeper portions of the mantle continues.  There are two broad end-member, mineralogical and compositional models for the Earth's mantle based on (1) a homogeneous model, with an upper and lower mantle of similar compositions (Hart and Zindler, 1986~\cite{hart}; McDonough and Sun, 1995~\cite{McDonoughSun}; All\'egre et al., 1995~\cite{allegre95}; Palme and O'Neill, 2003~\cite{palme}), and (2) a layered model, with an upper and lower mantle of distinctly different compositions (e.g., Anderson, 2002~\cite{anderson02}; Javoy et al., 2010~\cite{javoy}; Murakami et al., 2012~\cite{murakami}). Variants on these models that envisage lesser degrees of layering involve only parts of the lower mantle. These concepts include basal mantle cumulate layers resultant from early earth magma ocean conditions (Labrosse et al., 2006~\cite{labrosse}; Lee et al., 2007~\cite{lee}), or gravitationally sequestered layers of early-enriched crust (Boyet and Carlson, 2005~\cite{boyet}).

\begin{figure}[tb]
\begin{center}
\centering{\epsfig{file=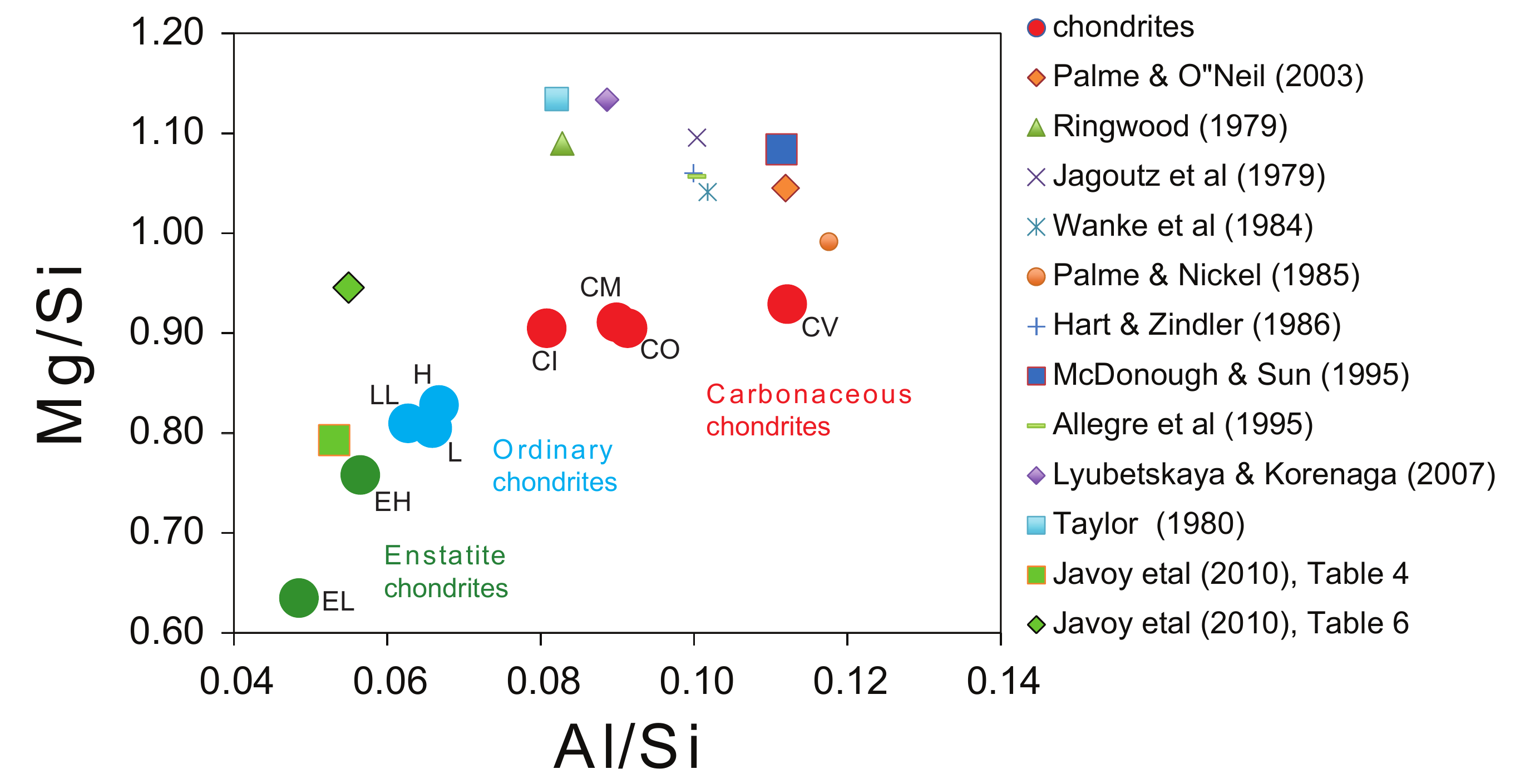,scale=0.5}}
\begin{minipage}[t]{16.5 cm}
\caption{A plot of the weight ratios of some major elements in chondrites (red, green and blue circles) and various models of the silicate Earth.  The Mg/Si value established the relative proportion of olivine (atomic Mg/Si of 2) to pyroxene (atomic Mg/Si of 1) in the upper mantle. The present day upper mantle has a Mg/Si weight ratio of about 1.1. Silicate Earth models that fall below this Mg/Si value of 1.1 (e.g., Javoy et al., 2010~\cite{javoy} and an assumed CI model of Murakami et al., 2012~\cite{murakami}) predict the lower mantle that is enriched in pyroxene and chemically distinct from the upper mantle.
\label{Fig:MgAloverSi}}
\end{minipage}
\end{center}
\end{figure}

Recently, Murakami et al.~\cite{murakami} reported new developments in obtaining sound velocity data for lower mantle minerals held at appropriate pressure and temperature conditions for the lower mantle and went on to critically evaluate analog compositional models of the deep Earth.  Examining a mixed assemblage of silicate perovskite and ferropericlase (see Fig.~\ref{Fig:geotherm} for an example of possible mineral proportions) they found that the best fit of their data was a lower mantle model that contained 93 volume percent of silicate perovskite or about half the ferropericlase to that shown in Fig.~\ref{Fig:geotherm}.  This higher proportion of perovskite requires that the lower mantle be enriched in silicon relative to the upper mantle and that the bulk composition of the silicate Earth follows that of a CI type chondrite (i.e., Mg/Si of about 0.9 in Fig.~\ref{Fig:MgAloverSi}).  The implication of this model is that the mantle is chemically stratified and that there is limited mass transport across the 660\,km seismic boundary layer.  

One of the grand challenges in Earth sciences is the integration of data from a wide variety of new technologies into a coherent picture that is constrained by the uncertainties of both the data and the models.  The example of the Murakami et al.~\cite{murakami} data is a case in point. This exciting new technological development provides unparalleled opportunities for mineral physicists to characterize sound wave speed in minerals at appropriate lower mantle pressures.  Results from this study, however, (1) were for shear wave velocities ($V_s$) only (not compressional wave velocities ($V_p$)), (2) were extrapolated to modeled lower mantle temperatures, (3) were conducted on simple analog compositional materials (e.g., not considering Ca, Al, and Fe$^{3+}$ contributions) and (4) were compared without uncertainties to the 1D PREM model, which is based on seismological data that have their own uncertainties.  It is also worth noting that constraints on mantle composition from $V_s$ data are significantly weaker than that based on $V_p$ data.  Likewise, models using density and bulk modulus data from other mineral physics experiments encounter similar issues with the full propagation of uncertainties in these systems.  

Similarly, claims by isotope geochemists that the compositions of Earth materials match a specific family of chondrites need to be placed in a greater context. Physically observable differences in shapes and sizes of components and the redox states of minerals lead to the classification of meteorite groups.  Recent strides in mass spectrometry have allowed us to identify small isotopic differences (i.e., a few parts in $10^4$ to $10^6$) between these petrographic groups of chondritic meteorites (e.g., Boyet and Carlson, 2005~\cite{boyet}; Gannoun et al., 2011~\cite{gannoun}; Warren, 2011~\cite{warren}; Zhang et al., 2012~\cite{zhang}; Fitoussi and Bourdon, 2012~\cite{fitoussi}). Javoy, 1995~\cite{javoy1995} highlighted the shared oxygen isotopic composition of the Earth and enstatite chondrites and later Javoy et al., 2010~\cite{javoy} suggested a compositional model for the Earth based on the same enstatite chondrites. More recently some scientists have continued to support the match between the Earth and enstatite chondrites (e.g., Gannoun et al., 2011~\cite{gannoun}; Warren, 2011~\cite{warren}; Zhang et al., 2012~\cite{zhang}), whereas others have highlighted differences between these bodies (Fitoussi and Bourdon, 2012~\cite{fitoussi}). The Javoy et al., 2010~\cite{javoy} model utilizing the enstatite chondrites postulates a gross chemical distinction between the upper and lower mantle, thus requiring convecting layering of the mantle.

A major constraint on the nature of the deep mantle has come from studies of the noble gas isotopic composition of basalts. Basaltic lavas erupted at deep ocean ridges, the boundaries of tectonic plates in the ocean, have a chemical and isotopic composition that are distinctly different from those erupted at places like Hawaii and Iceland, which are volcanic edifices distributed randomly at the Earth's surface without regard to tectonic plate boundaries.  The latter, ocean island basalts (often labeled OIB), have noble gas isotopic compositions (i.e., enrichments in primordial $^3$He and low $^{40}$Ar/$^{36}$Ar) that are indicative of un-degassed source regions. The former, mid-ocean ridge basalts (often labeled MORB), have noble gas isotopic compositions (i.e., enriched in $^4$He and $^{40}$Ar) consistent with their sources being degassed and enriched in the radiogenic products of $^{40}$K and $\alpha$ decays. Consequently, these observations lead many to conclude that the mantle is chemically layered with the lower mantle (depth not being constrained by these chemical arguments) containing a more primordial, under-degassed noble gas component that is tapped by focused upwelling plumes, while the degassed upper mantle is tapped by the whole sale sampling of 40,000\,km of mid-ocean ridges~\cite{allegre96}.

Increasingly, the end-member layered mantle models, which were greatly in favor some thirty years ago, have come under considerable scrutiny and disfavor. This is due mostly to seismic tomographic observations that show oceanic lithospheric plates plunging into the lower mantle, with some projecting steeply down from the subducting trench and others showing a stair-step transition of ponding in the Transition Zone and later laterally becoming unstable and then plunging into the deep mantle~\cite{grand,hilst}.  Collectively, these seismic images of subducting lithospheric plates traversing the mantle are taken as evidence of whole mantle convection with mass transfer occurring over the depth and breadth of the mantle.  Importantly, however, the interpretation of geological information requires a 4D integration of data, which has led some to accept this seismological evidence, while postulating that the mass transport condition is a relatively recent development in the Earth~\cite{allegre02, allegre04}.  Most recently, however, noble gas models of the mantle have reconciled the observational data with whole mantle convection~\cite{gonnerm}.

\subsection{Earth's thermal budget, heat producing elements, and geo-neutrinos}
\label{SubSec:heat}

Models that predict the major elemental composition of the silicate Earth (sometimes referred as the bulk silicate Earth, or BSE) also predict the Th and U content, given the assumption that the refractory lithophile elements are in chondritic proportions and were excluded from the core.  This assumption is applied to all of the terrestrial planets and it assumes, for example, that given an Al content and chondritic Ca/Al value (1.1), the Ca content is predicted to be 10\% greater. Likewise, chondritic ratios of Al/Th ($(2.9 \pm 0.2) \times 10^5$) and Th/U ratio (3.9 $\pm$ 0.3) can be used to establish the abundances of Th and U in models of the silicate Earth, particularly when considering differences between model compositions.  Table~\ref{Tab:EarthComp} presents a range of compositional models for the silicate Earth and included model U contents ranging between 12 and $26 \times 10^{-9}$\,kg/kg.  A more enriched model by Wasserburg et al., 1963~\cite{wasserburg}, which was based on chemical and isotopic observations of oceanic and continental rocks, proposed the Bulk Silicate Earth has a U content of $33 \times 10^{-9}$ kg/kg.  The most U and Th depleted model was presented by O'Neill and Palme, 2008~\cite{ONeill} and Campbell and O'Neill, 2012~\cite{campbell} and was based on a concept of collisional erosion in the early Earth, where a fraction of the Earth's surface crust (~2\% of the mass of the silicate Earth) enriched in the heat producing elements, is lost to space.  In this model, the Earth has lost about half of its budget of heat producing elements leaving the Earth with only $\sim$$10^{-8}$ kg/kg of U.

\begin{table}
\begin{center} 
\begin{minipage}[t]{16.5 cm}
\caption{Compositional models of the silicate Earth. Abundances of major elements are given in weight percent, those of Th and U in $10^{-9}$ kg/kg.}
\label{Tab:EarthComp}
\end{minipage}
\begin{tabular}{l|lllll|ll|ll}
\hline
Compositional Models	& Al &	Ca &	Mg	& Si &	Fe &	Al/Si &	Mg/Si &	Th &	U \\
\hline
Ringwood, 1979~\cite{ringwood}	&   1.75	& 2.22 &	23.0 &	21.1 &	6.22 &	0.083 &	1.09 & -  &  - \\
Jagoutz et al., 1979\cite{jagoutz} &  2.12	& 2.50 &	23.1 &	21.1 &	6.06	&      0.100 &	1.10	& 94	& 26 \\
Taylor, 1980~\cite{taylor}	&           1.75	& 1.89 &	24.1	&      21.3	&     6.22& 	0.082 &	1.13	& 70& 18 \\
Wanke et al., 1984~\cite{wanke}  &    2.17	& 2.50 &	22.2	&      21.3 &	5.83&	0.102 &	1.04 & - & - 	\\	
Palme and Nickel, 1985~\cite{PalmeNickel} & 2.54	& 3.14 &	21.4 &	21.6	&      5.99	&     0.118 &	0.99 & - & -\\
Hart and Zindler, 1986~\cite{hart} & 2.15	& 2.34 &	22.8	&      21.5	 &     6.22	&     0.100 &	1.06	& 79 & 21\\
Anderson, 1989~\cite{anderson89} &         1.69	& 2.43 &	19.7	&      21.0 &	12.2 &	0.080 &	0.94&- &	 -\\
McDonough and Sun, 1995~\cite{McDonoughSun}& 2.34 & 2.52 &	22.8	 & 21.0	& 6.25 &	0.111	& 1.08 &	80	& 20\\
All\'egre et al., 1995~\cite{allegre95}	&   2.16	& 2.31& 22.8	& 21.6 &	5.82	& 0.100	& 1.06 &	 - & -\\
Palme and O'Neil, 2003~{palme} &	2.38 & 	2.61 &	22.2	& 21.2 &	6.30&	0.112 &	1.04&	83 &	22\\
Anderson, 2007~\cite{anderson07} & 	2.02	& 2.20 &	20.5 &	22.4 &	6.11 &	0.090 &	0.92 &	77	& 20\\
Lyubetskaya and Korenaga, 2007~\cite{lyub} &	1.86	& 1.99	& 23.8 &	21.0&	6.20	& 0.089 &	1.13	& 63 &	17\\
Javoy et al., 2010, Table 4 in~\cite{javoy}  &	1.28 &	1.28	& 19.1	& 24.1	& 8.63	& 0.053	& 0.79	& 43	& 12\\
Javoy et al., 2010, Table 6 in~\cite{javoy}&	1.28 &	1.34	& 22.0 &	23.3	& 6.87&	0.055	& 0.95	&43	&12\\
\hline
\end{tabular}
\end{center}
\end{table}

Compositional models for the Earth predict a factor of 3 difference in the amount of U in the Earth.  Assuming the Earth has a chondritic Th/U of 4 and a planetary K/U of $1.4 \times 10^4$~\cite{wasserburg,jochum,arevalo}, one can calculate a total heat production from the decay of these elements, as well as a surface geo-neutrino flux.  \v{S}r\'amek et al., 2013~\cite{sramek} recently modeled this range of compositional space and showed that the heat production power ranges from 10 to more than 30\,TW relative to an estimate of the surface heat flux of 46-47\,TW~\cite{jaupart,davies}. Recent work by Huang et al.~\cite{huang} finds that the continental crust has $6.8^{+1.4}_{-1.1}$\,TW of radiogenic power and that the remaining power resides in the mantle, not the core. This then translates to as little as 3\,TW and to as much as 23\,TW of radiogenic power in the mantle to drive convection and plate tectonics.

As mentioned above, it has been proposed that the core contains U and/or K~\cite{rama}, but these arguments have been addressed here and in~\cite{mcdonough} where it is recognized that positing such models needs to be coupled to corroborating geochemical evidence that is also free of negating consequences.  Studies have shown that reducing U into the metallic state has a far greater effect on other elements (e.g., Ti) that show no evidence for a core extraction.  Likewise, there is limited potential for CaS and various REE to accompany a potassium sulfide into a core forming phase; again there is no evidence for such process.  Finally, Herndon~\cite{herndon} has suggested a U-driven georeactor in the Earth's core as a consequence of his model of Earth's formation that involved a highly reduced Earth. Herndon's compositional model is inconsistent with chemical and isotopic observations of the Earth's mantle presented in McDonough, 2003~\cite{mcdonough}, particularly given a core containing significant quantities of Ca, Mg, U, Th, and other lithophile elements based on analogies with enstatite chondrites (highly reduced meteorites).  

Estimates over the last 40 years for the Earth's surface heat flow are between 41 and 47\,TW, with recent estimates being $46 \pm 3$~\cite{jaupart} and $47 \pm 2$~\cite{davies}.  The latter estimate considers data from $>$38,000\,heat flow measurements from around the globe.  Measuring the temperature and temperature gradient in the Earth and then projecting this temperature condition into the body is a considerable challenge and it was this matter that confused Lord Kelvin when he took his observations and folded them into an estimate of the age of the Earth.  Determining the Earth's heat flow from measurements of gradients, heat production, and conductivity is both a surprisingly simple and complex concept at the same time (see also the discussion in \cite{Sramek2013} and \cite{Dye2012}). The recent recognition that the near surface gradient in a heat flow measurement also records climate change effects on millennia time scales has many going back to examine the original heat flow measurements to de-convolve this effect from the estimated surface flux.

Combining the present day surface flux ($46 \pm 3$\,TW) and estimates of the radiogenic heat production allows one to estimate the amount of primordial heat remaining in the Earth. Models envisaging "low-$Q$" heat production for the Earth, with as little as 10\,TW of radiogenic power (e.g.,~\cite{ONeill,javoy,campbell}), require that the Earth has a significant amount of primordial heat ($\leq$36\,TW), whereas the "high-$Q$" models (e.g.,~\cite{wasserburg,turcotte2001,turcotte2002}) project a limited primordial heat budget left in the Earth (e.g., $\leq$16\,TW).  Geophysical models of the Earth seek satisfactory solutions to the planetary thermal evolution by fitting the relative contributions of primordial heat and heat production, while being consistent with the Earth's secular cooling record~\cite{schubert}.  These models parameterize mantle convection in terms of the force balance between buoyancy and viscosity versus thermal and momentum diffusivities, while recognizing that the convective state of mantle greatly exceeds its critical Rayleigh number, which marks the onset of convection. Typically these are high-$Q$ models, requiring that more than 50\% of the present heat flow be due to radiogenic heating.

\section{Geo-neutrinos from the crust and mantle}
\label{Sec:GeoSignal}

\subsection{Geo-neutrino signal from the crust}
\label{SunSec:GeoCrust}

Analyzing the arrival times of the refracted and reflected elastic waves produced by an earthquake with its epicenter close to Zagreb, Andrija Mohorovi\v{c}i\'c, in 1909, provided the first evidence of a discontinuity between crust and mantle. Further measurements of the seismic waves confirmed the presence of this boundary, which separates rocks having P-wave velocities of 6-7\,km/s from those having velocities of about 8-9\,km/s. This change of mechanical properties of Earth materials is due to a compositional transition from mafic rocks of the lower crust to ultramafic rocks of the upper mantle~\cite{ref2.1,RudnickFountain}. The crust is that part of the Earth with highest concentration of Heat Producing elements (i.e., $\sim$$1 \times 10^{-6}$\,kg/kg) while their abundances drop rapidly (i.e., $\sim$$ 1 \times 10^{-8}$\,kg/kg) below the Mohorovi\v{c}i\'c discontinuity, often referred to as Moho.

The crust is divided in two main reservoirs: continental and oceanic crust. Although the mass of the continental crust is about 0.34\% of the Earth's mass, this crust contains approximately 40\% of the Earth's inventory of U and Th~\cite{huang}. Therefore, rocks of continental crust produce the highest rate of geo-neutrinos per unit of mass and they give the biggest contributions to geo-neutrino signals in the existing detectors.

The first estimations of the expected geo-neutrinos from U and Th in the crust are published in~\cite{ref2.4}. In this model uranium and thorium are distributed uniformly in a shell 30\,km thick having a mass of $2\times 10^{22}$\,kg with abundances $A$(U) = $4 \times 10^{-6}$ kg/kg and $A$(Th) = $19 \times 10^{-6}$\,kg/kg. The expected signal of $S$(U+Th)$\sim$32\,TNU is independent from the position of the detector on the Earth's surface. 

Further estimations have been published and are based on different geophysical crustal models. The authors of~\cite{ref2.5} take into account the spatial distribution of the continental and oceanic crust following the global crustal map CRUST5.1~\cite{ref2.6}, but not considering any crustal sub-layers. After 2004 a new generation of models for estimating geo-neutrinos from the crust was published by many authors~\cite{ref2.7,enomoto, ref2.9, ref2.10}, who all adopted the global crustal model on a $2^{\circ} \times 2^{\circ}$ grid as published by Laske et al.~\cite{ref2.11}. This geophysical model is made up of 16,200 tiles and describes 360 key 1D-profiles. The thickness, the density, and the velocities of compressional ($V_p$) and shear ($V_s$) waves traveling through are given explicitly for seven layers (ice, water, soft sediments, hard sediments, upper, middle and lower crust) in each tile. The accuracies of this model are not specified and they vary in different places since vast continental regions (large portions of Africa, South America, Antarctica and Greenland) lack direct measurements.

Table~\ref{Tab:GeonuSignal} shows the expected geo-neutrino signal in TNU at the Earth's surface from U and Th in the crust only, according to three different models published in the last decade~\cite{huang,ref2.7,ref2.10}. Kamioka and Gran Sasso are the locations where KamLAND and Borexino experiments are running from 2002 and 2007, respectively. The SNO+ experiment is the follow-up of the Sudbury Neutrino Observatory (SNO) and is in construction phase. The site of Hawaii is considered, due to its low geo-neutrino crustal signal. 

\begin{table}
\begin{center} 
\begin{minipage}[t]{16.5 cm}
\caption{Geo-neutrino expected signals in TNU from U and Th in the crust according to three different geophysical and geochemical models. All calculations are normalized to a survival probability $<P_{ee}> = 0.55$. The uncertainties of Mantovani et al.~\cite{ref2.10} correspond to the full range of the crustal models, while for Dye~\cite{ref2.7} and Huang et al.~\cite{huang} the $1\sigma$ errors are reported}
\label{Tab:GeonuSignal}
\end{minipage}
\begin{tabular}{l|c|c|c}
\hline
Site 	&   	Mantovani et al.~\cite{ref2.10} & Dye~\cite{ref2.7}	& Huang et al.~\cite{huang} \\
\hline
Kamioka	         & $24.7^{+4.3}_{-10.3} $  &	$23.1 \pm 5.5$   & 	$20.6^{+4.0}_{-3.5}$ \\
Gran Sasso	 & $29.6^{+5.1}_{-12.4}$     &	$ 28.9 \pm 6.9$  &	$29.0^{+6.0}_{-5.0}$ \\
Sudbury	         & $38.5^{+6.7}_{-16.1}$     & 	$34.9 \pm 8.4$   &	$34.0^{+6.3}_{-5.7}$ \\
Hawaii	         & $3.3^{+0.6}_{-1.4}$	    & $3.2 \pm 0.6$ & 	$2.6^{+0.5}_{-0.5}$ \\
\hline
\end{tabular}
\end{center}
\end{table}    

In Mantovani et al., 2004~\cite{ref2.10}, the radioactivity content of each layer of a $2^{\circ} \times 2^{\circ}$ global crustal model was calculated by averaging the abundances of U and Th values available in the GERM database (2003). The reported spread is obtained by using the maximal and minimal abundances of the compilations. The geo-neutrino signal from the crust reported in~\cite{ref2.7} differs from that of~\cite{ref2.10} for the composition of the crystalline crust. In this latter model the authors assign to each identifiable layer (upper, middle and lower crust) the U and Th abundances presented in the comprehensive review published by Rudnick and Gao~\cite{rudnick}. The uncertainty of the geo-neutrino signal for this model is the sum of the uncertainties due to $1\sigma$ error of U and Th abundances assigned to the crustal layers.

In reference~\cite{huang} the uncertainties of the expected geo-neutrino flux are calculated for the first time, taking into account the Th and U content of the crust and considering the geochemical and geophysical uncertainties associated with the input data. Observing a log-normal distributions of U and Th concentrations in crustal rocks, the median values are evaluated as the most representative number of the probability functions. The asymmetrical uncertainties are propagated from the non-Gaussian distributions of the abundances in the deep continental crust using a Monte Carlo simulation. The estimated signals from U and Th in the crust as calculated from this study are reported in Table~\ref{Tab:GeonuSignal}, with all values overlapping within the quoted uncertainties.

Due to the inverse-squared distance-dependence of the neutrino flux, the local and global reservoirs can provide comparable contributions to the geo-neutrino signal, at least for detectors sited in the continental crust. The boundaries of the local crust are a matter of convention. Following the reference~\cite{huang}, the crustal U and Th content in the 24 closest $1^{\circ} \times 1^{\circ}$ crustal voxels surrounding KamLAND, Borexino and SNO+ contribute 65\%, 53\% and 56\% of the total signal, respectively. Refined geochemical and geophysical models, that describe the Earth, have been developed for identifying with greater precision and accuracy the local contribution (circa 500\,km radius) surrounding each detector.
 
\subsection{Local geological model near the Kamioka site}
\label{subsec:kamioka}

The Japan island arc sits on a continental shelf situated close to the eastern margin of the Eurasian plate, one of the most seismically active areas of our planet. The Philippine tectonic plate is moving towards the Eurasia plate at about 40\,mm/year and ultimately, the Philippine plate is subducting beneath the southern part of Japan. The Pacific Plate is moving roughly in the same direction at about 80\,mm/year and is subducting beneath the northern half of Japan. Both subducting plates form deep submarine trenches and uplift areas parallel to the trench, and generate igneous activity, particularly the production of the volcanic island chain. The Sea of Japan is a typical marginal sea, which is incompletely bordered by islands and expanded basins on the back arc side (back arc basin), and is situated between the Japan island arc and the Asian continent. The geochemical and geophysical features of the Japanese crust, the effects of the subducting slab, and the intricate back-arc opening tectonics have been studied by Fiorentini et al.~\cite{fiorentini05} and Enomoto et al.~\cite{enomoto}, with the aim of estimating their effects on geo-neutrino signal.

\begin{figure}[tb]
\begin{center}
\centering{\epsfig{file=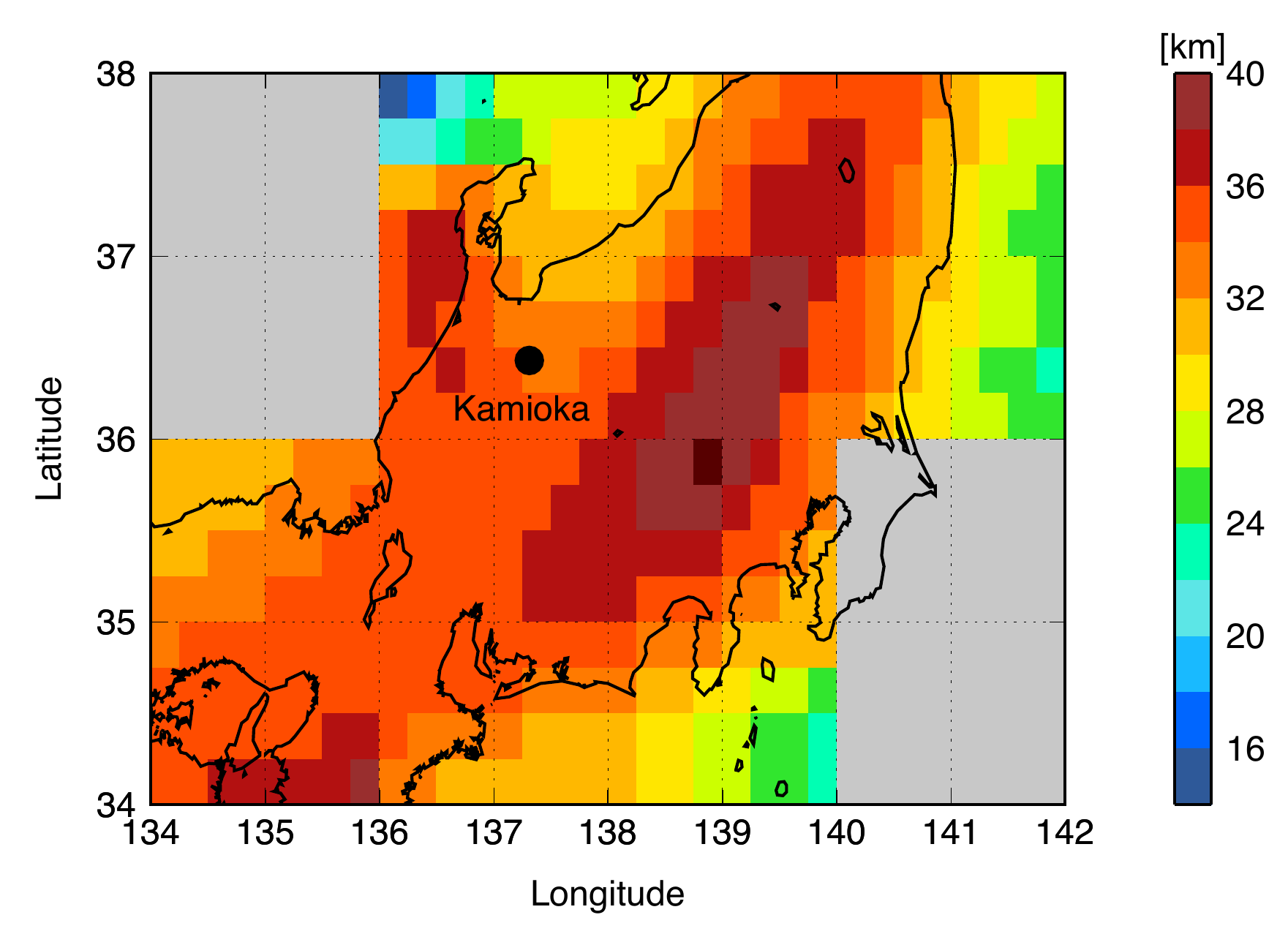,scale=0.6}}
\begin{minipage}[t]{16.5 cm}
\caption{Moho depth of the local refined model around KamLAND~\cite{fiorentini05}.
\label{Fig:Kamioka}}
\end{minipage}
\end{center}
\end{figure}

The six $2^{\circ} \times 2^{\circ}$ tiles around KamLAND produce $S$(U+Th) = 13.3\,TNU~\cite{huang}. A refined local model of the crust identifies two layers: an upper crust extending down to the Conrad discontinuity, and a lower part down to the Moho discontinuity. In~\cite{fiorentini05}, the map of Conrad and Moho depths beneath the Japan Islands is derived by Zhao et al.~\cite{zhao}, with an estimated standard error of $\pm1$\,km over most of Japan territory, see Fig.~\ref{Fig:Kamioka}. A detailed grid based on $0.25^{\circ} \times 0.25^{\circ}$ cells provided a sampling density for the study of the upper crust in the region near Kamioka that is equivalent to about one specimen per 400\,km$^2$.  Also, the vertical distribution of Th and U abundances in the crust provides even greater challenges because of the limited information on the chemical composition at scales smaller than the Conrad depth, which is generally about 20\,km deep. The chemical composition of the upper-crust of Japan was estimated by Togashi et al., 2002~\cite{togashi} and was based on 166 representative specimens, which can be associated with 37 geological groups, based on ages, lithologies, and provinces. In Fiorentini et al., 2005~\cite{fiorentini05}, a map of uranium abundance in the upper crust was built under the assumption that the composition of the whole upper crust is the same as that inferred in~\cite{togashi} from the study of the exposed portion.

The composition of the Japanese lower crust was assumed to be homogeneous and taken to be $A_{LC}$(U) = $(0.85 \pm 0.23) \times 10^{-6}$ kg/kg and $A_{LC}$(Th) = $(5.19 \pm 2.08) \times 10^{-6}$ kg/kg, based on the model of the lower continental crust reported in an extensive study of the Eastern China crust~\cite{gao98}. The expected U and Th geo-neutrino fluxes from the region surrounding KamLAND are $S$(U) = $11.17 \pm 0.65$\,TNU and $S$(Th) = $3.20 \pm 0.37$\,TNU, respectively (Table~\ref{Tab:KamiokaSignal})~\cite{fiorentini2012}. The maximal and minimal excursions of various inputs and uncertainties provide an estimate of the $3\sigma$ error range. Consequentially, and considering the measurement errors of the chemical analysis of the representative samples, a $3\sigma$ uncertainty of $\pm10$\% has been associated to the U and Th abundances in the Japanese upper crust. The full range of uncertainty, due to the unknown chemical composition of the lower crust, is the half-difference between the signals obtained for the extreme values of the estimated uranium abundances in the lower crust. All these effects are summarized in Table II of Fiorentini et al.~\cite{fiorentini2012}, together with the errors associated with the constraints of the model (discretization of the upper crust and the crustal depth).	
	
\begin{table}
\begin{center} 
\begin{minipage}[t]{16.5 cm}
\caption{Contributions to the geo-neutrino signal in KamLAND from the local geology. Quoted errors correspond to $1\sigma$~\cite{fiorentini2012}.}
\label{Tab:KamiokaSignal}
\end{minipage}
\begin{tabular}{l|c|c}
\hline
	  &   $S$(U) [TNU]	& $S$(Th) [TNU] \\
Six-tiles	& $11.17 \pm 0.65$ & 	$3.20 \pm 0.37$ \\
Subducting slab & 	$2.02 \pm 0.61$ & 	$0.90 \pm 0.27$ \\
Sea of Japan	& $0.34 \pm 0.10$ & 	$0.09 \pm 0.03$ \\
Local total & 	$13.53 \pm 0.90$ & 	$4.19 \pm 0.46$ \\
\hline
\end{tabular}
\end{center}
\end{table}

During the subduction of the Philippine and Pacific plates U and Th are carried down in marine sediments and in the oceanic crust. The potential exists for the lower part of the continental crust of Japan to be enriched in large-ionic lithophile elements via dehydration of the top of the subducting plate~\cite{ref2.18}. The degree of enrichment of U and Th into the overlying Japanese crust is still debated. Two extreme cases of this can be modeled: one assumes that the slab keeps its trace elements throughout the subduction process, and in the other, all the uranium from the subducting crust is dissolved in fluids and is transported to the base of the lower crust of Japan arc. Considering a single slab penetrating below Japan with a velocity $v$ = 60\,mm/year (the average of the two plates) on a time scale $T \sim$$10^8$\,year, the U and Th abundance in the lower crust can increase according to the extreme cases by a factor of 1.06 and 2.57, respectively~\cite{ref2.18}. Encompassing both scenarios at the $3\sigma$ level, the contribution to geo-neutrino signal in KamLAND from the subducting slab can be estimated as $S$(U) = $2.02 \pm 0.61$\,TNU and $S$(Th) = $0.90 \pm 0.27$\,TNU (Table~\ref{Tab:KamiokaSignal}). 

Although in the global crustal model CRUST 2.0~\cite{ref2.11} the crust beneath the Sea of Japan is classified as oceanic, its true nature remains uncertain. Tamaki et al.~\cite{tamaki} identify four distinctive crustal types: continental, rifted continental, extended continental, and oceanic. The minimal geo-neutrino signal, $S$(U+Th) = 0.06\,TNU, from the Sea of Japan was obtained assuming a homogeneous thin oceanic crust. On the other hand, a model based on a thick crust (up to 19\,km for the Oki bank) with U and Th abundances typical of continental crust overlain by a few km of sedimentary layer (up to 4\,km for the Ulleung basin), maximizes the geo-neutrino production to approximately $S$(U+Th) = 0.82\,TNU~\cite{fiorentini05}. In Table~\ref{Tab:KamiokaSignal}, we report the central values of these two extreme cases with $1\sigma$ uncertainties in order to encompass the extreme values with $3\sigma$.

\subsection{Local geological model near the Gran Sasso site}
\label{subsec:lngs}

The Borexino experiment is located under the highest mountains of the Apennines in Italy, in the Gran Sasso Range. This massif is the northern part of the Latium-Abruzzi carbonates, which is over-thrust onto the Umbria-Marche basin in the Apennines chain. The orogenesis of the Apennine belt began in the early Neogene (about $20 \times 10^6$\,years ago) and developed through the deformation of two major paleo-geographic domains: the Liguria-Piedmont Ocean and the Adria-Apulia passive margin. In particular the central Apennines are an arc shaped fold-and-thrust belt, with north-eastward convexity and vergence that plunges north-westward~\cite{carmignani}. These structures are clearly visible at the Gran Sasso massif, where a northern block is formed by Jurassic limestone over thrusting Miocene limestone and marls. The normal stratigraphic sequence is observed in the southern block where limestone and marls are stacked from Jurassic to Miocene in age.

\begin{figure}[tb]
\begin{center}
\centering{\epsfig{file=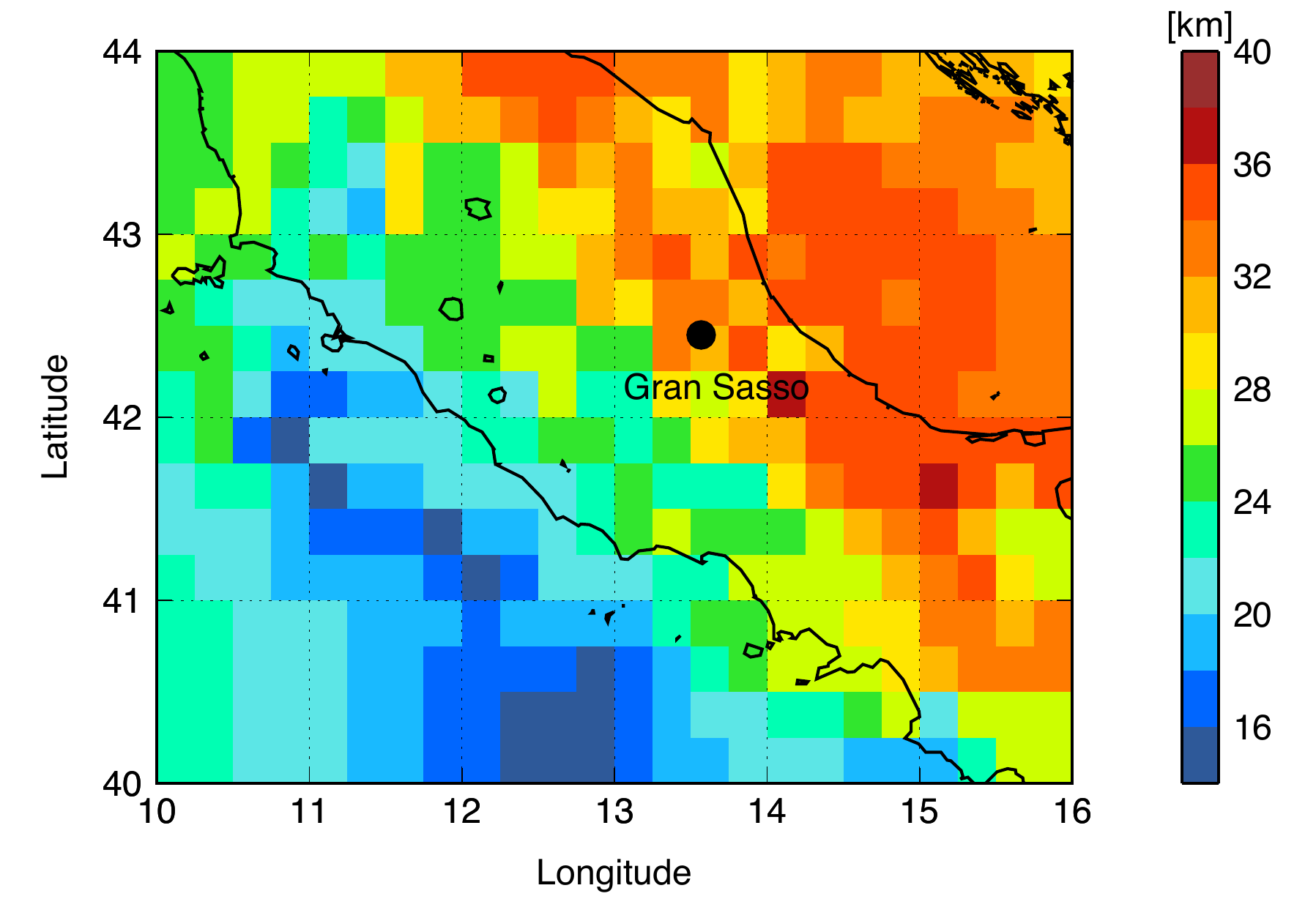,scale=0.6}}
\begin{minipage}[t]{16.5 cm}
\caption{Moho depth of the local refined model around Borexino~\cite{coltorti}.
\label{Fig:LNGS}}
\end{minipage}
\end{center}
\end{figure}

The refined reference model for the Gran Sasso area was developed by Coltorti et al., 2011~\cite{coltorti} and is based on two zones with different degrees of resolution: the central tile, a three dimensional geological model of the $2^{\circ} \times 2^{\circ}$ area centred at Gran Sasso National Laboratories, and the rest of the region, i.e. what remains of the six tiles after the central tile subtracted. In both areas, the crust is separated from the mantle by a well-defined Moho surface (Fig.~\ref{Fig:LNGS}) and is divided into three reservoirs: sediments, upper crust, and lower crust. 

In the central tile, the litho-stratigraphic framework of the 13\,km thick sedimentary deposit is geologically complex and composed of numerous sedimentary units characterized by different thicknesses, ages and depositional environments. Following the approach proposed in Plank et al., (1998)~\cite{PlankLangmuir}, Coltorti et al.~\cite{coltorti} assumed that sediments formed in similar depositional environments would have similar and somewhat homogeneous compositions. In particular, the sedimentary cover is modeled as having four types, characterized by different contents of radioactive elements. Uranium and thorium abundances have been measured on representative samples of geological formations using Inductively Coupled Plasma-Mass Spectrometry and Gamma Spectroscopy. 

The most massive sedimentary reservoir ($\sim$77\% of sedimentary mass) is composed of Mesozoic carbonate units set down in a shallow-water environment. The 12 samples of limestone, dolomite and evaporite, all show low and uniform values of U and Th abundances: $A$(U) = $(0.3 \pm 0.2) \times 10^{-6}$ kg/kg and $A$(Th) = $(0.2 \pm 0.2) \times 10^{-6}$ kg/kg, respectively. In the central tile approximately 16\% of the mass of the sedimentary cover is terrigenous deposits that progressively overlay the previous carbonate depositional systems, starting from the late Miocene onward. The main lithologies of these units are sandstones, silts and clays characterized by an average U and Th abundances of $a$(U) = $(2.3 \pm 0.6) \times 10^{-6}$ kg/ kg and $a$(Th) = $(8.3 \pm 2.5) \times 10^{-6}$ kg/kg, respectively. 

Neglecting the Meso-Cenozoic basinal carbonate units (less 2\% of sedimentary mass), the Permian clastic units (sandstones and conglomerates) are the result of the dismantling and erosion of the ancient Paleozoic crust. Since these units rarely outcrop within the entire Italian Peninsula and their geochemical nature is similar to the basement rocks, in the model of~\cite{coltorti} this reservoir is considered to have the same U and Th content of the upper crust (see below).

The mass weighted average abundances of the sediments of the central tile, and adopted for the rest of the region, are $A_{Sed}$(U) = $0.8 \times 10^{-6}$ kg/kg and $A_{Sed}$(Th) = $2.0 \times 10^{-6}$\,kg/kg. This is significantly lower than the world average for sediments from the global crustal model, $A_{Sed}$(U) = 1.7 and $S_{Sed}$(Th) = $6.9 \times 10^{-6}$ kg/kg~\cite{PlankLangmuir}. This is a consequence of large volumes of U- and Th-poor carbonates. The expected geo-neutrino signal from sediments in the six tiles is $S$(U+Th) = $2.93 \pm 0.25$\,TNU, corresponding to $\sim$30\% of the local contribution.

In Coltorti et al.~\cite{coltorti}, the three-dimensional structure of the crystalline basement is constrained by seismic profiles from the CROP Project~\cite{finetti} and by a Moho isopach map obtained from seismic and gravity data~\cite{finetti}. Approximately 62\% of the central tile volume is occupied by upper and lower crust having an averaged thickness of 13 and 9 km, respectively. Since basement rocks do not outcrop in Central Italy, an accurate sampling has been performed on representative outcrops of the upper crust in the Southern Alps and of the lower crust in Ivrea-Verbano Zone, which is the most classic and extensively studied, deep crustal section found in the Alps. In particular two rocks types were analyzed: U- and Th-enriched felsic rocks and intermediate/mafic rocks that are depleted of heat producing elements. In the case of upper crust, the average U and Th abundances for the two groups are calculated by fixing the relative proportion of these two lithologies, using seismic data. In the lower crust, the fraction of felsic and mafic rocks was estimated on the basis of geophysical and geochemical information, obtaining a felsic and mafic percentage of 40\% and 60\%, respectively, in agreement with what was proposed by~\cite{wedepohl, RudnickFountain, gao98}. 

The crystalline basement outside the central tile is modeled by Coltorti et al.~\cite{coltorti}, where they distinguished the upper and lower crust and treated them as separate, homogeneous layers. The abundances of U and Th calculated for the crust of the central tile are adopted for the rest of the region. Taking into account that the maximal and minimal excursions of various input values and uncertainties are taken as a proxy for the $3\sigma$ error range, the estimated geo-neutrino signal in the local upper and lower crust is $S_{UC}$(U+Th) = $6.15 \pm 1.2$\,TNU and $S_{LC}$(U+Th) = $0.59 \pm 0.22$\,TNU. In Table~\ref{Tab:LNGSSignal} the contributions to the geo-neutrino signal from U and Th within the three reservoirs of the local geological model near Gran Sasso massif are summarized.

\begin{table}
\begin{center} 
\begin{minipage}[t]{16.5 cm}
\caption{Contributions to the geo-neutrino signal in Borexino from the local geology (six tiles). Quoted errors are at $1\sigma$ ~\cite{fiorentini2012}.}
\label{Tab:LNGSSignal}
\end{minipage}
\begin{tabular}{l|c|c}
\hline
	  &   $S$(U) [TNU]	& $S$(Th) [TNU] \\
\hline
Sediments	& $2.53 \pm 0.21$ & 	$0.40 \pm 0.04$ \\
Upper crust	& $4.94 \pm 0.97$ &	$1.21 \pm 0.24$\\
Lower crust	& $0.34 \pm 0.11$ & 	$0.25 \pm 0.11$\\
Local total	& $7.81 \pm 0.99$ & 	$1.86 \pm 0.27$\\
\hline
\end{tabular}
\end{center}
\end{table}  

\subsection{Geo-neutrinos from the mantle}
\label{subsec:mantle}

The contribution to the geo-neutrino signal from mantle depends on the total amount of heat generating elements as well as on their distribution deep inside the Earth, since sources closer to the  detector contribute more to the signal. For each value of the total mass of Th and U in any Earth model, we construct distributions of abundances that provide maximal and minimal signals, under the condition that they are consistent with the geochemical and geophysical information of the globe~\cite{fiorentini2007}.

\begin{figure}[tb]
\begin{center}
\centering{\epsfig{file=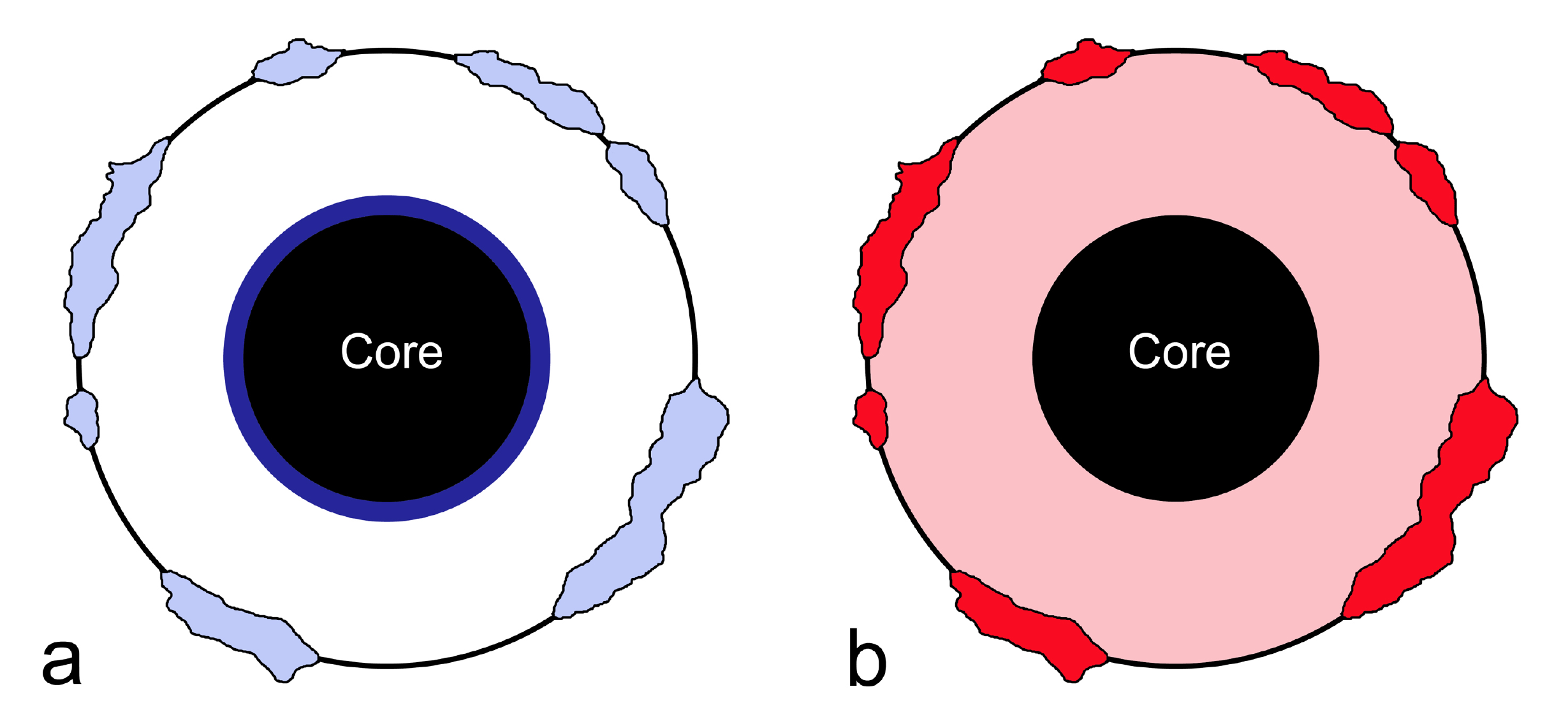,scale=0.6}}
\begin{minipage}[t]{16.5 cm}
\caption{Cartoons of two extreme distributions of U and Th in the Earth. The minimal geo-neutrino signal (a) is obtained by placing the heat-producing radiogenic elements as far inward as possible (i.e. in a thin layer at the bottom of the mantle), having as little radiogenic material as possible in the crust. An opposite condition is when there is a homogeneous mantle combined with the maximal amount of heat-producing radiogenic elements in the crust, which produces the highest geo-neutrino signal (b). 
\label{Fig:MantleSignal}}
\end{minipage}
\end{center}
\end{figure}

We can test mantle models using uranium in the mantle by assuming a spherical symmetry distribution of uranium throughout the mantle or a layered condition. It follows that, for a fixed uranium mass in the mantle $m_{M}$(U), the extreme predictions for the signal are obtained by: (1) placing uranium in a thin layer at the bottom and (2) distributing it with uniform abundance over the mantle (Fig.~\ref{Fig:MantleSignal}). These two cases give, respectively:
\begin{equation}
S_{M}^{min} = 11.3 ~ m_{M}({\rm U}) ~~ {\rm [TNU]}
\label{Eq:mantleMin}
\end{equation}
\begin{equation}
S_{M}^{max} = 16.2 ~ m_{M}({\rm U}) ~~  {\rm [TNU]}
\label{Eq:mantleMin}
\end{equation}

The relative geo-neutrino contributions from the crust and mantle can be combined so as to obtain predictions of the surface flux. For a total U mass fixed by a BSE model, assigning to the crust as much material as consistent with observational data and putting the rest in the mantle with a uniform distribution, the highest geo-neutrino signal is modeled. Similarly, the minimal signal is obtained for a minimal mass contribution in the crust and the rest in a thin layer at the bottom of the mantle. The total amount of radioactive elements should not produce more heat flow than $47\pm 2$\,TW~\cite{davies}; this limit represents the upper bound for a fully radiogenic Earth model. On the other hand the minimal geo-neutrinos signal is obtained with the minimal mass of uranium in the crust and a negligible amount in the mantle. On the basis of these arguments the two extreme total signals ($S_{high}$ and $S_{low}$) expected in one site can be plotted as a function of heat flow due to uranium and thorium in the Earth, considering a fixed chondritic ratio Th/U. The graph Signal vs Heat Power from U+Th is site dependent ($S$-$H$ plane); the intercept depends on the site, while the slope is universal.

On the $S$-$H$ plane we can identify the regions corresponding to three classes of BSE models extensively studied in \v{S}r\'amek et al., 2013~\cite{sramek}. The cosmochemical model~\cite{javoy} is based on an Earth composition that is similar to that observed in enstatite chondrites, which among the different types of chondrites shows the closest isotopic similarity with the mantle rocks and has sufficiently high iron content to explain the metallic core. This model is characterized by a relative low amount of U and Th, producing a total radiogenic power of $11 \pm 2$ TW. Taking into account that U and Th in the crust contribute 7\,TW\cite{huang}, the U and Th heat power of the mantle is approximately 4\,TW.

On the opposite end of the spectrum of proposed Earth compositions, the geodynamical model is based on the energetics of mantle convection and the observed surface heat loss~\cite{turcotte2002}. This model requires a high mantle $U_{rey}$ ratio as to prevent extremely high temperatures in Earth's early history. Assuming a mantle $U_{rey}$ ratio of $0.7 \pm 0.1$, the present radiogenic heat power produced by U and Th in the Earth is $33 \pm 3$\,TW, mainly (80\%) generated from the mantle.

A class of intermediate BSE models~\cite{hart,McDonoughSun, allegre95, palme} is based on the relative abundances of refractory lithophile elements in mantle samples. These models project back to their initial starting composition and are constrained by the relative abundances of the refractory lithophile elements in chondritic meteorites. Adopting the U and Th abundances of primitive mantle by McDonough and Sun~\cite{McDonoughSun} the power produced by these two elements corresponds to $16.6 \pm 3.0$\,TW.

\section{The detectors}
\label{Sec:detectors}

Only two geo-neutrino detectors, Borexino and KamLAND, are operating and have successfully measured the Earth's geo-neutrino flux. Both experiments have been primarily developed and constructed to achieve different goals than the measurement of the geo-neutrinos. Borexino, located in the underground laboratory of Gran Sasso, in central Italy, was designed to study the low-energy components of the solar neutrino flux. To this purpose, very high care has been devoted to keep at ultra-trace level the background due to the natural radioactivity. KamLAND, installed in the Kamiokande-Mozumi mine in central western Japan, was, on the other hand, designed to study the antineutrinos emitted by nuclear reactors. The physics justification for both experiments was focused on the study of the neutrino oscillation phenomenon, via two different approaches.

\subsection{Borexino}
\label{SubSec:BorexDetector}

The active detection medium of Borexino is a liquid scintillator which, with its intrinsic high luminosity ($\sim$50 times more than in the Cherenkov technique), is an ideal choice for massive calorimetric low-energy spectroscopy. In Borexino, the scintillator is a two component liquid, the solvent is Pseudocumene (PC, 1,2,4-trimethylbenzene) and the solute is a fluorescent dye (PPO, 2,5-diphenyloxazole) at a concentration of 1.5 g/l.

\begin{figure}[tb]
\begin{center}
\centering{\epsfig{file=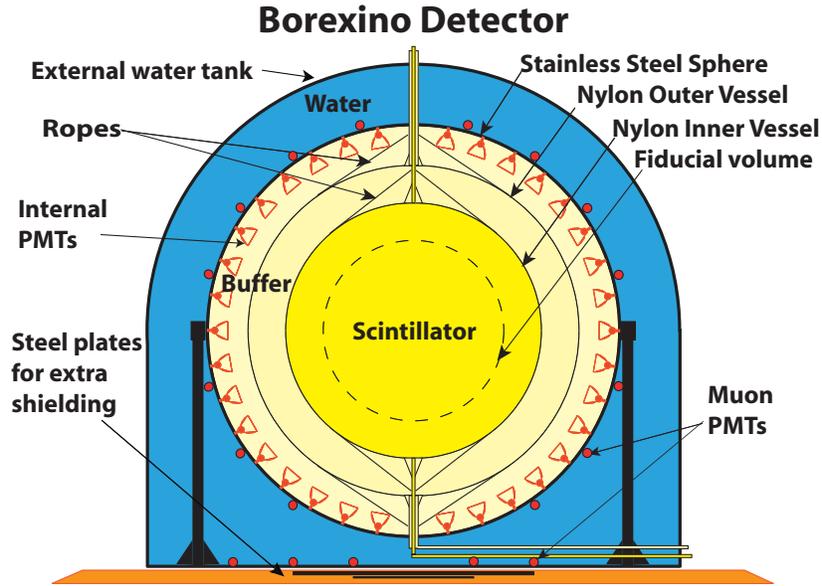,scale=0.4}}
\begin{minipage}[t]{16.5 cm}
\caption{Layout of the Borexino detector~\cite{Alimonti1}.
\label{Fig:DetectorBorex}}
\end{minipage}
\end{center}
\end{figure}

The Borexino design has been driven mostly by the need to keep the internal background as low as possible and to shield the external background as much as possible. Its layout (Fig.~\ref{Fig:DetectorBorex}) is based upon the principle of graded shielding: the detector structure consists of a set of concentric shells, more inner the shell, higher the radio-purity.

Starting from the outside, a cylindrical water tank (WT) with a radius of $\sim$9\,m and a height of 16.9\,m, filled with 2100\,m$^3$ of ultra-pure water, contains a stainless steel sphere (SSS) with a radius of 6.85\,m. The SSS supports Photomultiplier Tubes (PMT), and contains a spherical nylon Inner Vessel (IV), 4.2\,m of radius and 300\,m$^3$ of volume, surrounded by 1050\,m$^3$ of PC. The nylon wall of the IV is 125\,$\mu$m thick and, as a consequence, the buoyancy on it has to be very small: so the density of the external shielding liquid (buffer liquid) is only slightly different from the one of the internal scintillator. The buffer liquid is doped with 5\,g/l (later reduced to $\sim$2.5\,g/l) of a quencher (dimethylphthalate - DMP) in order to suppress the residual scintillation of the pure PC. Therefore the spectroscopic signals arise in a very large majority only from the interior of the IV. 

Two detectors are active in Borexino: an internal detector (ID) and an external one (OD), with the ID consisting of 2212 8-inch PMTs (ETL9351) distributed uniformly on the inner walls of the SSS. With the exception of 371 of them, the other PMTs are equipped with aluminum light concentrators in order to increase the light collection so that the optical coverage is $\sim$30\%. The optical concentrators are designed to collect only the light produced by the IV, minimizing the detection of the light originating in the buffer~\cite{Alimonti1, Alimonti2}. In the water tank another detector (outer detector - OD) consists of 208 PMTs, which detect the Cherenkov light produced by the muons in the shielding water. Despite the reduction of the cosmic muon flux by a factor of $10^6$, thanks to the 3800\,m.w.e. of the rock overburden, the muon flux in the hall C is still significant (1.2 muon m$^{-2}$ h$^{-1}$). Thus the muons crossing the whole detector are $\sim$8000/day.

The water contained in the water tank ($\sim$2\, m, at least, around the SSS in all directions; 2400\,m$^3$ in total) provides good shielding with respect to gammas and neutrons emitted by the rocks and by the surrounding laboratory environment.
The buffer liquid ($\sim$2.6\,m everywhere around the IV) has the duty to shield the nylon vessel IV from the radiations emitted by the PMTs and by the stainless steel of the SSS, and from the external residual gammas that survived through the shielding water of the water tank. The contribution to the background from the PMTs is relevant, even if they have been built keeping as low as possible the radioactivity of the components: this is especially true for the glass and ceramic components. 

However, this shielding is still not enough; the radiation emitted by the IV nylon walls, even if very thin, requires the interposition of a certain thickness of scintillator between detection volume and IV walls. Then a wall-less Fiducial Volume (FV) is software defined. In general, in the analysis of the low neutrino interactions, the FV is defined as a sphere of 3.0\,m radius with additional cuts on the $z$-vertical axis due to the emanations from the IV reinforcements at top and bottom. But in the case of geo-neutrinos, where the interactions are well tagged with a delayed coincidence, a larger FV is considered, defining a volume having its walls $\sim$20\,cm inward from the IV walls.

A second nylon balloon is installed in the buffer liquid with a radius of 5.5\,m. It functions as a barrier against the Radon, emitted mostly by the PMTs, and other gaseous contaminants originated in the SSS. (for more details see~\cite{Alimonti1}).

\subsubsection{The radioactivity issue}
\label{subsection:radioactivityBorex}

The most likely sources of contamination for the active detection core of Borexino can be summarized as follow: $^{238}$U and $^{232}$Th families and $^{40}$K, present in all materials, in the dust and particulate residuals; $^{39}$Ar, $^{85}$Kr and $^{222}$Rn present in the air, which can penetrate through possible air leaks during the filling operations; $^{210}$Pb and $^{210}$Po (produced by the $^{222}$Rn decays), probably plated on the vessel and plumbing surfaces. A careful study of these possible backgrounds was carried out in the preparation phase of Borexino, by means of the CTF (Counting Test Facility), a Borexino prototype developed as a Borexino benchmark.

All these contaminants are a source of background. They can reside directly in the scintillator, on the materials used to construct Borexino and to shield the active part of it. In addition, residuals of radiations emitted by the environment which cross the shielding materials may contribute.
The tools used in order to keep the background produced by the radioactive contaminants as low as possible are the following:
\begin{enumerate}
\item {selection of the materials and components;} 
\item {purification of the shielding liquids;}
\item {purification of the scintillator.}
\end{enumerate}

The selection of the materials has been carried out searching for very low radioactivity materials. This is the case of the stainless steel of the water tank and of the SSS. Samples of the construction materials have been checked in the Gran Sasso Low Radioactivity Laboratory. The PMTs were constructed with low radioactivity glass and ceramics. Selected components, as valves, pumps, lines, etc, have been used for the liquid handling plant.

Special attention was given to the nylon used in constructing the vessels and especially that of the IV, whose surface is in direct contact with the ultra-pure scintillating core of the detector. The raw nylon material was carefully selected, measuring the pellet radioactivity with an inductively coupled plasma mass spectrometer (ICP-MS). The pellets were then extruded in controlled area. The construction of the vessel and its assembly was finally carried out in a special clean room, whose atmosphere was strictly controlled, not only for the reduction of the dust (class 100), but also for the $^{222}$Rn. It was equipped for this purpose with cryogenic systems able to keep the radon present in the air at very low level. Finally, the assembled IV was covered with a shroud to stop alphas, produced by the decay of the $^{222}$Rn present in the atmospheric air, and low energy electrons. The installation of the vessel in the SSS was done in an atmosphere of air obtained by mixing N$_2$, purified via cryogenic systems, installed on purpose (see below), and O$_2$, stored in bottles.

In order to avoid dust and particulate, the surfaces of the SSS and of the liquid handling pipes and components were treated with a process involving electro-polishing, pickling and passivation, followed by a precision cleaning performed with detergents and high purity water. All the assemblies of the plants have been done in an N$_2$ atmosphere.

All the components and devices of the detector were previously cleaned in clean rooms of class 100, while the same SSS was converted in a clean room of class 10,000.

The shielding water was purified via an "ad hoc" plant, installed in hall C. The plant includes: a high purity deionizer, a water softening, inverse osmosis systems, ion exchange beds, ultra filtrations, and nitrogen stripping in vertical column. The water, treated as such, showed a conductivity of $\sim$18\,M$\Omega$/cm. This high-purity deionized water was used to fill the water tank and the inner zones of the detector in a preliminary filling to assure a further cleaning before the introduction of pseudocumene and the other components (fluor and quencher). In addition, it has also been used also for cleaning and rinsing the surfaces of the plants. 

\subsubsection{The purification of the scintillator components}
\label{subsec:purificationBX}

Special care was used to clean the liquid scintillator component: PC and PPO.  A purification system was developed to purify the PPO in a concentrated master solution obtained dissolving the PPO in PC at high concentration. Pre-purification is connected to the PPO property to solidify below 70$^{\circ}$C and, as a consequence, the possibility that it blocks the transfer lines. The pre-purification has been done first with a water extraction process and then with a distillation. In addition nitrogen degasification stripping was carried out directly in the PPO storage tank to remove noble gas impurities.

The PC has been prepared by Polimeri Europa in the Sardinia plant, treating the crude oil extracted from very old Libyan layers. The aim of this choice is to keep at very low level the contamination of $^{14}$C, which cannot be cleaned: the crude oil, in very old layers, remained millions of years shielded with respect to the cosmic rays, which produce $^{14}$C in collisions with the oil molecules.

Once produced, the PC was shipped to the Gran Sasso underground laboratory with specially designed transport tanks that were pre-cleaned and treated before use. The PC transport time was kept as short as possible ($\sim$20 h), avoiding long exposure to cosmic rays, thus minimizing the cosmogenic $^7$Be production.

The PC was stored underground in special vessels, with the internal surfaces treated similarly to the detector surfaces, and it was subjected to the cleaning processes during the filling operation of the SSS and the IV. The cleaning processes included three steps: distillation, ultrafiltration, N2 stripping. The distillation was carried out with a six-stage column operating at 100\,mbar  and 100$^{\circ}$C of temperature. This process has been done at a relatively low temperature just to avoid that contaminants embedded in the column components would be stripped and inserted into the distilled liquid.

Nitrogen stripping, using very pure nitrogen gas, removes noble gases and other gaseous contaminants. A special nitrogen producing facility was developed just to produce nitrogen with only ultra-traces of $^{222}$Rn, $^{39}$Ar and $^{85}$Kr that are present in the atmosphere. Purified nitrogen was cryogenically generated with a sub-boiling system; the distribution and the storage of this very pure nitrogen supply was decoupled from other nitrogen circuits. In addition we have to emphasize that all plants of the Borexino subsystems were constructed to reach a tightness of $10^{-8}$\,scc/s.

The radiopurity of the Borexino detector is summarized in Table~\ref{tab:BXradiopurity}, where it is compared with the typical radio-purity of the various materials. The low contamination levels achieved are unprecedented and surpass design specifications.

\begin{table}
\begin{center} 
\begin{minipage}[t]{16.5 cm}
\caption{Radiopurity of the Borexino detector after the filling in May 2007.}
\label{tab:BXradiopurity}
\end{minipage}
\begin{tabular}{l|l|l|l|l}
\hline
Name	  &   Source         &     Typical                       &    Required    & Achieved \\ 
\hline
$^{14}$C     & intrinsic PC  & $\sim 10^{-12}$ g/g   & $\sim$$10^{-18}$ g/g   & $\sim$$2 \times 10^{-12}$ g/g \\
\hline
$^{238}$U   & dust            & $10^{-5} - 10^{-6}$ g/g   & $<  10^{-16}$ g/g   & $ (5.0 \pm 0.9)  \times 10^{-18}$ g/g \\
$^{232}$Th   &            &                                           &                                 & $ (3.0 \pm 1.0)  \times 10^{-18}$ g/g \\
\hline
$^{7}$Be   & cosmogenic           & $\sim$$3 \times 10^{-2}$ Bq/ton   & $ <  10^{-6}$ Bq/ton   & not observed \\
\hline
$^{40}$K   & dust, PPO           & $\sim$$2 \times 10^{-6}$ g/g (dust)   & $ <  10^{-18}$ g/g   & not observed \\
\hline
$^{210}$Po   & surface contamination           &                      & $ < 7$ cpd /ton   & May07: 70 cpd/ton\\
   &            &                      &                                                                                & May09: 5 cpd/ton \\
\hline
$^{222}$Rn   & emanation, rock          &      10 Bq/l (air, water)     & $ < 10$ cpd /100 ton   &  $< 1$ cpd/ 100 ton\\
                      &                                   &  100 - 1000 Bq/kg (rock) &                                    & \\
\hline
$^{39}$Ar       & air, cosmogenic  & 17\,mBq/m$^3$ (air)     & $<$1\,cpd/100\,ton    & $<<$ $^{85}$Kr \\
\hline
$^{85}$Kr       & air, nuclear weapons &       $\sim$1\,Bq/m$^3$ (air)     &   $<$ 1\,cpd/100\,ton    &   $ 30 \pm 5$\,cpd/100\,ton \\
\hline
\end{tabular}
\end{center}
\end{table}  

A further purification campaign was carried out in 2010 - 2011 that was specifically devoted to reduce $^{85}$Kr and $^{210}$Bi in the liquid scintillation. The campaign reduced $^{85}$Kr contamination to negligible level and $^{210}$Bi to $18 \pm 1.5$\,cpd/100\,ton. Contamination from $^{210}$Po was not reduced, but, leaving untouched and undisturbed the detector, it is naturally decaying ($\tau$ = 200\,days);  as of April 2013, its contribution is already less than 180\,cpd/100\,ton.

\subsubsection{The antineutrino detection and the resolution in Borexino}
\label{Subsec:resolutionBX}

In Borexino antineutrinos are detected via the inverse beta-decay reaction, see Eq.~\ref{Eq:InvBeta} with a kinematic threshold at 1.806\,MeV. This reaction is very well tagged because (see Sec.~\ref{Sec:Results}) it produces a prompt signal from positron and a delayed signal, a 2.2\,MeV gamma.

Radiopurity of the detector significantly influences also the study of the antineutrinos, despite well tagged reactions, due mostly to two contributions: neutron production from $\alpha$'s (in particular via the reaction $^{13}$C($\alpha$, n)$^{16}$O) and the accidental coincidences due to the background rate. As discussed in the Sec.~\ref{Sec:Results}, these backgrounds are negligible in Borexino. 

The energy estimators used in Borexino are three: $N_p$, the number of the PMTs, which detected one or more hits; $N_h$, the number of hits; $N_{pe}$, the number of photoelectrons (p.e.) for each event. Each of these estimators have pro and cons: the first two are better only at low energy, when 1\,p.e./PMT is dominating, while the last one is used in general over 2\,MeV of released energy. 

A calibration campaign was carried out in 2009 and 2010,  placing 11 different sources in the center and in many off-center positions ($\sim$300) in the IV: $^{37}$Co, $^{139}$Ce, $^{203}$Hg, $^{85}$Sr, $^{54}$Mn, $^{65}$Zn, $^{60}$Co, $^{40}$K, $^{222}$Rn, $^{14}$C. In addition, an AmBe source, producing about 10 neutrons with energies up to 10\,MeV, was deployed in twenty-five positions to study the detector response to neutrons and to protons recoiling off neutrons. In addition a $^{228}$Th source was placed in the buffer region of the detector to study the detector response to the major external background source, 2.615\,MeV gamma rays of $^{208}$Tl, a daughter isotope of $^{228}$Th. These calibrations reduced the systematic error associated with all Borexino results and helped to optimize the Monte Carlo simulation of the detector response~\cite{Back}.

The light yield is $\sim$500\,p.e./MeV; the energy scale resolution is $5\% / \sqrt {E {\rm [MeV]}}$ in the range 200 - 2000\,keV; over 2\,MeV it is slightly worse because in that range the calibration was less accurate. The position reconstruction accuracy is $\sigma (x,y,z)$ = 10 - 12\,cm.

The stability of the ID is continuously checked by means internal signals, as $^{210}$Po, $^{14}$C, $^{11}$C, the shoulder of $^{7}$Be and, just after the filling when $^{222}$Rn was present, via the $^{210}$Bi-$^{210}$Po sequence, while the PMTs are monitored via pulsed laser light distributed via optical fibers~\cite{Back}.

\subsection{KamLAND}
\label{subsec:kamland}

\begin{figure}[tb]
\begin{center}
\centering{\epsfig{file=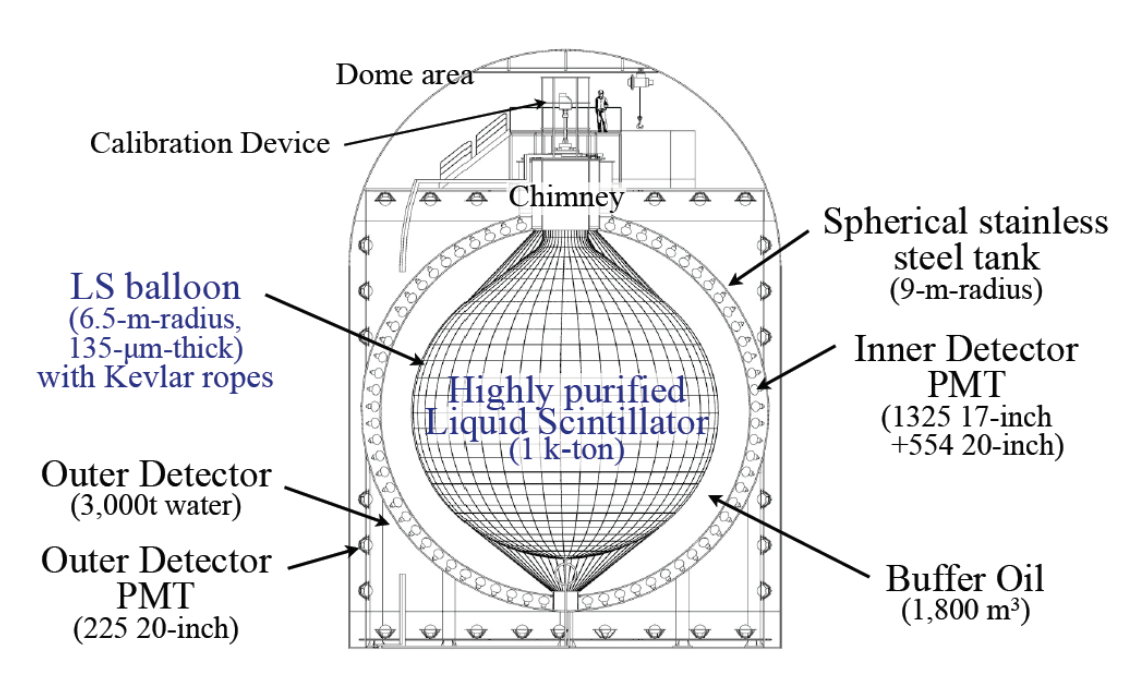,scale=1.4}}
\begin{minipage}[t]{16.5 cm}
\caption{Layout of the KamLAND detector~\cite{GandoT}.
\label{Fig:DetectorKL}}
\end{minipage}
\end{center}
\end{figure}

The geometrical structure of KamLAND is very similar to the Borexino's (Fig.~\ref{Fig:DetectorKL}), but the size in KamLAND is larger and it has a compositionally different scintillator and buffer liquids.  The KamLAND detector was designed to study nuclear reactor antineutrinos with a mean energy of prompt signals of $\sim$3\,MeV, and therefore the requirements of low radioactivity are less stringent than in Borexino.  

The liquid scintillator consists of 89\% dodecane and 20\% Pseudocumene, plus $1.36 \pm 0.03$\,g/l of the fluor PPO. About 1\,kton of this liquid scintillator is contained in a 6.5\,m radius spherical balloon made of transparent nylon-EVOH (ethylene vinyl alcohol copolymer) composite; this Inner Vessel (IV), supported by a network of Kevlar ropes, is the active core of the KamLAND detector. The IV has 135\,$\mu$m thick walls; its total volume is $1171 \pm 25$\,m$^3$.

The IV is surrounded by a buffer liquid consisting of 57\% isoparaffin and 43\% dodecane oils, which fills a 9 m radius stainless-steel sphere (SSS), which functions also as a support for the PMTs. The specific gravity of the buffer liquid is 0.04\%, lower than the one of the liquid scintillator, whose density is 0.78\,g/cm$^3$. A 3\,mm thick acrylic balloon, 8.3\,m of radius, functions as a barrier against the radon emitted by the PMTs. Finally 3.2\,kton of water surround the SSS and are contained in a cylindrical Water Tank (WT).

This sequence of the layers and different liquids, as in the case of Borexino, is designed to shield the innermost detector from the radiations emitted by the rocks and by the materials that make up the detector~(Fig.~\ref{Fig:DetectorKL}).

The signals, which are produced in the Inner Detector (ID), are read by an array of 1325 17-inch fast PMTs and 554 20-inch PMTs; this array, supported by the SSS, assures an optical coverage of $\sim$34\%. The Outer Detector (OD) processes the water-Cherenkov light, produced in the Water Tank and read by 225\,PMTs, mounted on the internal WT walls (for more details see~\cite{Abe2008, Abe2010}).
 
\begin{figure}[tb]
\begin{center}
\centering{\epsfig{file=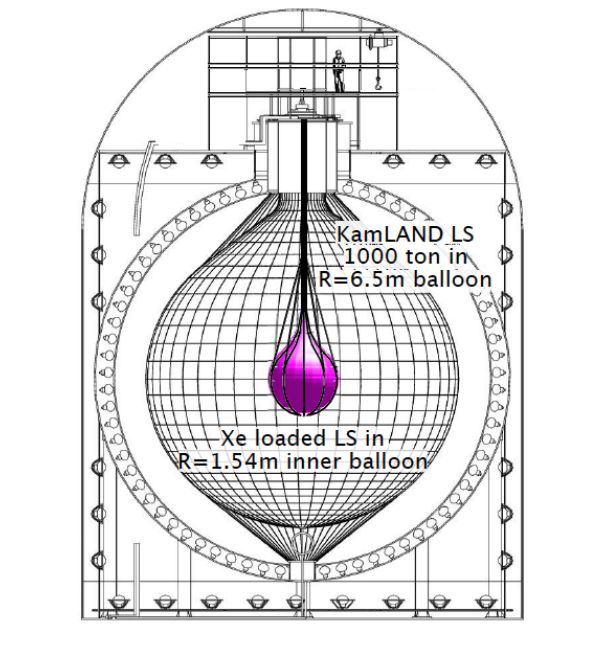,scale=1.3}}
\begin{minipage}[t]{16.5 cm}
\caption{Structure of the KamLAND-ZEN detector~\cite{Gando}.
\label{Fig:DetectorKLZen}}
\end{minipage}
\end{center}
\end{figure}

Having started data acquisition in January 2002 KamLAND has studied important aspects of neutrino oscillations using antineutrinos produced by nuclear reactors. In September 2011 KamLAND began the study of the $0\nu\beta\beta$-decay. To this purpose a transparent nylon balloon, 3.08\,m diameter, containing 13\,tons of Xe-loaded liquid scintillator was inserted into the center of the ID (KamLAND-ZEN, see. Fig.~\ref{Fig:DetectorKLZen}).

An important evolution of the geo-neutrinos study in the KamLAND-ZEN   experiment started in 2012, when the nuclear power plants have been switched off in Japan for check and maintenance (the last plant has been switched off at the beginning of May 2012). Thus a favorable period for the geo-neutrino study in Japan has started, because the near field reactor antineutrino background was suppressed. 
KamLAND is installed under the Ikenoyama peak, which assures an overburden of $\sim$2700\,m water equivalent.

\subsubsection{The radiopurity in KamLAND}
\label{subsec:KLradiopurity}

The purification of the liquid scintillator and the buffer oil has been done in two steps: initially the liquids have been submitted to a water extraction process during the detector filling period in March 2002; later each component of the liquid scintillator has been distilled in three separate towers and, after mixed, submitted to a high purity N$_2$ purging process (2007 - 2009 campaign). Each component has been distilled at optimal temperature and reduced pressure in order to prevent changes to the properties of the liquid~\cite{GandoT}. After the purification, $^{238}$U has been reduced to $(1.5\pm1.8) \times 10^{-19}$ g/g, $^{232}$Th to $(1.9 \pm 0.2) \times 10^{-17}$ g/g, and $^{40}$K to a limit $< 4.5 \times 10^{-18}$ g/g; $^{210}$Po is $\sim$2\,mBq/m$^3$, $^{210}$Bi $<$1\,mBq/m$^3$, $^{85}$Kr $\sim$0.1\,mBq/m$^3$. In addition, accidental background is due to the vessel walls (Inoue, private communication). 

The time range 2002-2007 of data taking defines the Period 1, while the period after the purification campaign until the fall 2011 is the Period 2. Since September 2011 the  events are collected with the Xe balloon already installed: this period is the Period 3.

A Fiducial Volume (FV) at KamLAND is software defined during the  analysis: it is a sphere of 6\,m of radius. In addition, in the period 3, only  events  in the FV and well outside the Xe balloon are considered~\cite{Gando}.

 The Kamioka mine underground water, used in the water extraction operations and in the shield of the detector, was cleaned by means a purification plant consisting of pre-filters, UV sterilizer, ion exchanger, vacuum degasser and reverse osmosis (RO) in order to remove dust, metal ions, natural radioisotopes, such as $^{222}$Rn, and to eliminate bacteria: the $^{238}$U and $^{232}$Th in water has been reduced to the order of $10^{-13}$\,g/g.

The two more important sources of background, due to the  radioactivity internal to the detector, are due to the reaction $^{13}$C($\alpha$,n)$^{16}$O and to accidental coincidences, which can mimic the reaction Eq.~\ref{Eq:InvBeta}. In addition, cosmogenic nuclides represent an additional possible background: the rate of cosmic muons in KamLAND  is $0.198 \pm 0.014$\,Hz (all this matter is discussed in Sec.~\ref{Sec:Results}). 

 \subsubsection{The antineutrino detection and the resolution in KamLAND}
\label{Subsec:resolutionKL}

KamLAND uses the reaction given in Eq.~\ref{Eq:InvBeta} for the detection of antineutrinos. A difference with respect to Borexino is the time interval between the prompt and the delayed signals, which in Kamland is $207.5 \pm 2.8$\,$\mu$s~\cite{Abe2008}.

The energy assessment processes the information stored in the digitized PMT signals by identifying individual PMT pulses in waveform. The integrated area ("charge") is computed from the individual pulses. 

Energy and spatial calibrations of the detector is done with 6 sources: $^{203}$Hg, $^{68}$Ge, $^{65}$Zn, $^{60}$Co, $^{65}$Zn, $^{241}$Am + $^9$Be and $^{210}$Po + $^{13}$C. Also in Kamland, as in Borexino, the residual background contaminants function as a continuous monitor of the detector response. The achieved vertex resolution is $\sim$$12$\,${\rm cm} / \sqrt {E {\rm [MeV]}}$, while the energy resolution is  $6.5\% / \sqrt {E {\rm [MeV]}}$~\cite{Abe2008}.

\section{Analysis and results obtained by Borexino and KamLAND}
\label{Sec:Results}

\subsection{Measuring geo-neutrinos}
\label{subsec:measuring}

The flux of geo-neutrinos at a given location on the surface of the Earth depends on the distribution of heat producing elements ($^{238}$U, $^{232}$Th, and $^{40}$K) within the mantle and the crust, as described in Sec.~\ref{Sec:GeoSignal}. Geo-neutrinos can travel as much as 12,000\,km to reach a detector location such as Kamioka or Gran Sasso. It has been established by a number of observations~\cite{GonzalesGarcia2010} that neutrinos traveling from a source to a detection point can oscillate between one flavor and another. Therefore, there is a finite survival probability, $P_{ee}$, for a given electron neutrino, for example, to be detected as such at some distance from its production position. The origin of neutrino oscillations is found in the difference between mass and flavor eigenstates. In the case of the three-flavor $(\nu_e, \nu_{\mu}, \nu_{\tau})$ and three mass eigenstates $(\nu_1, \nu_2, \nu_3)$ scenario the two sets of eigenstates are related by a mixing matrix which depends on three angles and one CP phase~\cite{GonzalesGarcia2010}. The survival probability is determined by the antineutrino energy, the distance traveled, and by the neutrino oscillation parameters: namely, the mixing angles and the mass-squared differences between mass eigenstates. In particular, for geo-neutrinos (thus, considering their energy spectra), the present determination of the neutrino oscillation parameters establishes a neutrino oscillation length on the order of 140\,km. By comparing the averaged distance traveled by the geo-neutrinos and the oscillation length, the survival probability is determined to be:
\begin{equation}
P_{ee} = \cos^4\theta_{13}\left(1-\frac{1}{2}\sin^2 2\theta_{12}\right)+\sin^4\theta_{13} \sim 0.55\pm0.01
\label{Eq:Pee}
\end{equation}
where $\theta_{12}$ and $\theta_{13}$ are the mixing angles between mass eigenstates. As a consequence of the oscillations the flux of geo-neutrinos at an experimental location will be reduced by about 45\%.

Geo-neutrinos, electron antineutrinos, interact with matter only through the weak interactions and thus the probability of such interaction is very low. In order to be able to detect the weak geo-neutrino signal it is therefore important to shield the experimental setup from cosmic radiations and to place the detector inside underground laboratories as it was described in Sec.~\ref{Sec:detectors}. In order to increase the number of detected events, large volume detectors are required. Borexino's liquid scintillator target has about 280\,tons while that of KamLAND has about 1\,kton. 

Electron antineutrinos are detected in liquid scintillator detectors by means of the inverse-beta decay interaction shown in Eq.~\ref{Eq:InvBeta}, having a kinematic threshold of 1.806\,MeV. This means that only a high energy tail from $^{238}$U and $^{232}$Th geo-neutrinos above this energy threshold can be detected. In particular, the fraction of detectable geo-neutrinos is 6.3\% for $^{238}$U  and 3.8\% for $^{232}$Th, while $^{40}$K geo-neutrinos are all below this threshold and cannot be detected. The emitted positron promptly comes to rest and annihilates emitting two 511\,keV\,$\gamma$-rays. The deposited positron's kinetic energy and the energy of the two $\gamma$-rays are detected in a single, so called {\em{prompt event}}, with a visible energy $E_{prompt}$ related to the energy of incident geo-neutrino $E_\nu$ by a simple relation
\begin{equation}
E_{prompt} = E_\nu - 0.784  ~~~~~~{\rm [MeV]}
\label{Eq:Evisible}
\end{equation}

The free neutron emitted is thermalized and then typically captured on protons with a meantime in the range of 200 - 300\,$\mu$s, depending on the scintillator type. The neutron capture results in the emission of a 2.2\,MeV de-excitation $\gamma$-ray, which provides a so called {\em{delayed event}}. In the thermalization process, the emitted neutron loses the memory of its original direction, and no directionality information about the incident antineutrino can be obtained by any of the two experiments measuring geo-neutrinos. The cross section of this detection interaction (see also Sec.~\ref{Sec:Intro}) is known with a high precision below 1\% and can be found for example in~\cite{strumia}.

The space and time coincidence of the prompt and the delayed event provides a unique tool to strongly suppress possible background sources that mimick antineutrino interactions. Of course, other electron antineutrinos having a different origin with respect to geo-neutrinos represent a possible background to geo-neutrino measurement as well. Antineutrinos emitted from nuclear power plants are the only relevant antineutrino background. 

The Borexino experiment was designed to measure solar neutrinos which are detected through a simple scattering off electrons, which does not provide a coincidence tag. Thus, neutrino interaction cannot be distinguished from an event due to the radioactive contaminants of the detector. Borexino succeeded in achieving an extreme level of radiopurity of the scintillator and of the construction materials, having as a consequence that backgrounds different from reactor antineutrinos are at almost negligible levels for the geo-neutrino measurement (see Sec.~\ref{subsection:radioactivityBorex}). In addition,  there are no nuclear power plants in Italy and the mean weighted distance of the reactors from the LNGS site is more than 1000\,km.
 
The KamLAND experiment was designed to study reactor antineutrinos which are detected by the same inverse beta decay interaction providing the same coincidence tag as in geo-neutrino detection. Therefore, such an extreme radiopurity as in Borexino is not strictly required in the KamLAND experiment and some non-antineutrino backgrounds, as accidental coincidences or ($\alpha$, n) interactions represent a non-negligible background source for KamLAND geo-neutrino measurements. In addition, proximity of nuclear power plants causes an increased rate of events due to reactor antineutrinos. Recently, after the Fukushima nuclear accident occurred in March 2011, all Japanese nuclear reactors were temporarily switched off providing a unique possibility for geo-neutrino measurement in KamLAND.

\subsection{Reactor antineutrino background}
\label{subsec:reactors}

Antineutrinos from nuclear power plants are a relevant background source for geo-neutrino measurements. It is therefore crucial to be able to calculate the expected number of events and the corresponding spectral shape of the electron antineutrinos from nuclear plants. There are four principal isotopes used as fuels in the cores of nuclear power plants: $^{235}$U, $^{238}$U, $^{239}$Pu, and $^{241}$Pu. They contribute to the total thermal power of the plant in the different, so called, power fractions, which depend on the reactor type and on the burn-up stage of the individual core. The most recent parametrization can be found in~\cite{mueller}. The overall spectral shapes are very similar to the older parametrization given in~\cite{huber} while it predicts a 3.5\% higher total flux. The energy spectrum of reactor antineutrinos overlaps with that of geo-neutrinos and extends to about 8\,MeV. In the analysis, the antineutrino candidates with energies above the geo-neutrino end point strongly constrain the contribution of the reactor antineutrinos.  

The oscillation phenomenon shapes the energy spectrum of the electron antineutrinos arriving from an individual core to the detector site. The survival probability depending on the neutrino oscillation parameters is also a function of the antineutrino energy and of the distance from the source to the detector. The oscillation length for antineutrinos of a few MeV is at the level of 100\,km and therefore in the calculation of the expected signal rate and the spectral shape it is important to consider all reactors individually. In the world there exist about 450 nuclear power plants concentrated mostly in Europe, North America, Japan, and Korea.  

\subsection{Non-antineutrino background sources}
\label{subsec:bgr}

Non-antineutrino background sources can be divided in two main categories: cosmogenic background and the background due to the radioactive contaminants of the scintillator and the construction materials of the detector, which includes the scintillator itself, the containment vessels, and the photomultipliers. 

The cosmogenic background is dominated by the cosmic muons and the spallation products, such as the fast neutrons and $^9$Li and $^8$He isotopes decaying in ($\beta$ + neutron) branches perfectly mimicking antineutrino interactions. Fast neutrons can penetrate through the construction materials of the detector and before their actual capture on protons (and thus before the 2.2\,MeV $\gamma$-ray emission) they can scatter off a proton, which can provide a prompt signal. Real coincidences of a scattered proton and a 2.2\,MeV gamma ray can therefore mimic a geo-neutrino interaction. The proton is a highly-ionizing particle and in principle, can be distinguished in liquid scintillators by the pulse-shape identification techniques from electron/positron or gamma ray interactions.  

Overburden rocks above the underground laboratories in which the detectors are placed reduce the muon flux by several orders of magnitude, see Sec.~\ref{Sec:detectors}. The remaining muons are detected by the Cherenkov Outer Detectors with high efficiency. A time veto after each detected muon suppresses muon-produced cosmogenic backgrounds. For muons passing only through the Outer Detector, it is sufficient to fight only the fast neutrons having a capture time of 200-300 $\mu$s since the charged nuclei as $^9$Li and $^8$He cannot penetrate to the scintillator volume. Instead, muons passing through the scintillator can produce such isotopes directly there. The isotopes of $^9$Li  and $^8$He have decay times of 26\, ms and 173\,ms, respectively, and longer vetoes are typically applied after such internal muons. 

The level of backgrounds due to non-detected muons can be estimated from a known muon detection inefficiency. The remaining contribution of the backgrounds due to fast neutrons and $^9$Li  and $^8$He isotopes produced by muons passing through the detector, which remain present in the tails even after the applied vetoes, can be estimated on the basis of the background events observed during the veto after the muons. A critical point is an estimation of the fast-neutron background due to the muon interactions within the rock walls surrounding the detector. This contribution has to be estimated by a careful Monte Carlo simulation.

An important background component is due to the accidental coincidences of non-correlated events from the interactions of radioactive contaminants of the construction materials and/or the scintillator. An optimized fiducial volume selection can strongly suppress the contribution of this background type. An optimization of the coincidence time window, $dt$, between the prompt and the delayed candidate can also contribute in maximizing the signal-to-background ratio. The contribution of the accidental background in the time correlated window can be determined by searching for the coincidences (with the same energy, position, pulse shape cuts) in an off-time window, typically much longer than $dt$ in order to improve the statistical precision. 

The ($\alpha$, n) interactions are another important source of background, namely the interaction $^{13}$C($\alpha$, n)$^{16}$O, as first investigated by KamLAND~\cite{Abe2008}. The $\alpha$-particles are mostly from $^{210}$Po contaminant of liquid scintillator. The prompt event can be due to the three different processes: 
\begin{enumerate}
\item{de-excitation 6.1\,MeV gamma ray of the $^{16}$O produced in the excited state;}
\item{4.4\,MeV $\gamma$ ray from the de-excitation of $^{12}$C excited by neutron; }
\item{proton scattered off by thermalizing neutron which, in principle, can be at least partially identified by the pulse shape identification techniques.}
\end{enumerate}
The neutron produced with energies up to 7.3\,MeV thermalizes and is captured on proton producing 2.2\,MeV gamma ray detected as a delayed candidate. The probability of the $^{210}$Po nucleus to give ($\alpha$, n) interaction in pure $^{13}C$ is discussed in~\cite{mckee}. The $^{210}$Po contamination of the scintillator is easily measurable by identification of a peak due to the $\alpha$s in the energy spectrum of single events. In Borexino, $^{210}$Po contamination is much lower than in KamLAND. The isotopic abundance of $^{13}$C in organic compounds is at the level of 1.1\%.

\subsection{KamLAND geo-neutrino analysis}
\label{subsec:ResultsKL}

\begin{figure}[tb]
\begin{center}
\centering{\epsfig{file=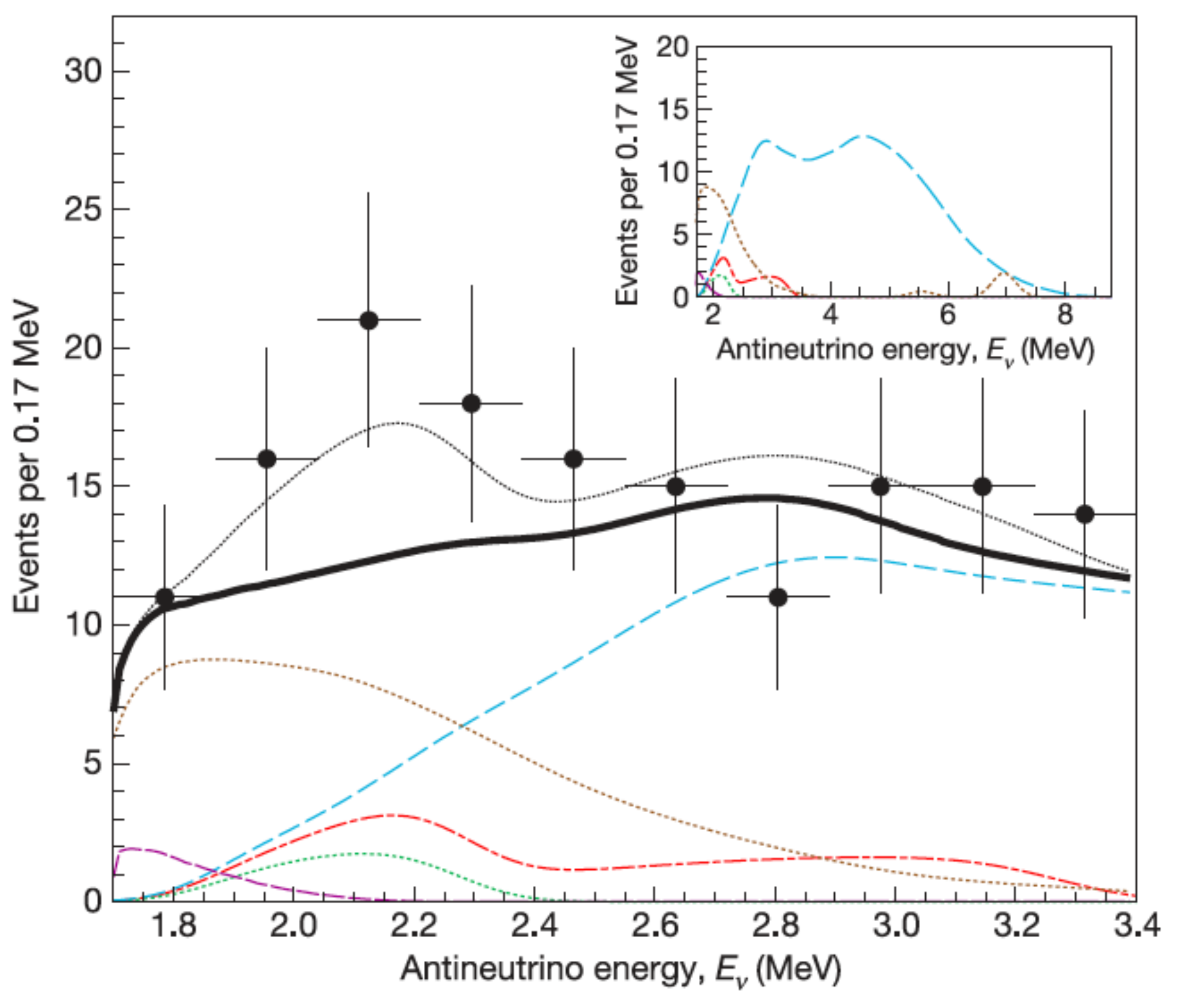,scale=0.5}}
\begin{minipage}[t]{16.5 cm}
\caption{Electron antineutrino energy spectra measured by KamLAND in 2005 taken from~\cite{araki2005}. The thick solid line shows the total expected signal excluding geo-neutrinos. The dashed light blue line is the signal from nuclear reactor antineutrinos. The dotted brown line corresponds to $^{13}$C($\alpha$,n)$^{16}$O background. In the inset, the expected spectra extended to\,8 MeV energy are shown.
\label{Fig:GeoFirstKL}}
\end{minipage}
\end{center}
\end{figure}

KamLAND provided in 2005~\cite{araki2005} the first experimental investigation of geo-neutrinos. In Fig.~\ref{Fig:GeoFirstKL} we show the energy spectrum of collected data and expectations. From this measurement it is shown clearly that the main background sources for the geo-neutrino detection in KamLAND are electron antineutrinos from nuclear reactors near the detector, and $\alpha$-particle induced neutron background from radioactive contaminants within the detector active mass. In particular, for this latter source of background the reaction $^{13}$C($\alpha$,n)$^{16}$O with the $\alpha$ particle from $^{210}$Po is the dominant contribution. For the 2005 data set the live time corresponds to $749.1 \pm 0.5$\,days after selection cuts with a total exposure of $(7.09\pm0.35)\times10^{31}$ target proton-year. In this first analysis the overall detection efficiency for geo-neutrino candidates with energy between 1.7 and 3.4\,MeV is determined to be $0.687 \pm 0.007$. The total number of observed electron antineutrinos is 152. A rate only analysis gives 25$^{+19}_{-18}$ geo-neutrino candidates from U and Th. An unbinned maximum likelihood fit assuming the shape of the signal and a Th/U mass ratio of 3.9 gives a best-fit consistent with the rate analysis. The significance of the geo-neutrino observation is at the level of 95\% C.L.

In 2008, KamLAND published new data for a total exposure equal to $2.44 \times 10^{32}$ target proton-year~\cite{Abe2008}. The geo-neutrino analysis is performed by fixing the Th/U mass ratio again at the chondritic value 3.9. The combined U + Th geo-neutrino signal corresponds to $73 \pm 27$ events or to a flux equal to $(4.4 \pm 1.6) \times 10^6$\,cm$^{-2}$s$^{-1}$, in agreement with the used reference model~\cite{enomoto}. This work shows again that the most important source of background comes from $^{210}$Po through $^{13}$C($\alpha$,n)$^{16}$O.

A more detailed study of geo-neutrinos is presented in Gando et al., 2011~\cite{gando2011}. The data collected correspond to a total live-time of 2,135\,days. The exposure is equal to $(3.49 \pm 0.07) \times 10^{32}$ target proton-year. The number of observed candidates in the geo-neutrino prompt energy range [0.9, 2.6]\,MeV is 841 against a predicted number of $729.4 \pm 32.3$\,events from reactors and background sources. Fixing the Th/U mass ratio to a chondritic value, the best-fit gives a signal of geo-neutrinos equal to 106$^{+29}_{-28}$, which corresponds to a flux on surface of 4.3$^{+1.2}_{-1.9} \times 10^6$ cm$^{-2}$s$^{-1}$. The null hypothesis for geo-neutrinos is disfavored at the 99.997\% C.L. 

	In 2013, new data from KamLAND~\cite{Gando} was published including a period with reactor-off activity following the Fukushima nuclear accident in March 2011. The data reported sum up to a total live time of 2991\,days from March 9, 2002 to November 20, 2012. The exposure is determined to be $(4.90 \pm 0.10) \times 10^{32}$ target proton-year. The data set is divided into three main periods: early KamLAND data-taking (1486\,days); after purification of the liquid scintillator (1154\,days); reactor-off period (351\,days). The number of reactor antineutrinos, which gives the largest contribution to the signal in KamLAND, is predicted from reactor data including thermal power variations and fuel exchange. The antineutrino emission spectra from Japanese commercial reactors are determined considering relative fission yields averaged over the live-time period given by: ($^{235}$U, $^{238}$U, $^{239}$Pu, $^{241}$Pu) = (0.567, 0.078, 0.298, 0.057). The contribution of Korean reactors, Japanese research reactors, and other world reactors is also included (the overall effect of these contributions accounts to about 6\%). After all selection cuts the expected number of events from reactors without oscillations is $3564 \pm 145$.

An unbinned rate + shape analysis is performed in the energy range [0.9, 8.5]\,MeV by means of the following $\chi^2$:
\begin{eqnarray}
           \chi^2  & = & \chi^2 \left(  \theta_{12}, \theta_{13},  \Delta m^2_{21}, N^{geo}_{U,Th}, N_{BG,1 \rightarrow 5},N_{sys,1 \rightarrow 4}  \right) - \nonumber \\
 &  & 2 \ln{ L_{shape}\left(\theta_{12}, \theta_{13},  \Delta m^2_{21}, N^{geo}_{U,Th}, N_{BG,1 \rightarrow 5},N_{sys,1 \rightarrow 4}   \right)} + \nonumber \\
                  &    & \chi^2_{BG} \left( N_{BG,1 \rightarrow 5} \right) + \chi^2_{sys} \left( N_{sys,1 \rightarrow 4} \right) + \chi^2 \left(\theta_{12}, \theta_{13},  \Delta m^2_{21} \right) 
\label{Eq:Chi2KL}
\end{eqnarray}
where $N_{BG}$ accounts for five different background sources (accidentals, Li-He cosmogenic, fast neutrons and atmospheric neutrinos, two $^{13}$C($\alpha$,n)$^{16}$O contributions depending on the final state of $^{16}$O). $N_{sys}$ accounts for four systematic errors (reactor spectrum, energy scale, event rate, and energy-dependent detection efficiency). The last three terms in the definition of the $\chi^2$ are penalty functions on background, systematics, and neutrino oscillation parameters. The likelihood, $L_{shape}$, takes into account the single event energy and the spectral shape. 

	In KamLAND electron antineutrinos are searched for by applying a number of selection cuts. In particular,
\begin{enumerate}
\item{prompt energy $E_p$ cut: $0.9 < E_p {\rm ~[MeV]} < 8.5$ (2.6\,MeV for geo-neutrinos);}
\item{delayed energy $E_d$ cut : $1.8 < E_d {\rm ~[MeV]} < 2.6$ (capture on proton) or $4.4 < E_d {\rm [MeV]} < 5.6$ (capture on carbon); }
\item{spatial correlation of prompt and delayed event: $\Delta R < 2$\,m;}
\item{time correlation between the delayed and prompt candidate: $0.5 < \Delta t ~[\mu$s$] < 1000$; }
\item{fiducial volume on $R_{p,d}$.}
\end{enumerate}
 Moreover, to maximize the sensitivity to electron antineutrinos, a figure of merit is made based on Monte Carlo pdf's for neutrinos, $f_\nu$, and accidental coincidences, $f_{acc}$. For each candidate one determines $L = f_\nu / (f_\nu + f_{acc})$ and compares it with a selection value to maximize the figure-of-merit given by $S/\sqrt{(S+B_{acc})}$, where $S$ and $B_{acc}$ are the expected signals for neutrinos and accidentals, respectively. In KamLAND, the vertex and energy reconstructions are determined by means of calibrations with radioactive sources, see also Sec.~\ref{Subsec:resolutionKL}.

\begin{figure}[tb]
\begin{center}
\centering{\epsfig{file=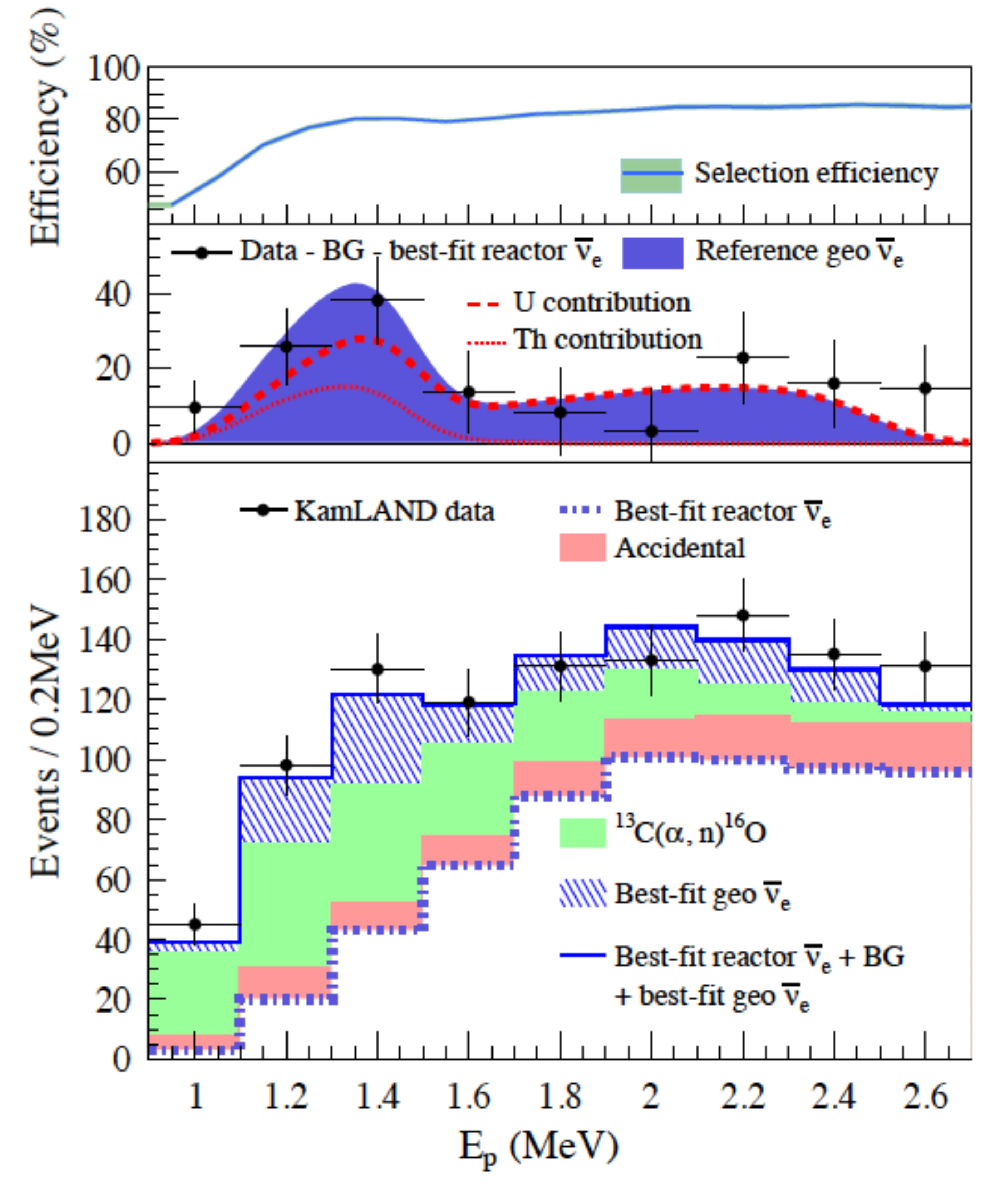,scale=0.5}}
\begin{minipage}[t]{16.5 cm}
\caption{Lower panel: The prompt candidate energy spectrum in the geo-neutrino energy window from KamLAND 2013 data~\cite{Gando}. Middle panel: observed geo-neutrino signal after subtraction of reactor antineutrinos and backgrounds. Top panel: detection efficiency as a function of energy.
\label{Fig:KL2013}}
\end{minipage}
\end{center}
\end{figure}

In the 2013 data set, the number of observed events passing all selection criteria is equal to 2611. The overall background is determined to be $364.1 \pm 30.5$ events. Some 57\% of the background is due to $^{13}$C($\alpha$,n)$^{16}$O. This background after purification (distillation) has been reduced by a factor of 10. The accidental background is also reduced by a factor of 5. The contribution of systematic uncertainties due mainly to the energy scale, fiducial volume, reactor power, and fuel composition is estimated to be 4\%. In Fig.~\ref{Fig:KL2013} we report the prompt candidate energy spectrum in the geo-neutrino energy window from Gando et al., 2013~\cite{Gando} and the selection efficiency as a function of energy. Assuming a Th/U mass ratio of 3.9, the total number of U + Th geo-neutrino events is determined to be 116$^{+28}_{-27}$, which corresponds to a flux of $(3.4 \pm 0.8)\times 10^6$\,cm$^{-2}$s$^{-1}$. This flux corresponds to a geo-neutrino signal $S_{geo}$ = 29.8 $\pm$ 7.0\,TNU, which is in agreement with the expected signal $S_{geo}$(U+Th) = $31.5^{+4.9}_{-4.1}$\,TNU calculated in~\cite{huang} on the base of the refined model of the region near KamLAND described in Sec.~\ref{subsec:kamioka}. The null hypothesis is disfavored at the level of $2\times 10^{-6}$.

\subsection{Borexino geo-neutrino analysis}
\label{subsec:ResultsBX}

Borexino has provided the first geo-neutrino observation at more than $4\sigma$ in 2010~\cite{BX2010} and then recently updated the measurement with 2.4 times more exposure~\cite{BX2013}. The first measurement was based on the data from December 2007 to December 2009, corresponding to the exposure of 252.6\,ton-year after cuts or $1.52 \times 10^{31}$ proton-year. The update from 2013 is based on the data from December 2007 to August 2012, corresponding to the exposure of $(613 \pm 26)$ ton-year or $(3.69 \pm 0.16) \times 10^{31}$ proton-year after the selection cuts.    

\begin{figure}[tb]
\begin{center}
\begin{minipage}[t]{8 cm}
\centering{\epsfig{file=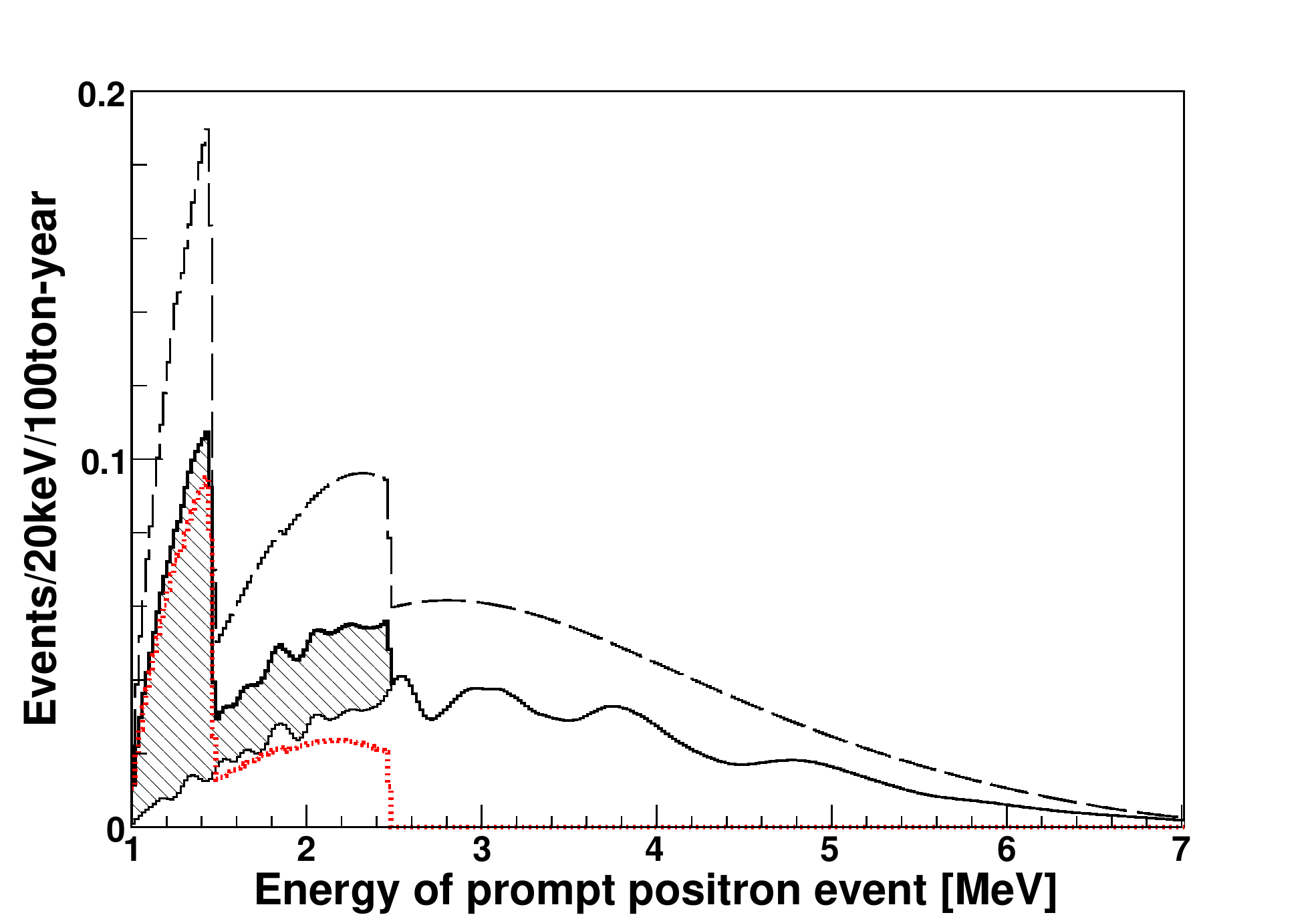,scale=0.43, angle = 90}}
\end{minipage}
\begin{minipage}[t]{8 cm}
\centering{\epsfig{file=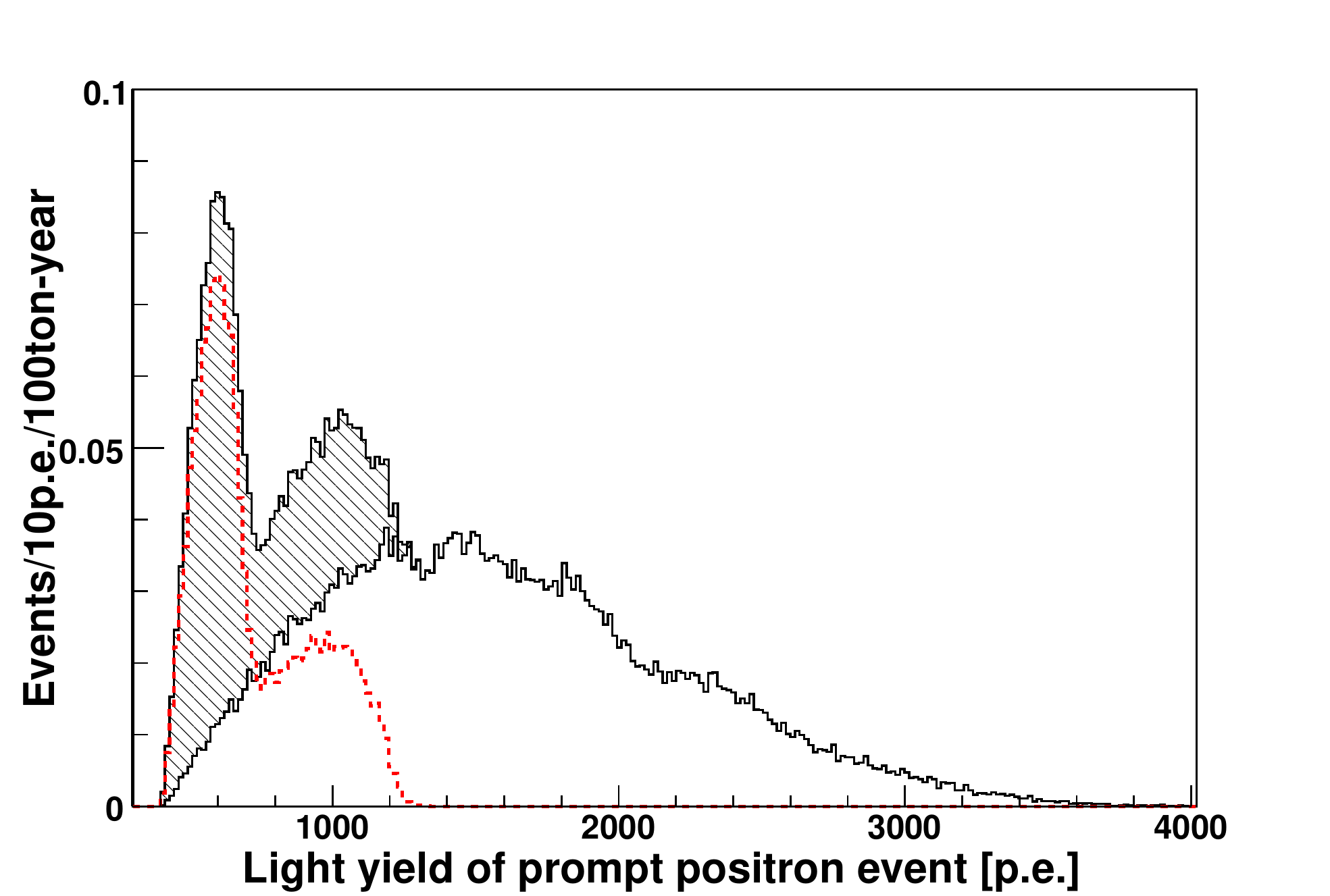,scale=0.45, angle = 90}}
\end{minipage}
\begin{minipage}[t]{16.5 cm}
\caption{Left: the expected energy spectrum of the prompt event (positron) due to the electron antineutrinos in Borexino~\cite{BX2010}.  The dashed/solid black lines show the total geo-neutrino plus the reactor antineutrino spectra without/with oscillations. The dotted red line shows the oscillated geo-neutrino spectrum, while the thin solid line is the oscillated reactor antineutrino spectrum.  The dashed area isolates the contributions of geo-neutrinos in the total oscillated spectrum. Right: Expected prompt positron event spectrum as obtained from the Monte Carlo simulation~\cite{BX2010} expressed in the light yield, e.g. in the number of detected photoelectrons. The spectra from the left part of this Figure are used as input for the Monte Carlo simulation. The event selection criteria described in the text were applied. Approximately, Borexino detects about 500 photoelectrons/MeV. 
\label{Fig:BXspectraAntinu}}
\end{minipage}
\end{center}
\end{figure}

The geo-neutrino and reactor antineutrino spectra expected in Borexino are shown in Figs.~\ref{Fig:BXspectraAntinu}. The left part of this Figure shows the energy spectrum of the prompt candidate, expressed in energy [MeV], without the effect of detector resolution. In the construction of the geo-neutrino spectrum, the Th/U mass ratio was fixed to a chondritic value of 3.9. For comparison, the spectra are shown with and without the neutrino oscillations. As it can be seen, for geo-neutrinos the oscillations change only the absolute normalization of the spectrum, while the spectral shape is not affected. Instead, the spectral shape of reactor antineutrinos is strongly changed by the oscillation phenomenon.

The expected geo-neutrino and reactor antineutrino energy spectra with oscillations are used as input for the Geant-4 based Monte Carlo. From the Monte Carlo output, the expected geo-neutrino and reactor antineutrino spectra, expressed in the number of measured photoelectrons (so called light-yield spectrum shown in right part of Fig.~\ref{Fig:BXspectraAntinu}), automatically incorporate the detector response function. Approximately, Borexino detects about 500\,photoelectrons/MeV. The detector response function was studied considering an extensive calibration campaign~\cite{Back}, see also Sec.~\ref{Subsec:resolutionBX}. These calibration data have been used to reduce the systematic error associated with all Borexino results and to optimize the Monte Carlo simulation of the detector response.

The expected event rate and the spectral shape of reactor antineutrinos were calculated by considering reactors all over the world.  The time variation of the thermal power of individual cores was considered by using the monthly mean load factor, a ratio of the actual and nominal power, provided by the International Atomic Energy Agency. The power fractions of $^{235}$U : $^{238}$U : $^{239}$Pu : $^{241}$Pu used in the calculations were 0.56  : 0.08 : 0.30 : 0.06, while for the European reactors using Mixed Oxide technology were 0.000 : 0.080 : 0.708 : 0.212, and finally for the reactors using heavy water moderator they were 0.542 : 0.411 : 0.022 : 0.0243. The different stages of the burn up process of the fuel contributes to the systematic error at 3.2\%. The matter effect due to the antineutrino propagation through the Earth was estimated to +0.6\%, while the contribution of the long-lived fission products in the spent fuel was set to 1\% based on Kopeikin et al., 2006~\cite{kopeikin}.

In Bellini et al., 2013~\cite{BX2013}, the three flavor neutrino oscillation was considered and the total systematic error on the expected signal was estimated to 5.8\%. For the exposure of $(3.69 \pm 0.16) \times 10^{31}$ proton-year for the period from December 2007 to August 2012, the expected number of events from reactor antineutrinos was $N_{react}$ = $33.3 \pm 2.4$ events corresponding to $90.2 \pm 6.5$\,TNU. The 33.3\% of the reactor antineutrino signal falls within the geo-neutrino energy window (below 1300\,photoelectrons when expressed in light yield). In the absence of neutrino oscillations the number of expected events due to reactor antineutrinos is $60.4 \pm 4.1$.

In Borexino, other (non antineutrino) background sources are reduced to almost negligible levels. In total, Borexino expected $0.70 \pm 0.18$ background events among all antineutrino candidates detected during the exposure of $(3.69 \pm 0.16) \times 10^{31}$ proton-year in the period from December 2007 to August 2012. From these, 65.7\% are expected in the geo-neutrino energy window below 1300 photoelectrons (about 2.6\,MeV). The dominant background sources are due to cosmogenic $^9$Li - $^8$He ($0.25 \pm 0.18$\,events), accidental coincidences ($0.206 \pm 0.004$ events), and events due to ($\alpha$, n) reactions ($0.13 \pm 0.01$ events). The background due to possible untagged muons is at the level of $0.080 \pm 0.007$ events, considering different combinations of neutrons, multiple neutrons, and muons mostly passing the buffer region and producing fake prompt and delayed signals of antineutrino coincidence. Borexino has also identified a background correlated with $^{222}$Rn~\cite{BX2013} having $\tau$ = 5.52 \,days. The $^{214}$Bi($\beta$) - $^{214}$Po($\alpha$) coincidence from the $^{222}$Rn chain has a time constant close to the neutron capture time. The highly ionizing particles, as alpha particles, have the visible energies shifted towards lower values in liquid scintillators and thus, normally, the alpha particles from $^{214}$Po (decay are well below the neutron energy window, so that the $^{214}$Bi($\beta$) - $^{214}$Po($\alpha$) coincidences cannot fake positron – neutron coincidences from antineutrino interaction. However, in $1.04 \times 10^{-4}$ and $6 \times 10^{-7}$ cases, the $^{214}$Po decays to excited states of $^{210}Po$ and the alpha particle is accompanied by the emission of prompt gammas of 799.7\,keV and of 1097.7\,keV, respectively. The signal from a gamma ray is less quenched with respect to the signal from the alpha particle of the same energy, and thus the signal of ($\alpha$ + $\gamma$) corresponds to the higher light yield with respect to pure alpha signal of the same $Q$-value. Such an increased value of visible energy can overlap the low energy tail of a neutron signal, especially at the large radii of the detector. Thus, in the Borexino analysis from Bellini et al., 2013~\cite{BX2013}, the low energy threshold of the neutron energy window was increased with respect to Bellini et al., 2010~\cite{BX2010}, since in this analysis the data set with increased $^{222}$Rn contamination due to the tests of purification of the scintillator have been included. To further suppress this background, a particle identification technique, the so called Gatti filter~\cite{gatti}, was applied in order to provide the distinction of the $^{214}$Po($\alpha$ + $\gamma$) signal from 2.2\,MeV gamma ray from neutron capture.

The selection criteria of the golden antineutrino candidates have been tuned as follows. First, all identified muons are removed from the analysis. A 2\,s veto is applied after each muon passing through the scintillator volume in order to suppress the $^9$Li-$^8$He background. After each muon passing only through the external water tank, a veto of 2\,ms is applied. The total loss of the exposure due to these vetoes is about 11\%. The energy window of the prompt candidate corresponds to the kinematic threshold of the inverse beta decay interaction considering the energy resolution broadening. No upper energy cut has been applied. The energy window of the delayed candidate was tuned to cover the peak of 2.2\,MeV $\gamma$ ray from the neutron capture in the whole FV used in the geo-neutrino analysis. On the delayed signal a slight cut of the Gatti filter was applied as explained above. No Gatti cut is applied on the prompt candidate. The relative time window between the prompt and the delayed signal was required to be between 20 and 1280\,$\mu$s by considering the neutron capture time of ($245.5 \pm 1.8$) $\mu$s~\cite{BXmuons}. The relative distance between the prompt and the delayed signal has to be below 1\,m. The total detection efficiency of these selection criteria was determined through the Monte Carlo simulation to be $0.84 \pm 0.01$. A 25\,cm minimal distance from the inner vessel containing the scintillator is required for the prompt candidate mostly in order to suppress the ($\alpha$, n) background due to the alphas from higher $^{210}$Po contamination in the buffer liquid surrounding the scintillator. The inner vessel shape is reconstructed on a weekly basis by means of events from its radioactive contaminants. The systematic errors of the reconstruction of the vessel shape (1.6\%) and of the position of the prompt candidate (3.8\%) are included together with the 1\% error on the efficiency of the other selection criteria in the overall error of the total exposure.

\begin{figure}[t]
\begin{center}
\centering{\epsfig{file=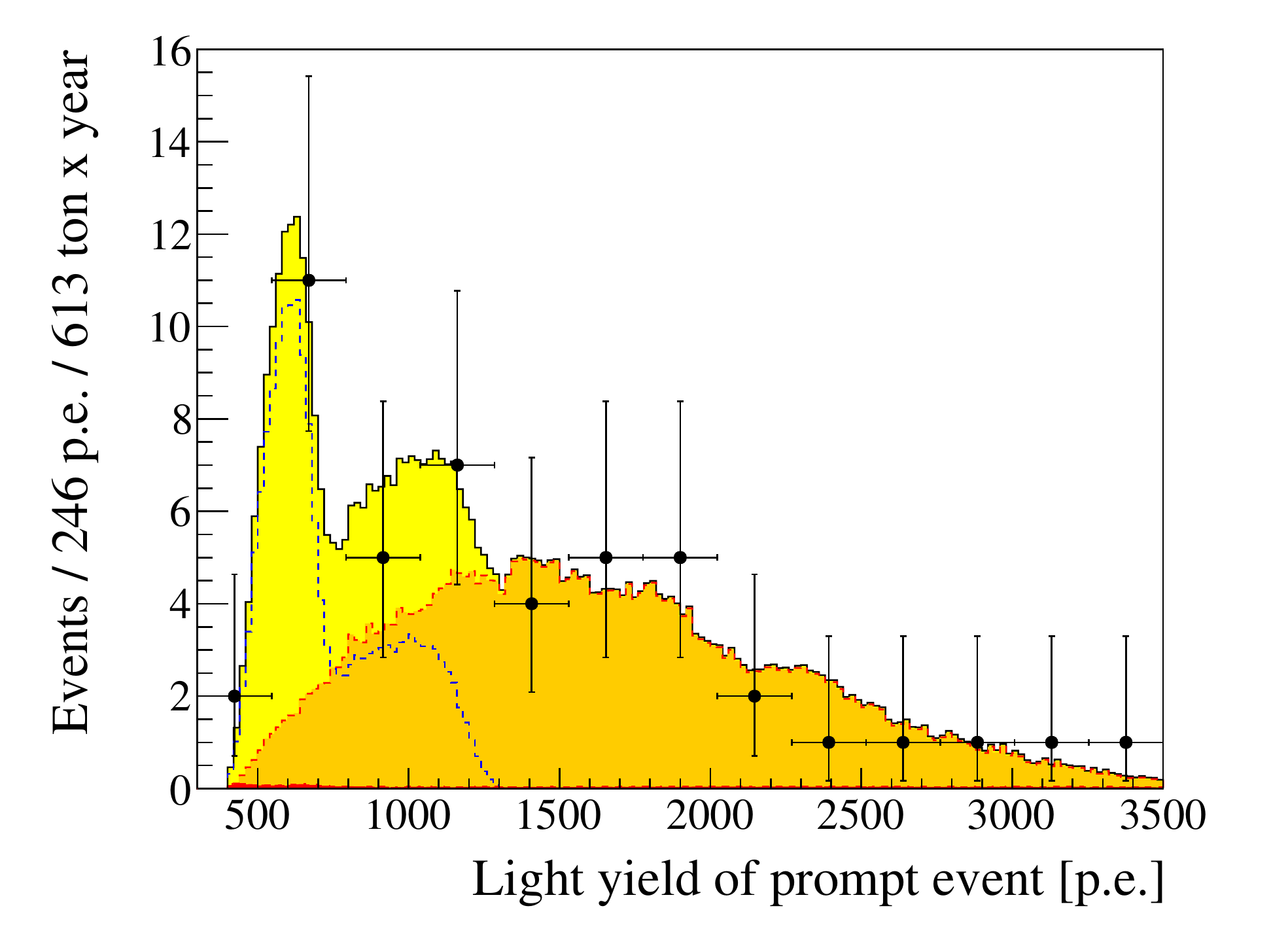,scale=0.5}}
\centering{\epsfig{file=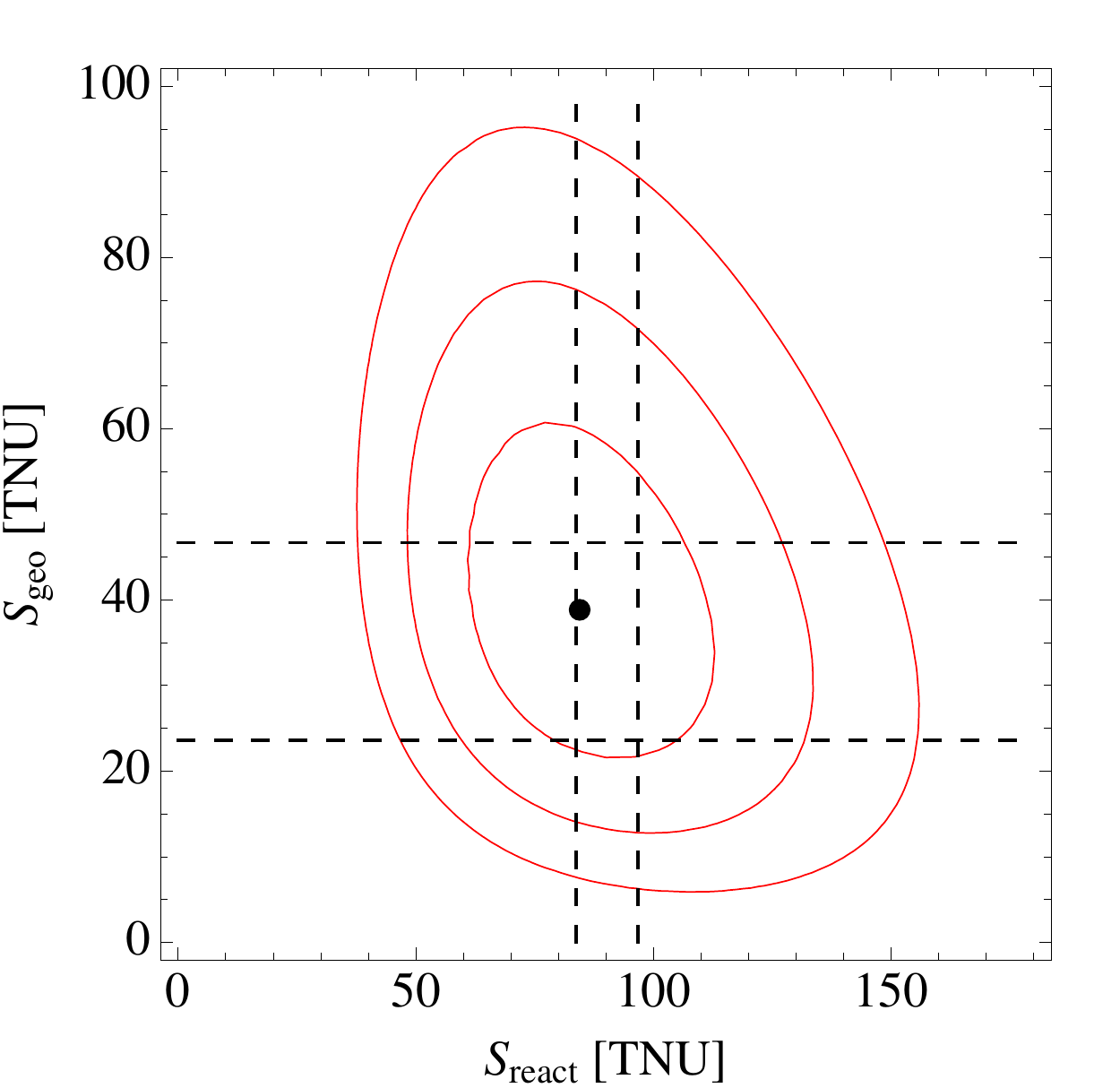,scale=0.6}}
\begin{minipage}[t]{16.5 cm}
\caption{Left: Light-yield spectrum of the 46 prompt events measured by Borexino from Bellini et al., 2013~\cite{BX2013}. The light yield is $\sim$500\,p.e./MeV. The geo-neutrino contribution in the total spectrum is showed in yellow, while the orange color (dashed red line) shows the reactor antineutrino spectrum. The dashed blue line shows the geo-neutrino spectrum. The other background contribution is almost negligible and is shown by a small red filled area in the lower left part. Right: The 68.2, 95.45, and 99.73\% C.L. contour plots from~\cite{BX2013} for the geo-neutrino and reactor antineutrino signal expressed in TNU resulting from the fit shown in left part of Fig.~\ref{Fig:BXSpc2013}. The black point represents the best fit. The vertical dashed lines represent the 1 sigma band of the expected signal from reactor antineutrinos, while the horizontal dashed lines represent the expectations based on different BSE models.
\label{Fig:BXSpc2013}}
\end{minipage}
\end{center}
\end{figure}

Borexino has identified 46 golden candidates passing all selection criteria (from these, 25 in the geo-neutrino energy window) during the exposure of ($613 \pm 26$) ton-year or $(3.69 \pm 0.16) \times 10^{31}$ proton-year after the selection cuts.  (In the 2010 measurement with 2.4 times less exposure, Borexino has detected 21 candidates, from which 15 were in the geo-neutrino energy window). The time and the radial distributions of the candidates are compatible with the expectations. The distribution of the time difference between the delayed and the prompt candidate is compatible with that of the neutron capture time. All prompt events have a negative Gatti parameter, confirming that they are not due to $\alpha$s or protons.

In order to determine the relative contributions of geo-neutrinos, antineutrinos from nuclear power plants, and from other background sources, an unbinned maximal likelihood fit of the light-yield spectrum of prompt candidates in the whole energy range was performed. All 46 candidates in the whole energy range were considered in the fit (the light yield of the prompt event of all detected candidates is below 3500 photoelectrons, so below about 7\,MeV). The contributions of geo-neutrinos and antineutrinos from nuclear power plants were left as free fit parameters without any constrains, using the Monte Carlo functions shown in Fig.~\ref{Fig:BXspectraAntinu} (considering neutrino oscillations) as the probability distribution functions.  The Th/U mass ratio was fixed to a chondritic value of 3.9. The background components were constrained within the $\pm$1$\sigma$ range around the expected values using either measured energy spectra (accidental coincidences), or the Monte Carlo ones ($^9$Li – $^8$He, ($\alpha$, n) background).
The light-yield spectrum of the 46 golden candidates with the best fit and the 68.27\%, 95.45\%, and 99.73\% C.L. contour plots of the geo-neutrino signal $S_{geo}$ and the signal from reactor antineutrinos $S_{react}$ expressed in TNU are shown in Fig.~\ref{Fig:BXSpc2013}, left and right, respectively. The best fit values are $N_{geo} = (14.3 \pm 4.4)$\,events and $N_{react} = 31.2^{+7.0}_{-6.1}$\,events, corresponding to signals $S_{geo} = (38.8 \pm 12.0)$\,TNU and $S_{react} = 84.5^{+19.3}_{-16.9}$\,TNU. This geo-neutrino experimental result can be compared to the expected signal $S_{geo}$(U+Th) = 34.9 $\pm$ 4.7\,TNU calculated by~\cite{coltorti} taking into account a refined model of the local geology of Gran Sasso area described in Sec.~\ref{subsec:lngs}. The measured geo-neutrino signal obtained in this analysis with fixed Th/U ratio corresponds to overall oscillated fluxes from U and Th decay chains of $\phi({\rm U}) = (2.4 \pm 0.7) \times 10^6$\,cm$^{-2}$s$^{-1}$ and $\phi({\rm Th}) = (2.0 \pm 0.6) \times 10^6$\,cm$^{-2}$s$^{-1}$. From the $\ln{\cal{L}}$ profile, the null geo--neutrino measurement has a probability of 6 $\times$ $10^{-6}$.

\subsection{Geological implications of geo-neutrino measurements}
\label{subsec:geoImpl}

The Kamland and Borexino results on geo-neutrinos have an impact on several aspects of our knowledge of the Earth's composition: radiogenic heat, U/Th ratio, radio-nuclides in the mantle, and the existence of U in and around the Earth's core.

{\it {\underline{Comparison of measured geo-neutrino signal with the expectations and the Earth radiogenic heat.}}}

Both Borexino and KamLAND geo-neutrino results are in a fairly good agreement with the geological expectations. This is of an extreme importance for this new interdisciplinary field, confirming both the validity of geological models and the fact that a new tool to study the deep Earth has arisen. This is valid even if the experimental results do not have sufficient precision (mostly limited by statistics) in order to discriminate among different geological models. 

\begin{figure}[t]

\begin{center}
\begin{minipage}[t]{0.49 \textwidth}
\centering{\epsfig{file=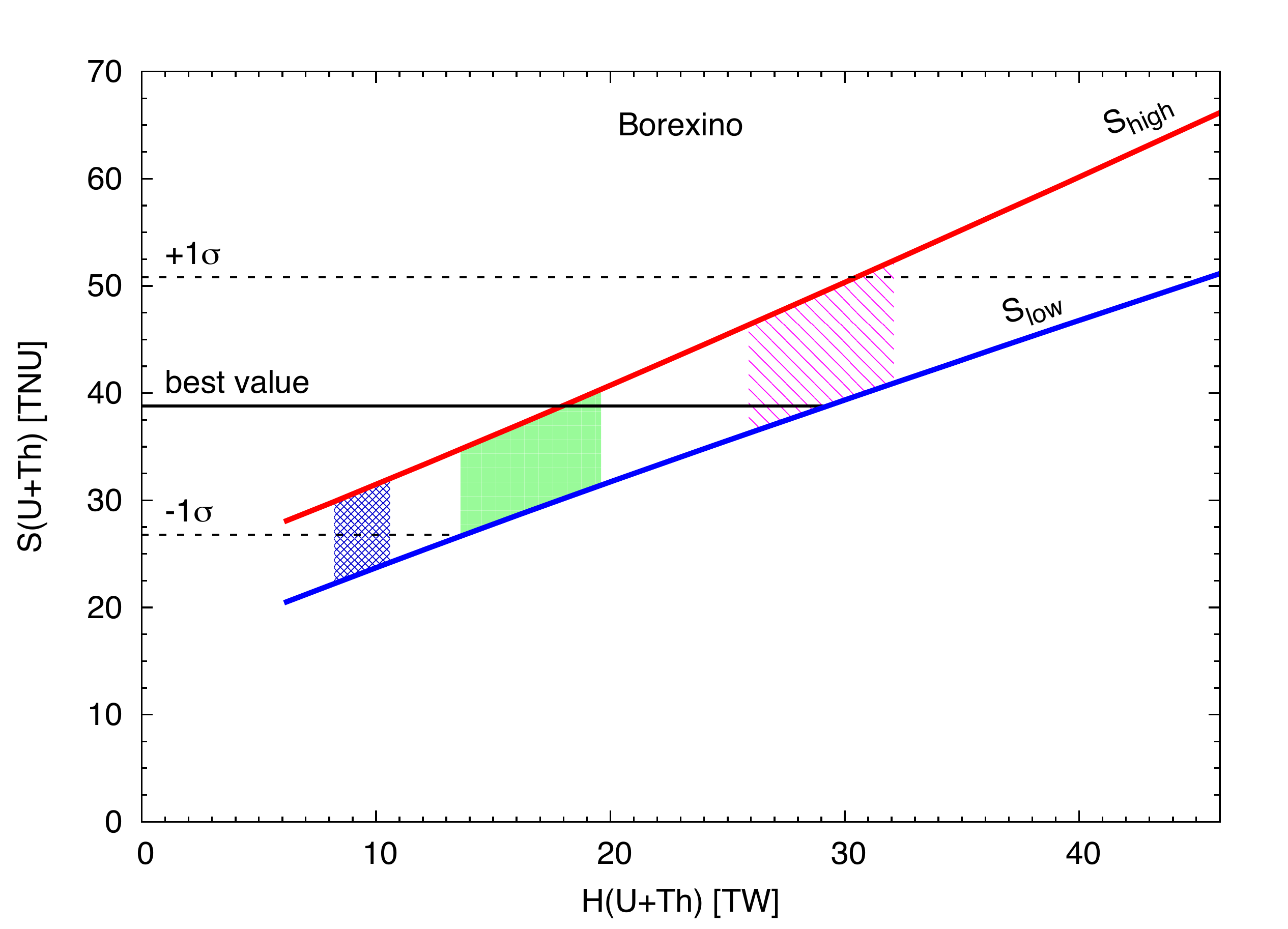,scale=0.37}}
\end{minipage}
\begin{minipage}[t]{0.49 \textwidth}

\centering{\epsfig{file=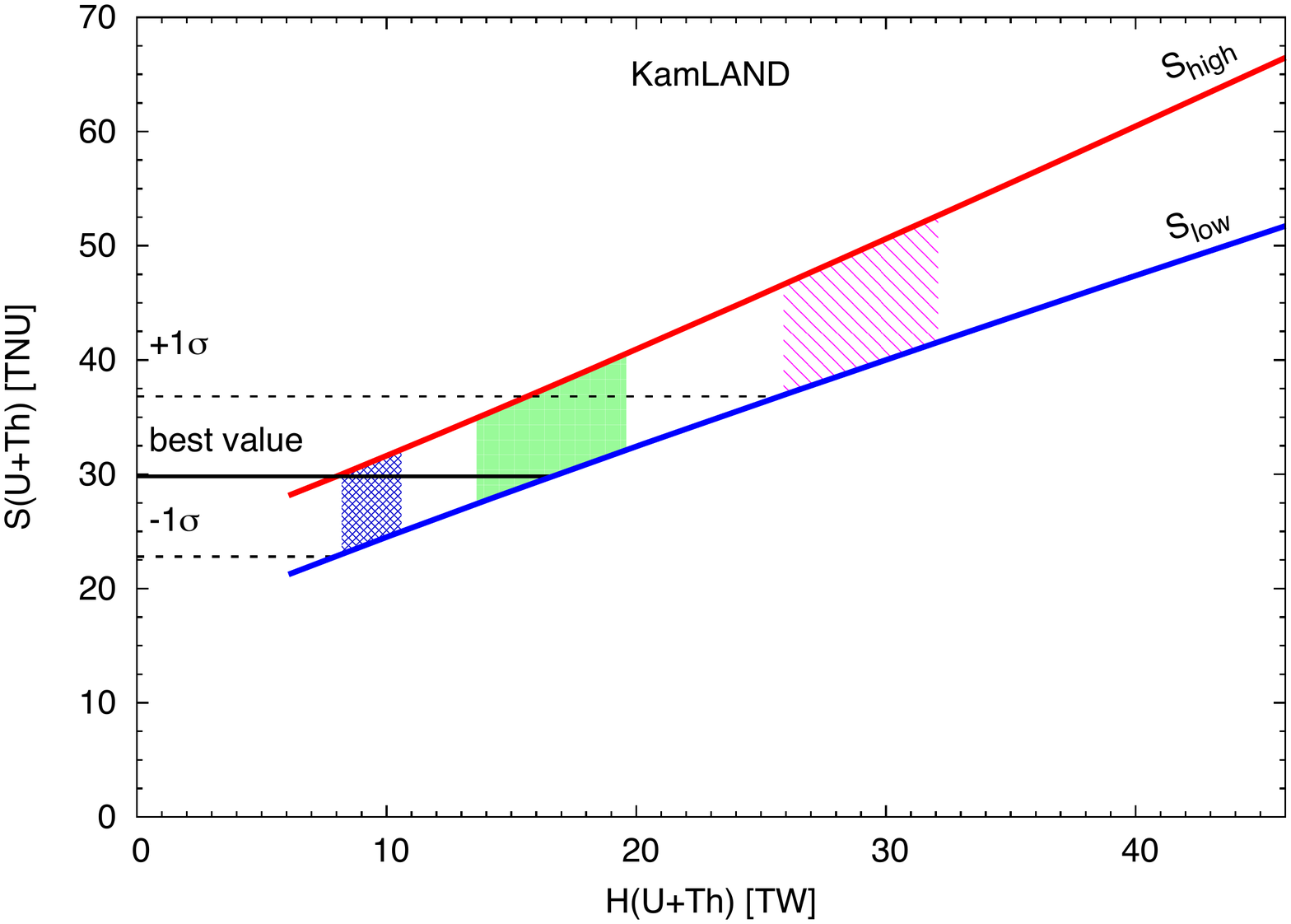,scale=0.37}}
\end{minipage}
\begin{minipage}[t]{16.5 cm}
\caption{The expected geo-neutrino signal in Borexino (left) and in KamLAND (right) from U and Th as a function of radiogenic heat released in radioactive decays of U and Th. The Borexino and KamLAND results from~\cite{BX2013} and \cite{Gando} are indicated by the horizontal lines, respectively. The three filled regions delimit, from the left to the right, the cosmochemical, geochemical, and geodynamical BSE models~\cite{sramek}, respectively.
\label{Fig:GeovsBSE}}
\end{minipage}
\end{center}
\end{figure}

It is not straightforward to extract the radiogenic heat power from U and Th decays from the measured geo-neutrino flux.  As a matter of fact, the measured geo-neutrino signal does not depend only on the absolute mass abundances of U and Th, but also on their distribution throughout the Earth. Therefore, the radiogenic heat power extracted from a measured $S_{geo}$ is model dependent. In Fig.~\ref{Fig:GeovsBSE}  the expected geo-neutrino signal in Borexino (left) and KamLAND (right) are shown, respectively, as a function of the produced radiogenic heat. The red and blue lines consider the high and low models described in Sec.~\ref{subsec:mantle}, where the error in the prediction of the crustal signal is taken into account as well as different U and Th distributions through the mantle, as it was illustrated in Fig.~\ref{Fig:MantleSignal}. The three filled areas in Fig.~\ref{Fig:GeovsBSE}  represent the three classes of BSE models: cosmochemical, geochemical, and geodynamical, according to the classification from \v{S}r\'amek et al., 2012~\cite{sramek}. The horizontal lines represent the 2013 results of Borexino~\cite{BX2013} and KamLAND~\cite{Gando}, respectively. Borexino is compatible with all BSE models within 1$\sigma$, while KamLAND is compatible within 2$\sigma$.

\begin{figure}[h]
\begin{center}
\centering{\epsfig{file=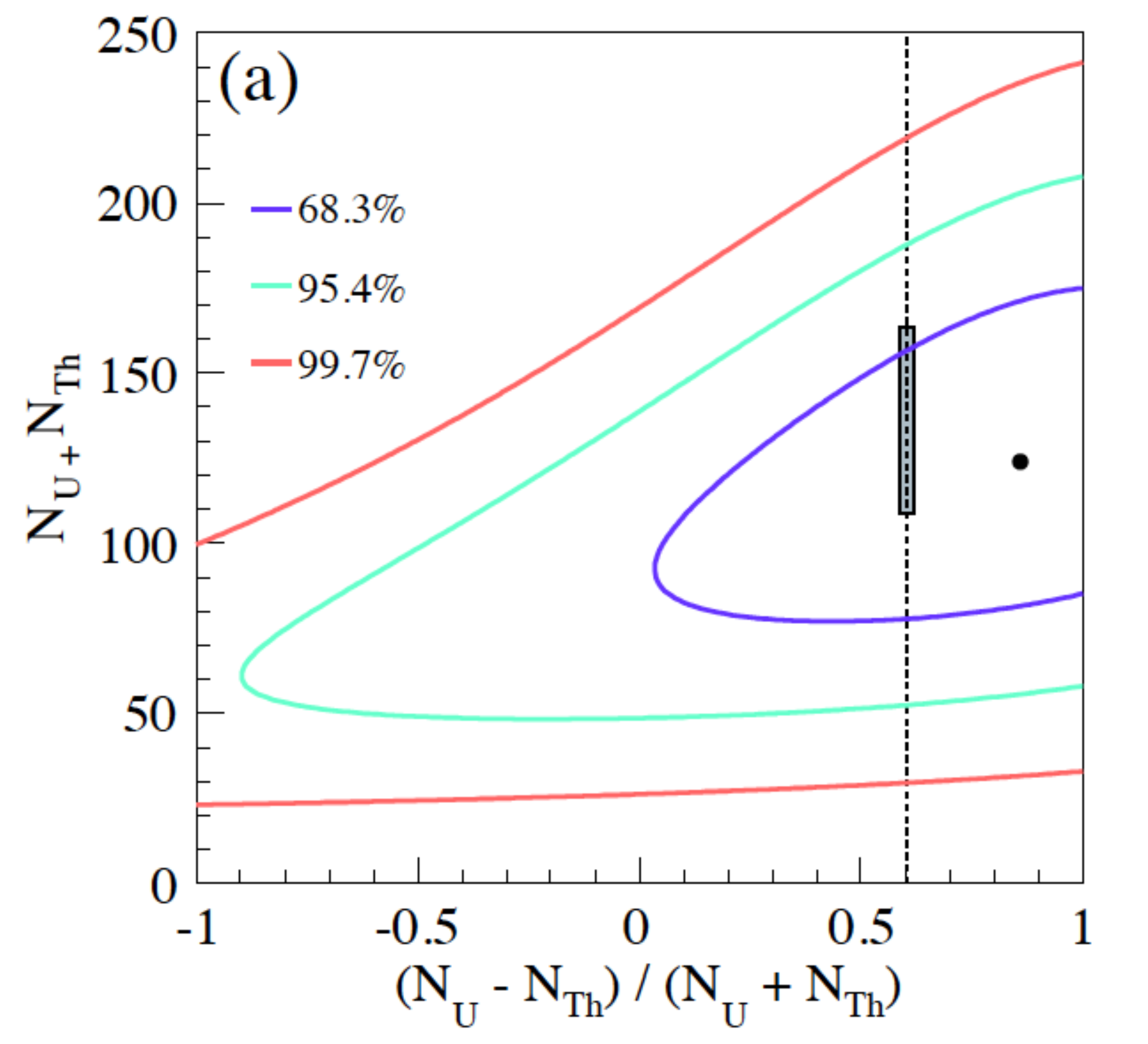,scale=0.35}}
\begin{minipage}[t]{16.5 cm}
\caption{Confidence level contours for the observed number of geo-neutrino events in KamLAND, taken from~\cite{Gando}. The small shaded region represents the prediction of the reference model of~\cite{enomoto}. The vertical dashed line represents the value of $(N_{\rm U} - N_{{\rm Th}}) / (N_{\rm U} + N_{{\rm Th}})$ expected for a Th/U mass ratio of 3.9. 
\label{Fig:KLUTh}}
\end{minipage}
\end{center}
\end{figure}

\begin{figure}[t]
\begin{center}
\centering{\epsfig{file=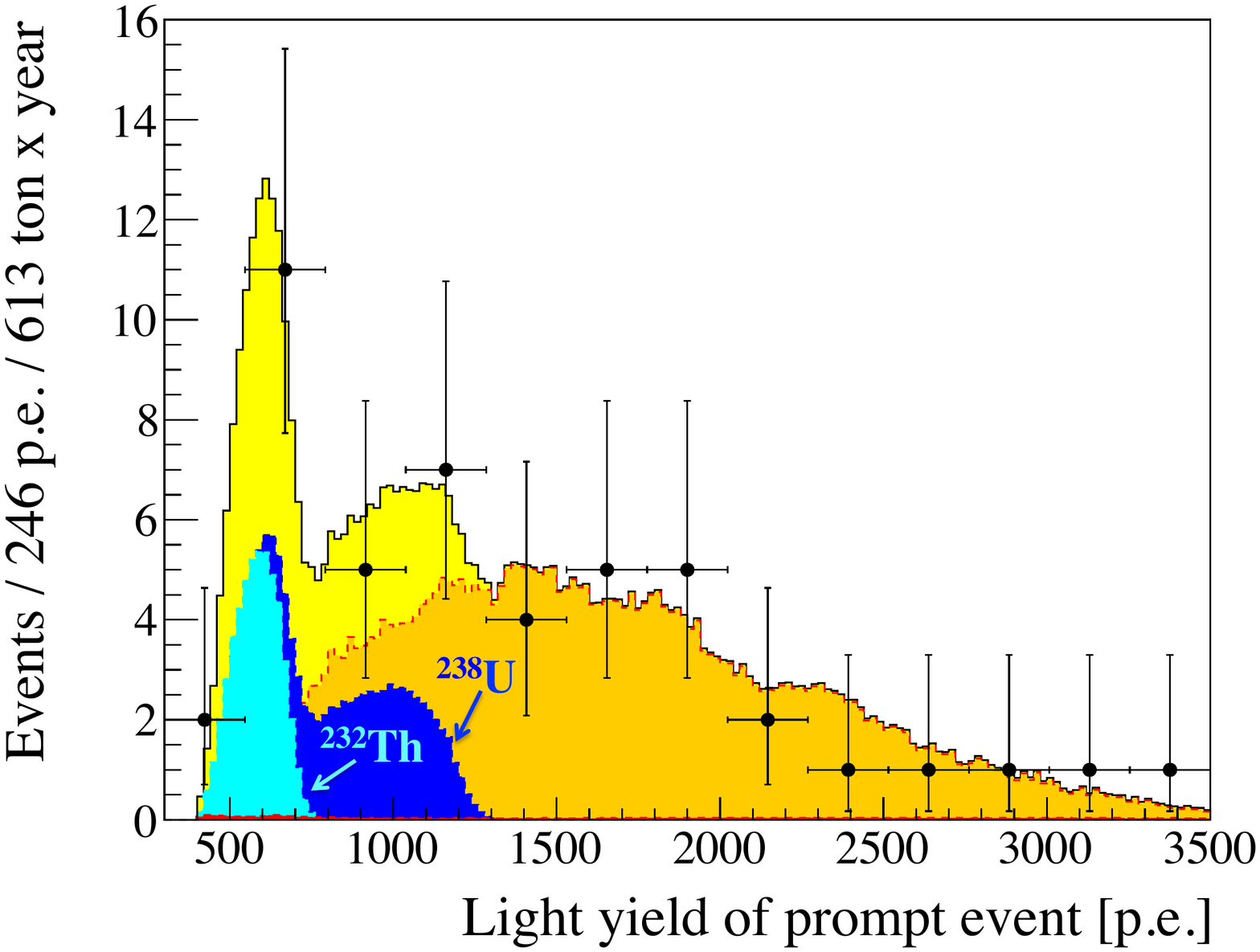,scale=0.5}}
\centering{\epsfig{file=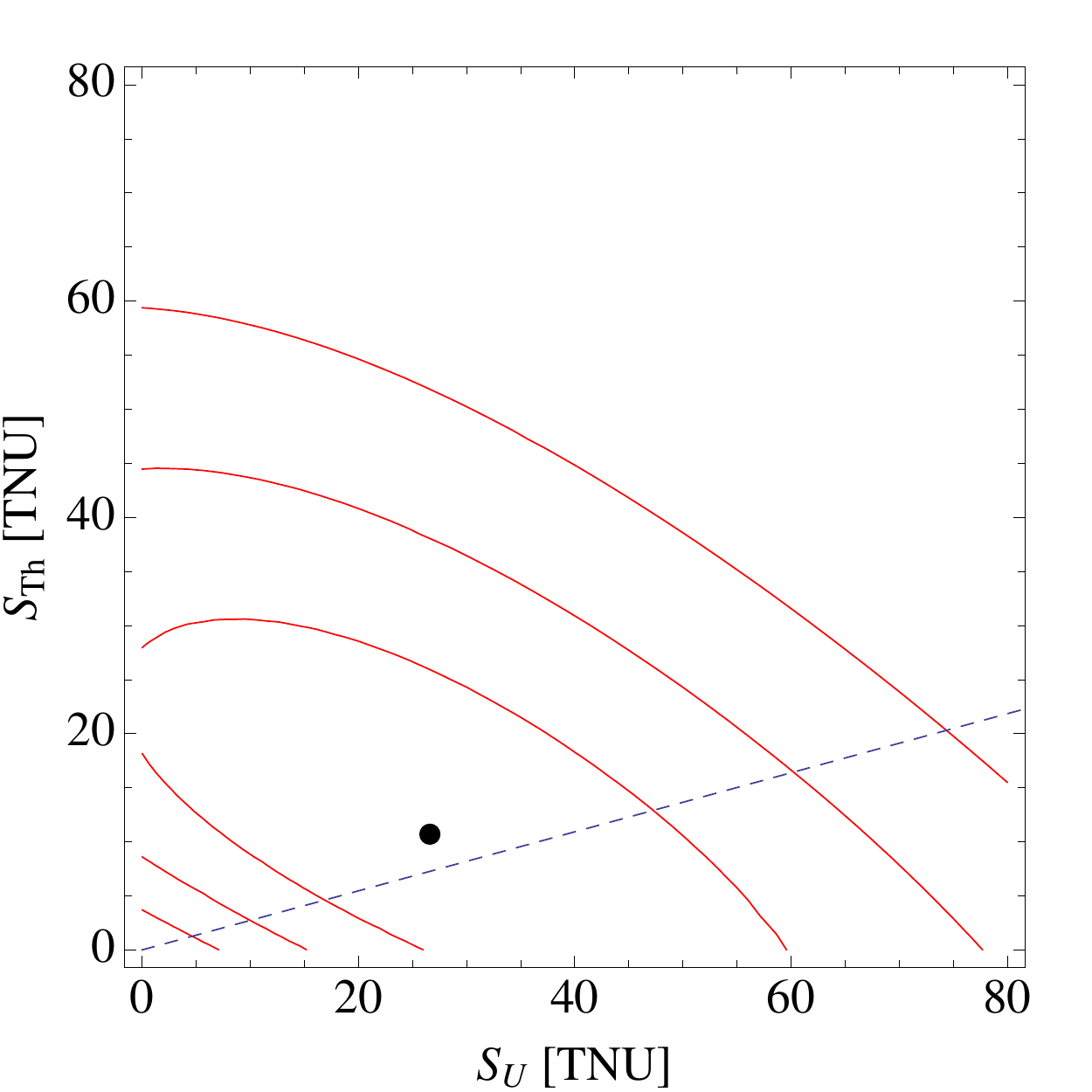,scale=0.6}}
\begin{minipage}[t]{16.5 cm}
\caption{Top: The light-yield spectrum of 46 Borexino antineutrino candidates as in Fig.~\ref{Fig:BXSpc2013}, taken from~\cite{BX2013}. The difference is that the Th and U contributions to geo-neutrino signals are left as free fit parameters and are shown in cyan and blue, respectively. Bottom: The 68.27\%, 95.45\%, and 99.73\% contour plots for the Th and U geo-neutrino signal expressed in TNU, from~\cite{BX2013}. The black dot shows the best fit point.  
\label{Fig:BXUThfree}}
\end{minipage}
\end{center}
\end{figure}

 {\it {\underline{Th/U ratio}}} 

In order to study the individual contributions of U and Th to the total geo-neutrino signal, an unbinned maximal likelihood fit similar to the ones described above can be performed. The only difference is that the Th and U ratio is not fixed according to chondritic mass ratio but both contributions are left as free  individual fit components. 

In Fig.~\ref{Fig:KLUTh} the result of such KamLAND analysis is shown~\cite{Gando}. Here we show the confidence level contours for the sum $N_{\rm U} + N_{{\rm Th}}$ and the asymmetry factor $(N_{\rm U} - N_{{\rm Th}}) / (N_{\rm U} + N_{{\rm Th}})$. The vertical dashed line corresponds to the chondritic Th/U mass ratio. The vertical shaded region shows the prediction from the reference model~\cite{enomoto}. The fit determines an upper limit at 90\% C.L. for the Th/U mass ratio equal to 19. 

In Fig.~\ref{Fig:BXUThfree} the result of a similar Borexino 2013 analysis~\cite{BX2013} is shown. The best fit values are $N_{{\rm Th}} = 3.9 \pm 4.7$\,events and $N_{\rm U} = 9.8 \pm 7.2$\,events, corresponding to signals $S_{{\rm Th}} = 10.6 \pm 12.7$\,TNU and $S_{\rm U} = 26.5 \pm 19.5$\,TNU and the oscillated total fluxes of $\phi({\rm Τh}) = (2.6 \pm 3.1) \times 10^6$\,cm$^{-2}$ s$^{-1}$ and $\phi({\rm U}) = (2.1 \pm 1.5) \times 10^6$\,cm$^{-2}$s$^{-1}$. Although these data are compatible within 1$\sigma$ with either only $^{238}$U signal (and $S_{{\rm Th}}$ = 0) or only $^{232}$Th signal (and $S_{{\rm U}}$ = 0), the best fit of the Th/U ratio is in very good agreement with the chondritic value.

{\it {\underline{Mantle geo-neutrinos}}}

The measured geo-neutrino signal has its component from the crust and mantle, while no contribution is expected from the core, as discussed in Sec.~\ref{Sec:GeoModels}. Therefore, by subtracting the relatively well known crustal contribution from the total measured signal, it is, in principle, possible to extract the mantle signal.

Borexino alone has inferred the mantle contribution to be $15.4 \pm 12.3$\,TNU by subtracting the crustal contribution of $23.4 \pm 2.8$\,TNU from its measured signal~\cite{BX2013}. By assuming a homogeneous mantle and thus the same signal from the mantle geo-neutrinos on the Earth surface, the Borexino and KamLAND results from Gando et al., 2011~\cite{gando2011} have been combined in Bellini et al., 2013~\cite{BX2013} and the mantle geo-neutrino signal of $14.1 \pm 8.1$\,TNU has been extracted. Using KamLAND 2013 data and subtracting out the crust contribution determined by the reference model from Enomoto et al., 2007~\cite{enomoto}, in the hypothesis that U and Th are uniformly distributed throughout the mantle, the total mantle radiogenic heat production is calculated to be $11.2^{+7.9}_{-5.1}$\,TW~\cite{Gando}.

{\it {\underline{Georeactor}}}

The hypothesis of a georeactor present in the very innermost core of the Earth was described in Sec.~\ref{Sec:GeoModels}. Both Borexino and KamLAND experiments have been able to test this hypothesis based on their geo-neutrino data. No positive evidence of its existence has been found. 

Borexino sets the upper limit on the power of a georeactor with the composition of $^{235}U : ^{238}U \simeq 0.76 : 0.23$~\cite{herndon1} to 4.5\,TW at 95\% C.L. The analysis is performed by adding a Monte Carlo spectrum corresponding to this hypothetical georeactor in yet another maximal likelihhod fit and by constraining the signal from the nuclear power plants to the expected value of $33.3 \pm 2.4$\, events.

KamLAND 2013 data have been also used to search for a georeactor. For the georeactor a fission ratio $^{235}$U:$^{238}$U $\simeq 0.75:0.25$ is used~\cite{Gando}. In particular, using a constrain on the oscillation parameters, that is on antineutrinos from nuclear power plants, and setting free the contribution from geo-neutrinos and from the georeactor, an upper limit for the power of this latter is determined to be $< 3.7$\,TW at 95\% C.L.

\section{Future prospects}
\label{Sec:Future}

Geo-neutrinos have been measured with high statistical significance by two different experiments placed in two different geological settings on two sides of the globe. Both experiments have seen the signal, which is in agreement with the geological expectations. Unfortunately, the existing results are not sufficient in order to firmly discriminate among several geochemical and geophysical models. The first attempts of combined analysis have arisen and have shown the importance of multi-site measurements. The first indications of measurements of geo-neutrinos from the mantle, the indicative exclusion of the fully radiogenic Earth model, the invalidation of the georeactor in the Earth's core with power greater than few TW, and the indication of chondritic Th/U ratio, are examples of the first geologically important results of this new inter-disciplinary field of Neutrino Geoscience. All of these measurements need further confirmations with much higher statistical significance. This means that the future projects having geo-neutrinos among their scientific goals should be detectors even bigger than the current ones, at the scale of several to few tens of kton. Another key point is the selection of the geological setting for the future experiment. 

The most exciting question is the measurement of the geo-neutrino signal from the mantle, as it is discussed in Secs.~\ref{Sec:GeoModels} and \ref{Sec:GeoSignal}. Ideally, such an experiment should be placed there where the crustal contribution is minimal and easily estimated. This is the case of the ocean floor, where the signal is expected to be largely dominated by the mantle geo-neutrinos. 
Another interesting point is to test if the mantle is homogeneous or not. There have been recent seismic-tomographic measurements proving that a large inhomogeneity~\cite{wang, wen} does exist below Africa and below Pacific. However, it is not clear if they are also related to compositional inhomogeneity. Geo-neutrinos are recently a unique tool able to obtain information on this problem, if measured at several locations distributed around the globe~\cite{sramek}. 

In this Section we shortly describe the main future projects having geo-neutrinos among their scientific goals.  

\subsection{SNO+ (Sudbury Neutrino Observatory+)}
\label{subsec:SNO}

SNO+~\cite{maneira,chen} is a revised version of the SNO detector, which had an important role in fixing the Solar Neutrino Problem, and thus in studying the neutrino oscillations. Its structure is based upon an active volume, a 12\,m diameter acrylic sphere, which in SNO was filled with $\sim$1000\,tons of heavy water, replaced in SNO+ with $\sim$780\,tons of liquid scintillator. The active volume is shielded by ultra pure water inserted between the acrylic sphere and 9000 read out photomultipliers, which take place within the scintillator, and between the geodesic structure, supporting the PMTs, and the walls of a cylindrical water tank. The shielding water is in total $\sim$7000\,tons; it suppresses or reduces the radiations emitted by the rocks and by the construction materials of the detector itself.

\begin{figure}[h]
\begin{center}
\centering{\epsfig{file=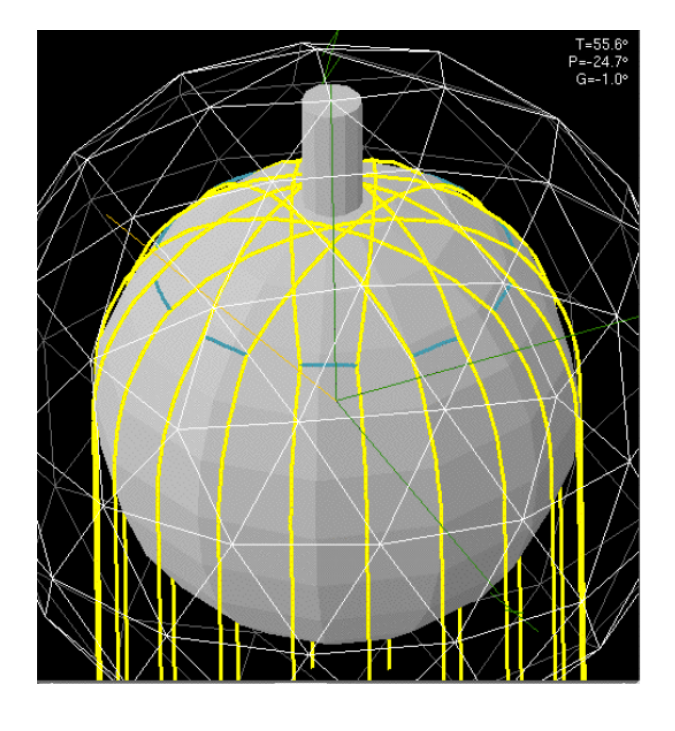,scale=1.2}}
\begin{minipage}[t]{16.5 cm}
\caption{Sketch of the SNO+ detector~\cite{maneira}.
\label{Fig:SNO+}}
\end{minipage}
\end{center}
\end{figure}

The liquid scintillator is the CH$_2$ based linear alkylbenzene with the addition of a PPO fluor. The energy resolution is expected to be $\sim$5\% at 1\,MeV and 3.5\% at $\sim$3\,MeV. Due to the scintillator density, 0.86 with respect to water, a rope system will be installed to hold down the acrylic vessel, which replaces the old support ropes. The scintillator will be purified to reach a good radiopurity (the goal would be $\sim$$10^{-17}$\,g/g), but because the radiopurity of the vessel is not very high, a definition of a fiducial volume, lower than the scintillator sphere, seems needed. The detector is installed underground with an overburden of 6080\,m water equivalent. We do not discuss here the insertion in the scintillator of a  $0\nu \beta \beta$ decay candidate as tellurium for a double beta decay study of SNO+. In Fig.~\ref{Fig:SNO+} a sketch of the SNO+ layout is shown~\cite{maneira}.

The ability of SNO+ to study the geo-neutrinos can be summarized as follow: the geo-neutrino detected rate will be in principle 20 per year, efficiencies included, possible fiducial volume effect excluded; the reactor antineutrinos flux at the Sudbury site is limited: $\sim$44.3\,TNU (geo-neutrino expected rate: $51 \pm 10$\,TNU)), to be compare with Kamioka: $\sim$152\,TNU ($34 \pm 14$\,TNU expected geo-neutrinos) and with Gran Sasso: $\sim$23.1\,TNU ($41 \pm 8$\,TNU geo-neutrinos)~\cite{fiorentini2007}.
 
The expected geo-neutrino signal in SNO+ is dominated by the contributions from the crust and the lithospheric mantle of the Canadian Shield. According to Huang et al.~\cite{huang} the lithosphere produces a total signal $S_{LS}({\rm U+Th} )$ =  36.7$^{+7.5}_{-6.3}$\,TNU, which comes mainly from U and Th of the local Precambrian rocks and Paleozoic sediments. \v{S}r\'amek~\cite{sramek} recently proposed an exhaustive analysis of geo-neutrino rates from different mantle structures based on three classes of BSE models (i.e. cosmochemical, geochemical and geodynamical). A present depleted mantle produces in Sudbury the minimum geo-neutrino signal, corresponding to 2.3 - 3.7 TNU. This contribution is less than 10\% of the crustal geo-neutrinos and it will be unlikely investigable in SNO+. On the other hand, a mantle model having a high Urey ratio (e.g. 0.6 - 0.8) could be an intensive source of geo-neutrinos, which can reach rates of 11.2 - 16.1\,TNU in Sudbury. Such high contribution will not be hidden even in SNO.

On this ground the regional contribution to the geo-neutrino flux and its uncertainties needs to be determined including the geological, geochemical and geophysical information of the Canadian Shield. On the base of the refined reference crustal model~\cite{huang} the U and Th in the crust of $6^{\circ} \times 4^{\circ}$ region, surrounding the detector, gives a signal of $S_{LOC}({\rm U+Th})$ = $18.9^{+3.5}_{-3.3}$\,TNU, which is more than that expected by the whole mantle based on a cosmochemical BSE model. The main reason of such high contribution is the crustal thickness ranging between 44.2\,km and 41.4\,km: this reservoir is approximately 40\% thicker than the crust surrounding the Gran Sasso and Kamioka sites. Moreover, excluding a thin layer ($<$3\,km) of Paleozoic sedimentary rocks southward the Grenville front, a large portion of crystalline basement (e.g. Grenville Province, Yavapai and Mazatzal Terranes) contains significant quantities of felsic rocks, which are enriched in U and Th comparing to most other lithologies. A possible enhancement of crustal geo-neutrinos is due to high concentration of U and Th in the Sudbury basin as reported in~\cite{perry} on the base of geothermal arguments. In particular these authors focus on heat flux anomaly of 43 - 60\,mW m$^{-2}$ which is significantly higher than the average for the Canadian Shield (42\,mW m$^{-2}$). Assuming a homogeneous Moho heat flux throughout the Canadian Shield~\cite{mareschal} the higher heat flux measured in this area can be explained as a local enrichment in crustal radio-elements within a 50\,km radius from SNO Lab. It could strongly affects the geo-neutrino signal expected in SNO+. In this framework a detailed calculation of the local geo-neutrino flux, relying on direct summation of the individual contribution of all the geological units, is desirable and partially anticipated by Huang~\cite{huang}.
SNO+ is expected to start the data taking in 2014-1015.

\subsection{LENA (Low Energy Neutrino Astronomy)}
\label{subsec:LENA}

Lena is an ambitious proposal for a large, 50 kton liquid scintillator detector~\cite{laguna} having the geo-neutrino measurement among one of their scientific goals. The project is a part of the European LAGUNA design study and it identifies itself as a multipurpose neutrino observatory. A combination of an unprecedented volume with the high radio-purity comparable with that, which has been reached by Borexino, would provide a unique tool for a wide variety of measurements both in neutrino physics and in testing possible physics beyond the Standard Model.

The detector should be a vertical cylinder containing the target of 100\,m height and 26\,m diameter. The liquid scintillator would be separated from a 2\,m thick not scintillating buffer liquid by a thin nylon vessel. The scintillation light would be viewed by 30 to 50,000\,PMTs mounted on a steel cylinder containing the organic liquids. In the 100\,kton water Cherenkov detector, a muon veto system would surround the steel cylinder, being equipped with about 3000\,PMTs, and an array of plastic scintillator panels necessary for the reconstruction of muon tracks in such a huge volume. The mostly discussed possible future locations are Pyh\"asalmi in Finland and Fr\'ejus in France.  

Considering the $2^{\circ} \times 2^{\circ}$ crustal map and the mean chemical compositions of the main crustal layers, and the BSE models for the estimation of the mantle contribution, the expected geo-neutrino signal at Pyh\'asalmi is $51.3 \pm 7.1$\,TNU, while for the Fr\'ejus site is $41.4 \pm 5.6$\,TNU. In both cases, a detailed analysis of the contribution from the local crust surrounding the detector would be important in order to further constrain the expected signal and to better interpret the possible measured signal.

	LENA would detect about 1000 geo-neutrino events per year.  The main antineutrino background, namely that from the nuclear power plants, would be several times smaller in the Finland location: this would make it a preferable location from the point of view of the geo-neutrino measurement.  In the geo-neutrino energy window the expected reactor antineutrino signal would be about 20 to 37\,TNU (depending on the construction of several new power plants in Finland), while it would be about 145\,TNU at Fr\'ejus, based on the thermal power of nuclear power plants as reported in 2009.  

	Assuming $2.9 \times 10^{33}$ target protons, a light yield of 400\,photons/MeV, the chondritic Th/U mass ratio of 3.9, and the Borexino radiopurity (no-background approximation), the geo-neutrino flux would be determined with a few percent precision within the first few years, an order of magnitude improvement than the current experimental results. Thanks to the high statistics, LENA would be able to measure the Th/U ratio of the local geo-neutrino signal with unprecedented precision, reaching, after 3 years, 10-11\% precision in Phyahsalmi and 20\% precision in Fr\'ejus.

\subsection{Hanohano}
\label{subsec:Hanohano}

Hanhano is a proposed 10\,kton liquid scintillator detector designed to be deployed in the deep ocean at 3 to 5\,km depth~\cite{learned}. A tank of 26\,m in diameter and 45\,m tall would be placed vertically on a 112\,m long barge with 32\,m beam. This proposal aims to measure geo-neutrinos and to have a potential to measure neutrino mass hierarchy if the  neutrino mixing angle is relatively large (as it was recently proven to be~\cite{daya}).

Since the oceanic crust is thin and depleted in U and Th with respect to continental crust (on which all other existent and proposed projects able to measure geo-neutrinos are placed), the dominant contribution of about 75\% of a measured geo-neutrino signal would come from the mantle.  Thanks to a large volume the expected rate would be about 100 detected geo-neutrinos per year. Since the ocean sites are far away from nuclear power plants, only about 12 reactor antineutrinos per year would be detected, so the signal-to-background ratio would be high. This would in turn make possible to measure also the Th/U ratio of the measured mantle signal at the level of 10\% precision within few years. In addition, this detector should be portable in a sense that after an operation at one site, it could be brought to the surface and transported to a new site where it would be again lowered to ocean floor. As extensively supported by recent papers~\cite{Sramek2013, Dye2012, jocher} a movable deep-ocean detector could be the next challenge for measuring anthropic and terrestrial antineutrinos, especially for testing mantle lateral homogeneity in composition and the thermochemical evolution of the Earth.

\section*{Acknowledgments}

The authors wish to thank Ved Lekic for discussions on seismology and his production of a great figure, Kristi Engel for production of figures and support on edits, Kunio Inoue for discussion on KamLAND and Mark Chen on SNO+ project. We express our gratitude for useful discussions to G. Fiorentini, Y.~Huang, and O. \v{S}r\'amek. In addition, the authors acknowledge the Istituto Nazionale di Fisica Nucleare and the National Science Foundation (i.e., NSF EAR0855791, EAR-1067983, and EAR1321229) for their support. Finally, the authors are grateful to the Borexino and KamLAND collaborations which kindly allowed the use of figures from their documents and publications in this work.

\end{document}